\documentclass{svjour3}                     
\smartqed  
\usepackage{graphicx,amssymb,amsmath,multirow,array,tabularx}
\makeatletter
\renewcommand{\thefigure}{\ifnum \c@section>\z@ \thesection.\fi
 \@arabic\c@figure}
\@addtoreset{figure}{section}
\makeatother

\makeatletter
\renewcommand{\thetable}{\ifnum \c@section>\z@ \thesection.\fi
 \@arabic\c@table}
\@addtoreset{table}{section}
\makeatother

\makeatletter
\renewcommand{\theequation}{\ifnum \c@section>\z@ \thesection.\fi
 \@arabic\c@equation}
\@addtoreset{equation}{section}
\makeatother

\renewcommand{\thesection}{\Roman{section}}    

\begin{document}

\title{Shear-induced transitions and instabilities in surfactant wormlike micelles}


\author{Sandra Lerouge       \and
        Jean-Fran\c cois Berret 
}


\institute{S. Lerouge, J.-F. Berret \at
              Laboratoire Mati\`ere et Syst\`emes Complexes (MSC), UMR 7057 CNRS-Universit\'e Paris-Diderot, 10 rue Alice Domon et L\'eonie Duquet, F-75205 Paris Cedex 13, France \\
              \email{sandra.lerouge@univ-paris-diderot.fr; jean-francois.berret@univ-paris-diderot.fr}    
}
\date{Received: date / Accepted: date}

\maketitle

\tableofcontents

\vspace{0.5cm}

\begin{abstract}
In this review, we report recent developments on the shear-induced transitions and instabilities found in surfactant wormlike micelles. The survey focuses on the non-linear shear rheology and covers a broad range of surfactant concentrations, from the dilute to the liquid-crystalline states and including the semi-dilute and concentrated regimes. Based on a systematic analysis of many surfactant systems, the present approach aims to identify the essential features of the transitions. It is suggested that these features define classes of behaviors. The review describes three types of transitions and/or instabilities : the shear-thickening found in the dilute regime, the shear-banding which is linked in some systems to the isotropic-to-nematic transition, and the flow-aligning and tumbling instabilities characteristic of nematic structures. In these three classes of behaviors, the shear-induced transitions are the result of a coupling between the internal structure of the fluid and the flow, resulting in a new mesoscopic organization under shear. This survey finally highlights the potential use of wormlike micelles as model systems for complex fluids and for applications.
\end{abstract}

\section*{Abbreviations and notations}
\begin{tabbing}
*******************\=*********************\kill
Al(NO$_3)_3$\>aluminum nitrate\\
AlCl$_3$\>aluminum chloride\\
CP/Sal\>cetylpyridinium salicylate\\ 
CPCl\>cetylpyridinium chloride\\
CPClO$_3$\>cetylpyridinium chlorate \\
C$_8$F$_{17}$\>perfluorooctyl butane trimethylammonium bromide \\
C$_{12}$E$_{5}$\>penta(ethylene glycol) monododecyl ether\\
C$_{12}$TAB\>dodecyltrimethylammonium bromide \\
C$_{14}$TAB\>tetradecyltrimethylammonium bromide \\
C$_{14}$DMAO\>tetradecyldimethylamine oxide  \\
C$_{16}$TAB\>hexadecyltrimethylammonium bromide  \\
C$_{16}$TAC\>hexadecyltrimethylammonium chloride\\
C$_{18}$TAB\>octadecyltrimethylammonium bromide \\
C$_{18}$-C$_{8}$DAB\>hexadecyloctyldimethylammonium bromide \\
C$_{n}$TAB\>alkyltrimethylammonium bromide\\
CTAHNC\>cetyltrimethylammonium 3-hydroxy-2-naphthalenecarboxylate\\
CTAT\>hexadecyltrimethylammonium \textit{p}-toluenesulfonate \\
CTAVB\>cetyltrimethylammonium benzoate\\
Dec\>decanol\\
DJS\>diffusive Johnson-Segalman \\
DLS\>dynamic light scattering\\
DR\>drag reduction\\
EHAC\>erucyl bis(hydroxyethyl)methylammonium chloride \\
FB\>flow birefringence\\
FI\>Faraday instability\\
Gemini 12-2-12\>ethane diyl-1,2-bis-(dodecyl dimethylammonium bromide)\\
Hex\>hexanol\\
HPC\>hydroxypropyl cellulose\\
I/N\>isotropic-to-nematic\\
KBr\>potassium bromide\\
LAPB\>laurylamidopropyl betaine \\
LSI\>light scattering imaging\\
LCP\>liquid crystalline polymer\\
NaCl\>sodium chloride\\
NaClBz\>sodium chlorobenzoate\\
NaClO$_3$\>sodium chlorate\\
NaNO$_3$\>sodium nitrate\\
NaSal\>sodium salicylate \\
NaTos\>sodium \textit{p}-toluenesulfonate or sodium tosylate\\
NH$_4$Cl\>ammonium chloride\\
NMR\>nuclear magnetic resonance\\
PBLG\>poly(benzyl-L-glutamate)\\
PEO\>poly(ethylene oxide)\\
PIV\>particle image velocimetry\\
PTV\>particle tracking velocimetry\\
SANS\>small-angle neutron scattering\\
SALS\>small-angle light scattering\\
SAXS\>small-angle X-ray scattering\\
SDBS\>sodium dodecyl benzyl sulfonate\\
SDES\>sodium dodecyl trioxyethylene sulfate\\
SdS\>sodium decylsulfate\\
SDS\>sodium dodecyl sulfate \\
SIP\>shear-induced phase\\
SIS\>shear-induced structure\\
TTAA\>tris(2-hydroxyethyl)-tallowalkyl ammonium acetate \\
USV\>ultrasonic velocimetry\\

\end{tabbing}
\section{\label{part0}Introduction}
\indent

Wormlike micelles are elongated and semiflexible aggregates resulting from the self-assembly of surfactant molecules in aqueous solutions. Wormlike micellar solutions have received considerable attention during the past decades because of their remarkable structural and rheological properties.

Sixty years from now, Debye and his group  in Cornell had undertaken an extensive study of surfactant solutions using the light scattering technique. The goal of these investigations was the measurement of the dissymmetry of scattered light, in order to gain information regarding the molecular weight and thereby the shape of surfactant aggregates. The dissymmetry of scattered light was defined as the intensity ratio at two scattering angles far apart from each other. Would this ratio be one, the micelles were assumed to be spherical ; would it increase, the micelles were assumed to grow in size. In a famous paper, Debye and Anacker had discovered that the addition of an inorganic salt, potassium bromide to aqueous solutions of hexadecyltrimethylammonium bromide caused the colloidal aggregates to increase in size \cite{Deby51}. Based on these dissymmetry experiments, it was suggested that the micelles undergo a morphological transition, from spherical aggregates at low salt content to rodlike aggregates at high salt content. More than half a century later, the very same systems, now known as wormlike micelles continue to attract interest from a broad scientific community. 

Going from the structure to the rheology was not a straightforward path. 
One important contribution after that of Debye  was that of Nash who mapped viscoelastic regions of surfactant solutions using again hexadecyltrimethylammonium bromide and various naphtalene derivatives. The viscoelasticity was determined visually by looking at how fast a swirl applied by hand to a solution decayed with time \cite{Nas58}. Some years later, Gravsholt established that other additives, such as salicylate or chlorobenzoate counterions could be solubilized by the micelles and promote efficiently their uniaxial growth \cite{Gra76}. It was proposed that the viscoelasticity of these solutions had the same origin as that of polymer solutions, namely entanglements and reptation.

\begin{figure*}[t]
\begin{center}
\includegraphics[scale=0.15]{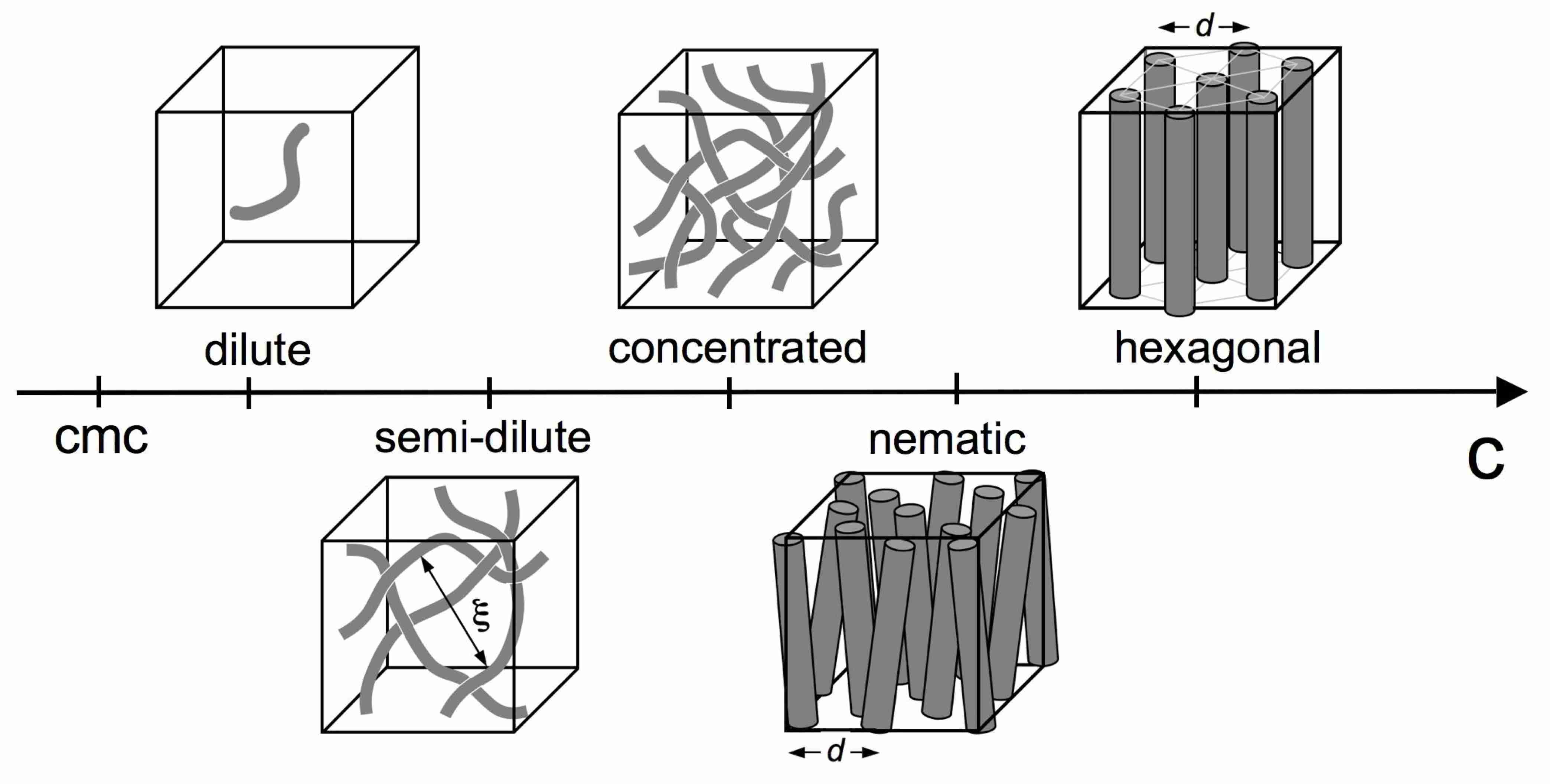}
\caption{Illustrations of the different concentrations regimes encountered in wormlike micellar solutions with increasing concentration. $\xi$ is the mesh size of the entangled network in the semidilute regime and $d$ denotes the average distance between colinear micelles in the concentrated isotropic, nematic and hexagonal phases. An estimate of $d$ can be gained from the position of the structure peak in the scattering function.}
\label{figI1}
\end{center}
\end{figure*}

In the early 80's, as more and more groups were involved in this research, discoveries were made at a faster pace. By a combination of light scattering, rheology and magnetic birefringence, it was first shown that under certain conditions, cylindrical micelles could be very long, up to micrometer in contour length, and flexible \cite{Port80,Port81,Iked84,Cand85,Imae84,Imae85,Imae86}. 
The terminology introduced was that of giant \cite{Port81,Port84,Por86} or worm-like \cite{Iked84,Imae84,Imae85,Imae86} micelles, instead of rodlike aggregates some years before. For several surfactants, Ikeda and collaborators reported electron microscopy images showing threadlike and tortuous filaments, later referred to as worms \cite{Imae84,Imae85}. Again using light scattering experiments, Candau and his group demonstrated the existence of a cross-over between dilute and semidilute regimes and of scaling laws as a function of the concentration, two features that were known from polymers \cite{Cand85,Can84,Cand88}. These authors pointed out a formal analogy between surfactant wormlike micelles and polymer solutions. This analogy was completed by the extensive investigations of phase behaviors of surfactant aqueous solutions, and the evidence of isotropic-to-nematic and a nematic-to-hexagonal transitions at high concentrations \cite{Por86,Ama87,Gom87,Ama88,Her88}. Fig.~\ref{figI1} provides a schematic  illustration of the different concentration regimes that will be surveyed in the present review. The analogy with polymers, as well as a marked viscoelasticity attracted attention from rheologists, who were at first interested in the linear mechanical response of these fluids. 

A decisive step towards the description of the micellar dynamics was taken with the first quantitative measurements of the linear viscoelastic response of these solutions. The pioneering works were those of Rehage, Hoffmann, Shikata and Candau and their coworkers \cite{Cand88,Hoff85,Shik87,Rehage88,Shik88,Shi88,Shik89,Cat90a,Shik90,Ker91,Reh91,Cla92,Kha93a,Kha93,Ber93,Ber94}. The most fascinating result was that the viscoelasticity of entangled wormlike micelles was characterized by a single exponential in the response function. The stress relaxation function $G(t)$ was found of the form $G(t)=G_0\exp{(-t/\tau_{\scriptscriptstyle{R}})}$  over a broad temporal range, where $G_0$ denotes the elastic modulus and $\tau_{\scriptscriptstyle{R}}$ is the relaxation time. Since then, this property was found repeatedly in semi-dilute wormlike micellar solutions. 
This rule has become so general that it is now recognized that a single relaxation time in the linear rheology is a strong indication of the wormlike character of self-assembled structures. A simple viscoelastic behavior, together with the fact that micellar solutions are easy to prepare, and not susceptible to aging or degradation have incited several groups to utilize wormlike micelles as reference for the testing of new rheological techniques \cite{Smo2003,Bell2002,Buc2005,Wil2007,Capp2007,Raud2008}.

On the theoretical side, the challenge was to account for this unique time of the mechanical response. This was done by Cates and coworkers in the late 1980's with the reptation-reaction kinetics model \cite{Cat88}. The reptation-reaction kinetics model is based on the assumption that the breaking and recombination events of the chains are coupled to the reptation \cite{Doi86} and as such accelerate the overall relaxation of the stress. In the fast breaking limit, a given micelle undergoes several scission and recombination reactions on the time scale of the reptation. Thus, all initial deformations of the tube segments relax at the same rate, this rate being driven by the reversible scission. \\

In the present review, we focus on the shear-induced transitions and instabilities that were disclosed in wormlike micellar systems during the last decade or so. The thermodynamics, structure and rheology of the aggregates at rest or under small deformation were reviewed many times in the past \cite{Hoff85,Cat90a,Reh91,Lequ94,Wal2001,Berret}, and they will not be treated here. Our survey of the non-linear rheology covers now all concentration regimes, from the dilute to the liquid-crystalline states, and including the semidilute and concentrated regimes (Fig.~\ref{figI1}). The present approach aims to demonstrate that the features of the shear instabilities are specific to a concentration regime. Sometimes, the characteristics of a transition extend over a broader concentration range. This is the case for the shear-thickening that was evidenced in the dilute and semi-dilute regimes.  Another goal is to establish correspondences between the shear-induced and the equilibrium phases. A good illustration is that of the isotropic-to-nematic transition, for which the induced nematic exhibits the same orientation and rheological properties than the nematic phase found in the equilibrium phase diagram at high volume fraction.

\indent

\begin{figure*}[t]
\begin{center}
\includegraphics[scale=0.12]{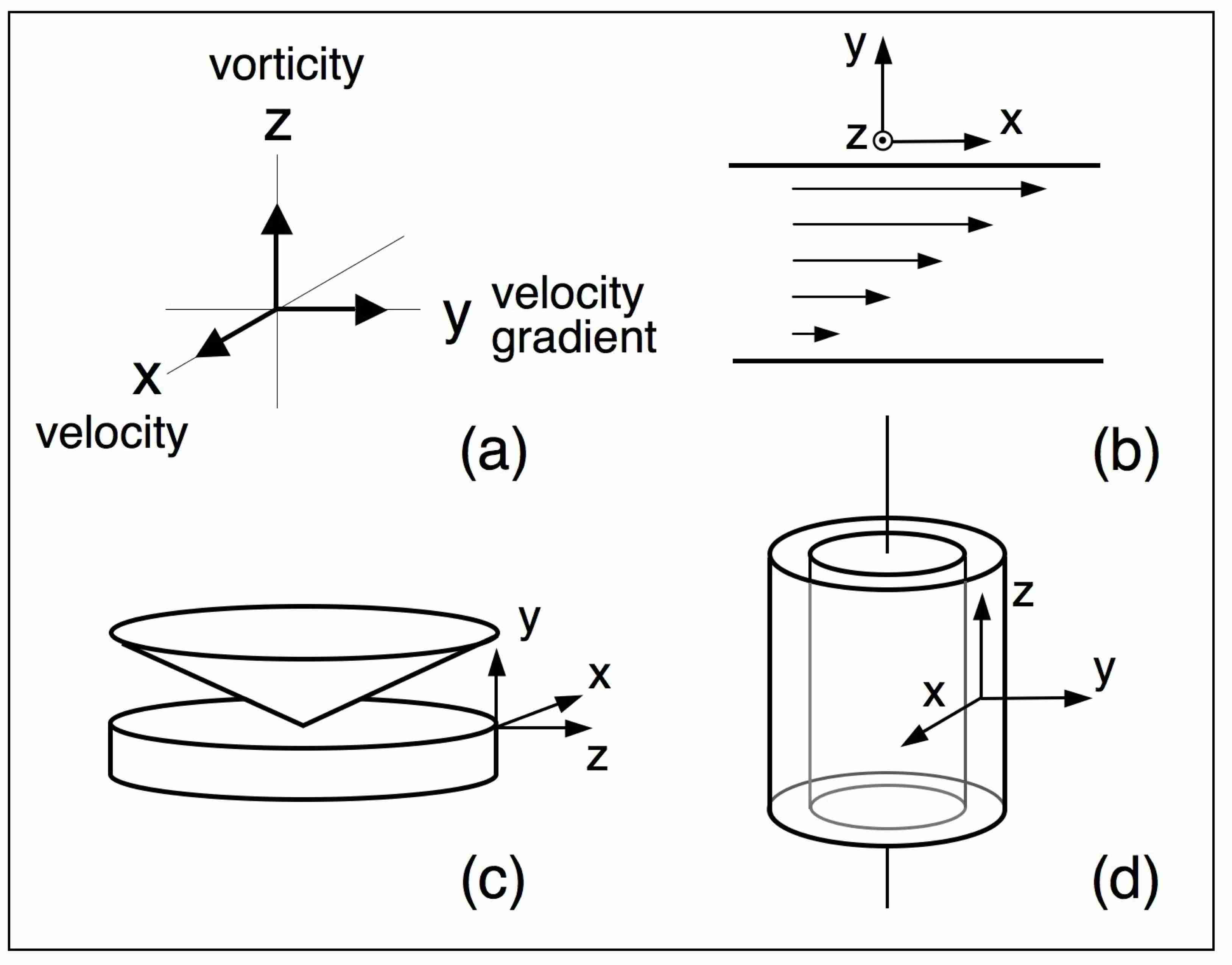}
\caption{(a) Conventions adopted in this review for the velocity, velocity gradient and vorticity axes characterizing shear flows. Most common geometries of shearing devices producing shear flows : (b) planar Couette ; (c) cone-and-plate, (d) cylindrical Couette.}
\label{figI2}
\end{center}
\end{figure*}

Although elongational flows have been also imposed to semi-dilute solutions \cite{Prud94,Wal96,Che97,Rot2003,Yes2006}, the review will focus essentially on shear flows. The most common devices for shear are the cylindrical Couette and the cone-and-plate geometries (Fig.~\ref{figI2}). 
In a cylindrical Couette system, the sample is contained between two concentric cylinders and the shear is applied by rotating the inner or outer cylinder. If the inner cylinder is rotated, inertia effects may cause a transition from laminar flow to Taylor vortex flow at high shear rates \cite{Swin86}. As for the results discussed in the review, the Reynolds numbers remained below or even much below than that of the Taylor instability. Compared to Couette systems, cone-and-plate devices have more uniform stress, provided the cone angle is small. Fig.~\ref{figI2} also specifies the geometrical conventions used throughout the review for the velocity, velocity gradient and vorticity directions. Another convention concerns the surfactant concentrations. Due to the fact that wormlike micelles were studied by scientists from various research fields, the surfactant concentrations appear in the literature with different units, including molar concentration, volume fraction, weight percent or weight per volume. In order to allow comparison between different surfactant systems, we have adopted the following rule. We have kept the units used by the authors in reference to their work, and we have added, when necessary, the value of the weight percent concentration $c$ (units wt.~\%). For ternary systems made of a surfactant and an additive (this additive being an hydrotope, a cosurfactant or an alcohol molecule), $c$ denotes the total surfactant and additive concentration. The molar ratio between additive and surfactant is expressed as $R$.

The review is organized as follows : Part \ref{part1} deals with the shear-thickening behavior found in dilute and very dilute surfactant solutions. Part \ref{part2} examines the shear-banding instability and the isotropic-to-nematic transition revealed in the semi-dilute and concentrated regimes, respectively. The last Part focuses on the wormlike micellar nematics under shear, and emphasizes the analogy with liquid-crystalline polymers. 
\section{\label{part1}Shear-thickening in dilute micellar solutions}
\subsection{\label{introst}Introduction}
\indent

\begin{figure*}[t]
\begin{center}
\includegraphics[scale=0.1]{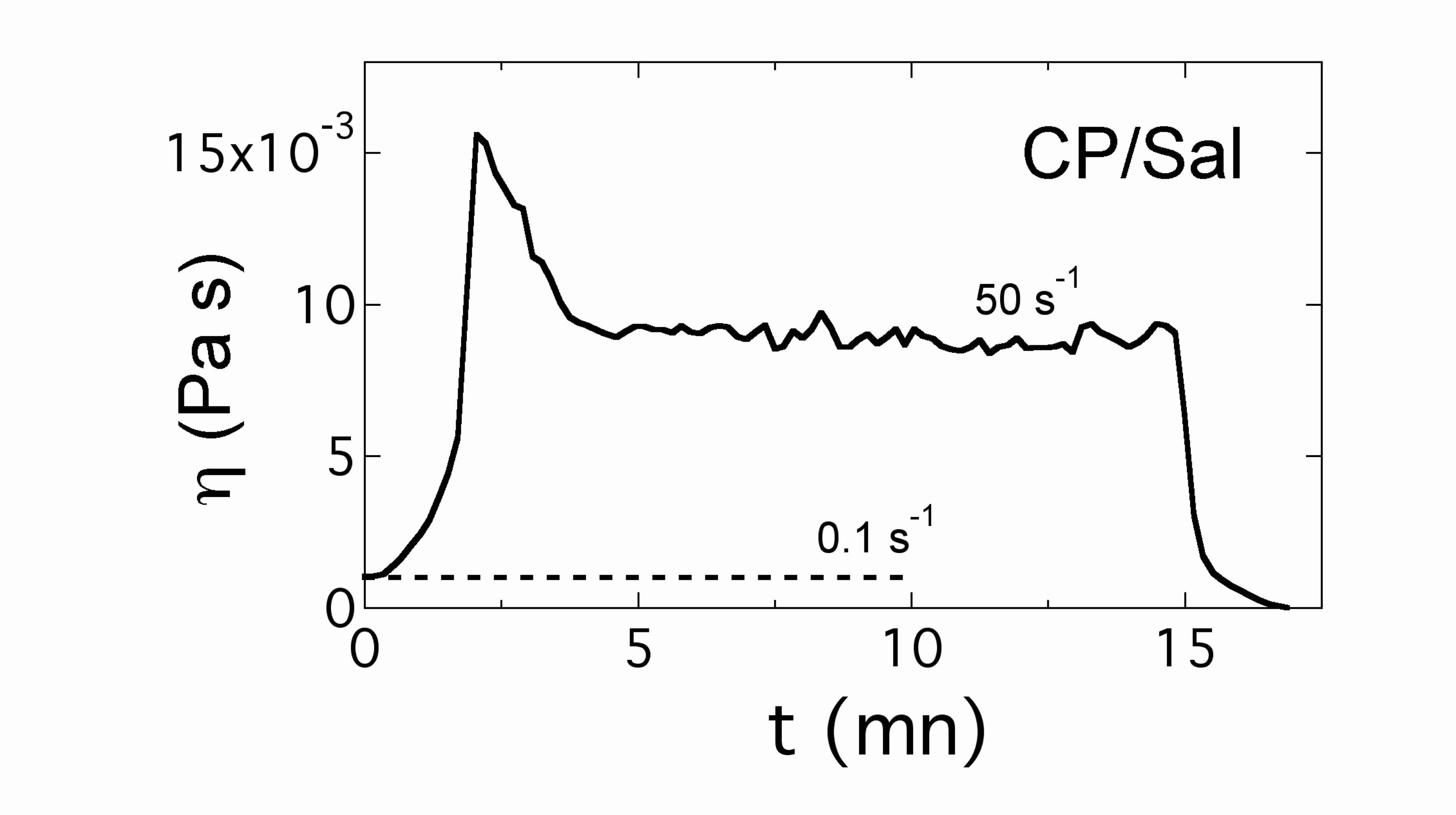}
\caption{Shear viscosity as a function of time for equimolar cetylpyridinium salicylate (CP/Sal) at 0.9 mM ($c=0.04$ wt. \%) and temperature $T=20$\r{ }C. The transient response at $\dot\gamma=0.1$ s$^{-1}$ exhibits a regular steady behavior for such dilute solution, whereas the response at 50 s$^{-1}$ shows a period of induction during which the viscosity increases and then stabilizes around 10 times the viscosity of water. Figure reprinted with permission from Ref.~\cite{Reh82}.}
\label{figII1}
\end{center}
\end{figure*}

Among the rich variety of shear-induced instabilities and transitions encountered in surfactant systems, one of the most puzzling is the shear-thickening effect observed in dilute or very dilute solutions. This transition was first noticed by Rehage and Hoffmann \cite{Hof81a} in 1981 for the system cetylpyridinium salicylate (CP/Sal) at a molar concentration of 0.9 mM ($c=0.04$ wt.~\%). In their original work, the shear stress was recorded as a function of time over several minutes, and it revealed an unexpected behavior. Above a critical shear rate, the transient stress exhibited a period of induction during which the viscosity increased and then stabilized around 10 times the viscosity of water. This early evidence of the shear-thickening has been reproduced in Fig.~ \ref{figII1}.  The phenomenon was explained by postulating the formation of a supramolecular structure during flow \cite{Hof81a,Reh82}. 

In the same decade, shear-thickening solutions have attracted much interest because of their potential applications in fluid mechanics. In a number of practical situations such as fire-fighting operations, transportation of fluids along cylindrical pipes, turbulence occurs near the solid surface and increases the energy losses associated to the flow. It was suggested that additives dispersed in water e.g. polymers or surfactant could diminish considerably the turbulent skin friction \cite{She84}. Bewerdorff and coworkers have shown that some surfactants reduced effectively the friction factor in turbulent pipe flows \cite{Bew86,Ohl86,Lin90}. Using tetradecyltrimethylammonium with salicylate counterions at concentrations as low as 0.1 wt.~\%, these authors were able to correlate the drag reduction to the increase of viscosity at high shear rates. Since these early studies, the interest for this transition has increased, especially in the context of the study of complex fluids under shear flow. 

In this part, we are dealing with the shear-thickening transition in dilute surfactant solutions. Only solutions with zero-shear viscosity close to that of the solvent and for which no apparent viscoelasticity is observed at rest will be considered. Systems showing both viscoelasticity and shear-thickening have been also found out, and will be evoked in the part devoted to the semi-dilute regime. 

The features of the shear-thickening transition are summarized as follows~:
\begin{enumerate}
\item Shear-thickening occurs for surfactants that self-assemble into cylindrical micelles. 
\item Under steady shear, above a critical shear rate, the shear viscosity increases as a new and more viscous phase develops. This shear-induced state is called SIS for shear-induced structure \cite{Ohl86} or SIP for 
shear-induced phase \cite{Bol97a} in the literature. 
\item An induction time is necessary to induce the SIS. At steady state, the stress displays fluctuations that are larger than the instrumental noise response. 
\item In the shear-induced state, the solutions are birefringent and exhibit a strongly anisotropic scattering in light and neutron experiments. This anisotropy is compatible with a strong alignment of the shear-induced 
structures in the flow.
\item The shear stress \textit{versus} shear rate curve depends on the geometry of shearing cells, and also of the thermal and shear histories experienced by the fluid prior to rheological testing. 
\end{enumerate}
In the following (section \ref{sts}), we will first provide a list of surfactant molecules that exhibit shear-induced structures in accordance with points 1-5, and then describe the phenomenology of the transition. The rheology (section II.3), the orientation properties (section II.4) and the structure of the flow field (section II.5) for these fluids will then be examined.
\subsection{\label{sts}Shear-thickening surfactants}
\indent

In comparison with the total number of surfactants available by now, only a few of them exhibit a shear-thickening transition in accordance with points 1-5 above. Table \ref{tabstk} lists these systems, with their counterions and/or their hydrotopes. In Table \ref{tabstk}, it can be seen that all the surfactants are cationic and that the number of carbon atoms ranges from 13 to 18 per elementary charge. Most of the rheological studies have been performed on systems of the alkyltrimethylammonium bromide class (C$_{n}$TAB), using strongly binding counterions or hydrotopes. Well-known examples of hydrotopes are salicylate, \textit{p}-toluenesulfonate and chlorobenzoate, which all contain an aromatic phenyl group. 
\begin{table}[t]
\caption{List of surfactants and additives found to exhibit shear-thickening described by points 1-5. Column 4 makes an inventory of the experimental techniques employed to investigate the SIS. (*) For the system hexadecyltrimethylammonium \textit{p}-toluenesulfonate without monovalent counterions, the notation CTAT is prefered to C$_{16}$TAB/Tos. (a) \cite{Hof81a,Reh82,Sun2003}, (b) \cite{Bew86,Ohl86,Lin90,Wun87,Koc98,Deh2007}, (c) \cite{Ohl86,Koc98,Wun89,Hu93a,Liu96,Vas2008,Cre98,Har98,Har97a,Ros89,Hu94,Har97,Kim2000,Kim2001,Hu2002}, (d) \cite{Har97a,Har97,Gam99,Tru2002,Berret2002,Ber2000a,Berret2001,Ban2000,Mac2003,Tor2007}, (e) \cite{Lu98,Qi2002}, (f) \cite{Deh2007}, (g) \cite{Hof91,Hu93,Pro97}, (h) \cite{Sch95a,Oda97,Oda2000,Web2002,Web2003,Oel2002,Oda96,Oda98}, (i) \cite{Bol97a,Hu98a,Mys94,Bol97,Hu98}, (j) \cite{Oel2002}.}
\label{tabstk}       

\begin{tabular}{ccccc}
\hline\noalign{\smallskip}
Surfactant & Additive & Salt & Experiment & refs. \\
\noalign{\smallskip}\hline\noalign{\smallskip}
CPCl 						 & NaSal 	& 	&	 FB &(a)			 \\
C$_{14}$TAB 		 &	NaSal	&	  & SANS, FB, DR &(b) \\
C$_{16}$TAB 		 & NaSal 	& 	& FI, FB, DR, SALS, PIV &(c) \\
C$_{16}$TAB 		 & NaTos (*)	& 	&	SANS, FB &(d) \\
C$_{16}$TAC 		 & NaClBz	& 	& DR &(e)\\
C$_{18}$TAB 		 & NaSal 	& 	& & (f) \\
C$_{14}$DMAO 		 & SDS 		& 	& SANS, FB, SALS & (g) \\
Gemini 12-2-12   & 				& Br$^{-}$	& SANS, FB, SALS &(h) \\
TTAA 						&	NaSal 	&		& LSI, PIV, DR & (i)\\
C$_8$F$_{17}$  	&					& Br$^{-}$&	& (j)\\
\noalign{\smallskip}\hline
\end{tabular}
\end{table}

In Table \ref{tabstk}, for sake of simplicity, we mentioned the abbreviations of the surfactants and hydrotopes with their monovalent counterions. Tetradecyltrimethylammonium bromide with sodium salicylate thus becomes in short C$_{14}$TAB/NaSal. In some cases \cite{Hof81a,Reh82,Wun87,Koc98}, the small monovalent counterions have been removed by ion exchange procedures, yielding a surfactant salt that is now abbreviated C$_{14}$TA/Sal \cite{Wun87}. In the following, the abbreviations will take into account these variations. Systems with hydrotopes were ge\-ne\-ral\-ly prepared at equimolar 1:1 conditions. It is interesting to note that according to Lu \textit{et al.} \cite{Lu98}, only chlorobenzoate isomers with the chlorine in the para-position yields significant shear-thickening and drag reduction, when put in combination with alkytrimethylammonium surfactants (in contrast to the ortho- and meta-isomers). More recently, surfactant systems without hydrotopes were uncovered. The double tail gemini 12-2-12 (ethanediyl-1,2-bis(dodecyl di\-methyl\-am\-mo\-nium bromide)) \cite{Sch95a,Oda97,Oda2000,Web2002,Web2003} and the partially fluorinated surfactant (perfluorooctyl butane trimethylammonium bromide) \cite{Oel2002} are among the most surveyed systems of this kind. Concerning the class of gemini surfactants, some molecules with specific architecture were also shown to self-assemble into micelles with more complex topologies, such as ring-like \cite{In99} and branched \cite{In99a} structures. Note finally a system made from oppositely charged surfactants, tetradecyl dimethylamine oxide (C$_{14}$DMAO) and dodecyl sulfate (SDS), which displays the above properties only for mole fractions [C$_{14}$DMAO]/([C$_{14}$DMAO]+[SDS]) between 0.5 and 0.8 \cite{Hof91,Hu93,Pro97}. In Table \ref{tabstk}, again for simplicity, we have omitted commercial surfactants showing a polydispersity of the aliphatic tails, or chemical structures that are less well characterized \cite{Lin2000}. 
\subsection{\label{str}Rheology}
\indent

The shear-thickening transition in dilute surfactant solutions was investigated using both strain- and stress-controlled  rheometry. Due to the low viscosity of the solutions, Couette geometries either with single or double Couette walls were preferred (Fig.~ \ref{figI2}). Due to the long transients in the kinetics of the SIS formation, the shear stress \textit{versus} shear rate curves were determined by measuring the time dependence of the stress, and by recording its stationary value. 
The flow curves were then constructed point by point so as to ensure that they were corresponding to the stationary state of flow. 
\subsubsection{\label{strate}Strain-controlled rheometry}
\indent
\begin{figure*}[t]
\begin{center}
\includegraphics[scale=0.115]{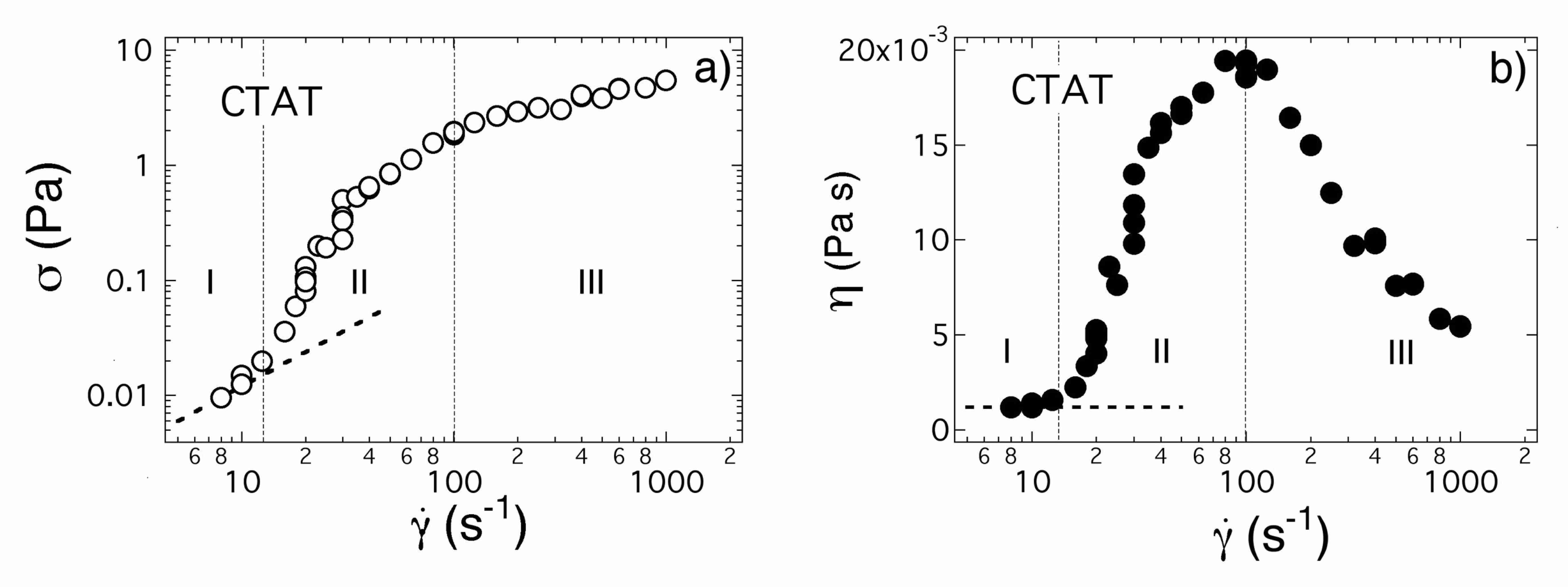}
\caption{Steady state shear stress (a) and viscosity (b) versus shear rate for hexadecyltrimetylammonium \textit{p}-toluenesulfonate (CTAT) at $c=0.41$ wt.~\% and $T=23$\r{}C. With increasing shear rates, three flow regimes are encountered. At low shear rates (Regime I), the stress increases linearly with the rate with a constant slope $\eta_0$, indicating a Newtonian behavior (dashed lines). At  $\dot\gamma_{c}=14$ $\pm$ 2 s$^{-1}$, the viscosity increases and deviates progressively from the Newtonian behavior. In Regime III, the viscosity passes through a maximum, and shear-thinning is observed. Figures adapted from Ref.~\cite{Gam99}.}
\label{figII2}
\end{center}
\end{figure*}
Fig.~\ref{figII2} displays the general behavior of the shear-thickening transition observed with imposed shear rate. The steady shear stress $\sigma (\dot {\gamma })$ and the steady apparent shear viscosity $\eta (\dot {\gamma })$ are shown as a function of the applied shear rate for the  hexa\-de\-cyl\-tri\-me\-thyl\-am\-mo\-nium \textit{p}-toluenesulfonate (CTAT) at $c=0.41$ wt.~\%. For this system, the overlap concentration was estimated at $c^{*}=0.5$ wt.~\% and the shear-thickening to be present over the range 0.05 - 0.8 wt.~\% \cite{Gam99,Tru2000}. 

In Fig.~\ref{figII2}, three flow regimes can be distinguished : 
\begin{itemize}
\item \textbf{Regime I} : At low shear rates, the stress increases linearly with the rate, indicating a Newtonian behavior.
\item \textbf{Regime II} : At $\dot\gamma_c$, the viscosity increases and deviates progressively from the Newtonian behavior. The transition toward the shear-thickened state is continuous. 
\item \textbf{Regime III} : The apparent viscosity passes through a maximum, at a level that is several times that of the solvent viscosity, and then shear-thinning is observed. 
\end{itemize}
The three regimes are indicated in the figures by dashed vertical lines. Similar data, and in particular the observation of a continuous viscosity increase in Regime II were obtained on various systems of Table \ref{tabstk}, namely on C$_{16}$TA/NaSal \cite{Wun89,Hu93a,Liu96,Vas2008,Lee2002}, ge\-mi\-ni 12-2-12 \cite{Sch95a,Oda97}. 
Note that for TTAA/Sal, a discontinuous transition between regime I and regime III was reported with controlled strain rates, as illustrated in Fig.~\ref{figII3} and discussed in details below \cite{Hu98}. 
\subsubsection{\label{ststress}Stress-controlled rheometry} 
\indent
\begin{figure*}[b]
\begin{center}
\includegraphics[scale=0.1]{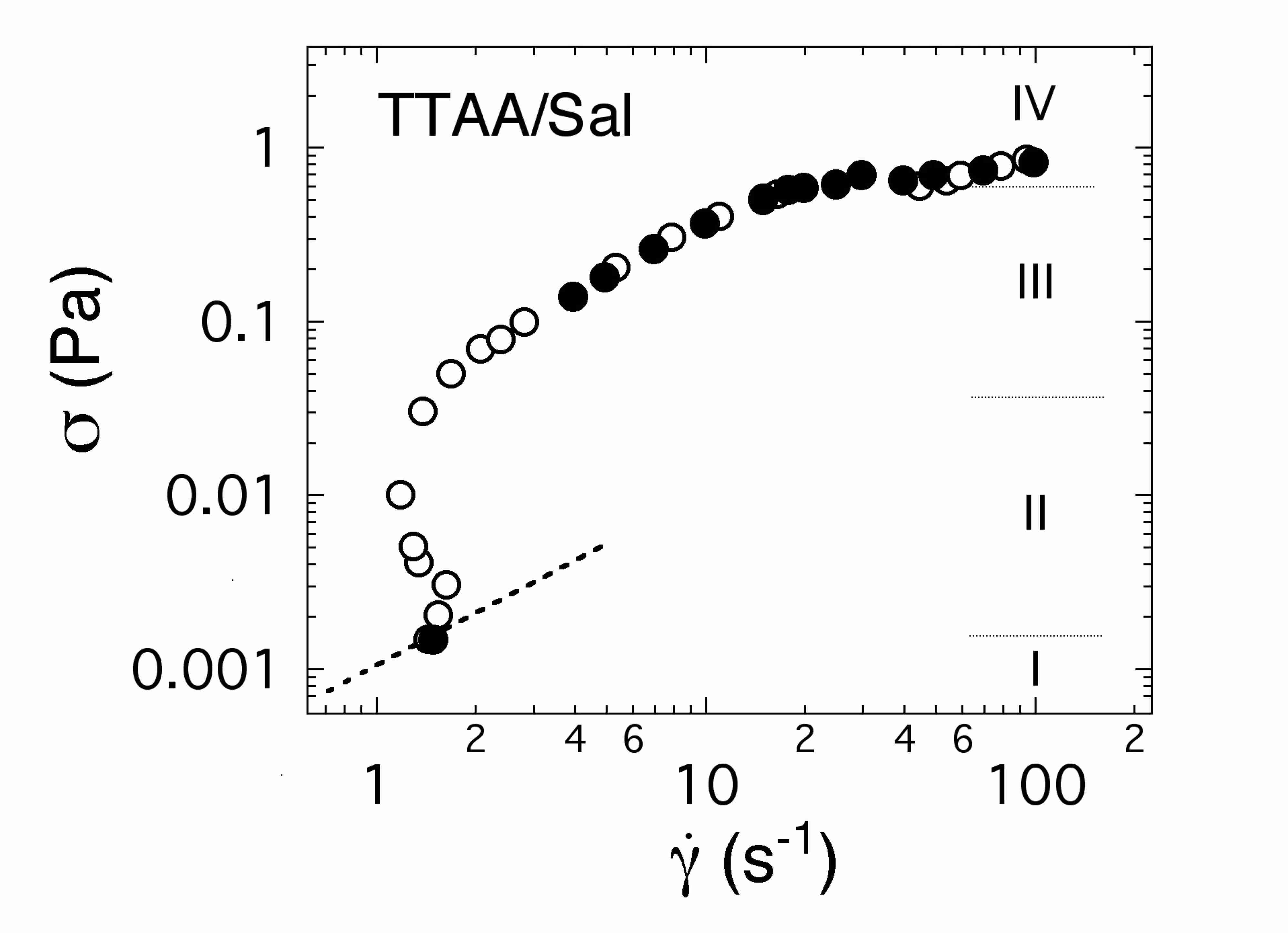}
\caption{Steady state rheological behavior of 1.7/1.7 mM tris(2-hy\-droxy\-ethyl)-tallo\-walkyl\-ammonium\-acetate/sodium salicylate (TTAA/Sal), corresponding to a total weight concentration $c=0.10$ wt.~\%. For this system, four flow regimes were reported, as indicated. Note the re-entrant shear stress versus shear rate curve for the stress-controlled data (open symbols), and the discontinuity for the strain-controlled data (closed symbols). Figure reprinted with permission from Ref.~\cite{Hu98}.}
\label{figII3}
\end{center}
\end{figure*}
Stress-controlled rheometry has been operated on fewer systems, as compared to the strain-controlled rheometry. In C$_{16}$TAB with salicylate and \textit{p}-toluenesulfonate counter\-ions, Hartmann \textit{et al}. observed a phenomenology close to that reported in Fig.~ \ref{figII2}, that is the occurrence of the sequence Newtonian (I) - shear-thickening (II) - shear-thinning (III) \cite{Har97a}. Pine and coworkers were the first to notice the existence of a re-entrant flow curve above the critical shear rate (Fig.~ \ref{figII3}). These results were obtained on a 1.7/1.7 mM tris(2-hydroxyethyl)-tallowalkyl ammonium acetate/sodium salicylate (TTAA/Sal), in which TTAA represents a mixture of C$_{16}$ and C$_{18}$ alkyl chain surfactants \cite{Bol97a,Hu98a,Mys94,Bol97,Hu98,Kel97}. The total weight concentration of this sample was $c=0.10$ wt.~\%. Careful transient measurements allowed to confirm the existence of stationary flows at shear rate below $\dot\gamma_c$ (re-entrant behavior). 
These findings were interpreted as a strong evidence that a more viscous phase was building up under constant stress. With increasing shear stress, the $\sigma(\dot\gamma)$-flow curve of TTAA/Sal was found similar to that of most compounds, showing a transition toward a shear-thinning at high shear stress. Unlike most rheological characterizations, Pine and coworkers reported four flow regimes, noted I to IV on the figure. There, regimes I, II and IV correspond to the three regimes of Fig.~ \ref{figII2}, whereas regime III sets the limits of a range where the  viscosity stays constant as a function of $\dot\gamma$. 
Using stress-controlled rheometry, Walker and coworkers also observed a slightly re-entrant behavior above the critical stress in the CTAT dilute solutions \cite{Tru2002}. 
\subsubsection{\label{sttr}Transient rheology}
\indent

As already mentioned, considerable care was taken by experimentalists in order to ensure the actual determination of the steady state. Most procedures used start-up experiments, which consisted to impose the shear rate (resp. stress) on freshly poured solutions, and to measure the stress (resp. rate) as a function of time. This approach was suggested already by the work by Hoffmann and Rehage (see Fig.~ \ref{figII1}). Start-up experiments have revealed two major results that were later corroborated on most systems: 

\begin{itemize}
\item In Regimes II and III, the shear-thickening state was reached after an induction time noted $t_{ind}$. 
\item As noticed by most of the earlier reports, this induction time was varying as 1/($\dot\gamma$ - $\dot\gamma_c$) \cite{Berret2002}, or as 1/$\dot\gamma$ far from the critical conditions \cite{Hu93,Pro97,Oda97,Bol97,Hof81}. In other words, the closer the shear rate was from the critical value, the longer was the time to reach stationary state. This result was interpreted as an indication that the relevant quantity for the induction of the SIS was the total deformation $\dot\gamma t_{ind}$ applied. These findings were observed for CP/Sal \cite{Hof81a,Reh82}, C$_{16}$TAB/NaSal \cite{Hu93a} and TTAA/NaSal \cite{Bol97}. 
\end{itemize}

More recently, a closer inspection of the transient stress rheology for thickening systems has revealed more complicated patterns, such as structural memory effects. Berret \textit{et al.} \cite{Ber2000a} and Oeschlager  \textit{et al.} \cite{Oel2002,Oel2002a} have observed that the transient mechanical response was also depending on the thermal and shear histories. Samples having been treated thermally, e.g. heated up to 90 \r{ }C for two hours behaved very differently from samples freshly prepared or already sheared. The induction time could last several hours, and was not proportional to the inverse shear rate, as mentioned previously. It was concluded that the lack of reproducibility under certain thermal and shear conditions might indicate that these surfactant solutions were characterized by long-lived metastable states.

Other transient experiments commonly carried out on these solutions were stop-flow measurements. When sheared in the thickening regime, at the abrupt arrest of the shearing cell, the shear stress was found to relax via a double exponential decay, the shortest time being of the order of one second (associated to the reorientation dynamics) \cite{Oda97}, and the longest time being of the order of seconds or minutes. Concerning this longer time, values in the range 1 - 1000 s, 5 - 500 s and 5 - 40 s were observed for C$_{14}$TA/Sal \cite{Ohl86}, gemini 12-2-12 \cite{Oda97} and CTAT \cite{Berret98} dilute solutions respectively. The above ranges correspond to different conditions of temperature and/or concentration. Because of the monoexponential character of the long-time relaxation, and also because semi-dilute micellar solutions respond to stop-flow similarly \cite{Rehage88}, the vanishing of the stress was ascribed to the relaxation of entangled wormlike micelles. Such a conclusion implicitly assumes that the micelles have grown under shear, although this was not formulated in such terms in the literature. We will come back to this point later. 
\subsubsection{\label{stconctemp}Concentration and temperature dependence}
\indent

With increasing concentrations, all the reported surfactants exhibit a transition between a dilute and a semi-dilute regime at $c^*$. Below $c^*$, micelles are short and do not overlap, whereas above $c^*$ chain entanglements slow down considerably the dynamics of the network and the zero-shear viscosity increases sharply \cite{Wun87,Deh2007,Vas2008,Cre98,Gam99,Hu98a,Hu98,Tru2000}. The shear-thickening transition has been observed for concentrations below and above the overlap concentration. Shear-thickening in solutions with viscosities up to 1000 times that of water were reported \cite{Hu98}. Concerning the concentration dependence of the critical shear rate $\dot\gamma_c$, no universal behavior could be evidenced. $\dot\gamma_c$ was found to increase in CTAT (with D$_{2}$O as a solvent) \cite{Gam99} and in TTAA/NaSal \cite{Hu98}, and to decrease in C$_{14}$DMAO/SDS \cite{Hof91}, in gemini 12-2-12 \cite{Oda97} and in C$_{16}$TAB/NaSal \cite{Har98}. In some other systems, it was found to remain concentration independent \cite{Gam99}.

Much stronger dependences were observed as a function of the temperature. All systems investigated exhibited an Arrhenius-type behavior for the critical shear rate $\dot\gamma_c$, \textit{i.e.} : 
\begin{equation}
\dot\gamma_c(T)\sim \exp\left(-\frac{E_a}{k_BT}\right)
\label{eqn1}
\end{equation}

\noindent
where $E_{a}$ is an activation energy, $k_{B}$ the Boltzmann constant and $T$ the absolute temperature. Activation energies $E_{a}$ were found in the range 20 - 120 $k_{B}T$ ($T=300$ K), or equivalently between 50 - 300 KJ mol$^{ - 1}$. Concomitant to the shift of $\dot\gamma_c$ to larger values, the amplitude of the shear-thickening effect diminished with increasing temperature \cite{Ohl86,Wun89,Hu93a,Gam99,Oda97}. Ultimately, above 50\r{ }C, the shear-thickening vanished. The origin of the underlying activated process in shear-thickening systems has not yet been identified.

In addition to concentration and temperature, other parameters capable of mo\-di\-fying the transition have been studied. These were \textit{i)} the shearing cell geometry, and in particular the gap of the Couette cell \cite{Ohl86,Sun2003,Wun89,Hu98a,Hu98,Lee2002}, \textit{ii)} the ionic strength \cite{Har97a,Hof91} and \textit{iii)} the addition of polymeric additives, such as PEO (poly(ethylene oxide)) or HPC (hydroxypropyl cellulose) \cite{Tru2000}. Among these parameters, the geometry effect is certainly the most intriguing. Already present in the work by Ohlendorf and Wunderlich \cite{Ohl86,Wun89}, it was noticed that smaller gaps shifted the critical shear rate towards higher values, and reduced the amplitude of the viscosity jumps. The gap effects were later interpreted by Pine and coworkers as a consequence of the slipping of the SIS along the walls, through the presence of a thin lubricating layer (see Fig~\ref{figII6} for details) \cite{Hu98a}.
\subsection{\label{ststruct}Structure and orientation under shear}
\subsubsection{\label{stsas}Small-angle scattering under shear}
\indent

Thanks to an excellent neutron scattering contrast of hydrogenated surfactants in deuterated water (D$_{2}$O), small-angle neutron scattering (SANS) has become a privileged tool for the investigation of dilute shear-thickening solutions. Most of the systems in Table \ref{tabstk} have been investigated by SANS in quiescent conditions \cite{Bew86,Lin90,Gam99,Tru2002,Hof91,Oda2000,Tru2000,Berret98}. All these studies have revealed a unique behavior : the surfactants self-assemble into cylindrical micelles. The radius of the micelles was also determined, and found around 2 nm  \cite{Bew86,Lin90,Gam99,Tru2002,Hof91,Oda2000,Tru2000,Berret98}. The other feature revealed by SANS was the occurrence of a structure factor indicative of strong repulsive interactions between the micellar threads. These interactions were attributed to the electrostatic charges at the surfaces of the rods. Electrostatic structure factors were observed on salt-free solutions of C$_{14}$DMAO/SDS \cite{Hof91}, gemini 12-2-12 \cite{Sch95a,Oda2000} and CTAT \cite{Gam99,Berret2001,Berret98}. 
\begin{figure*}[t]
\begin{center}
\includegraphics[scale=0.105]{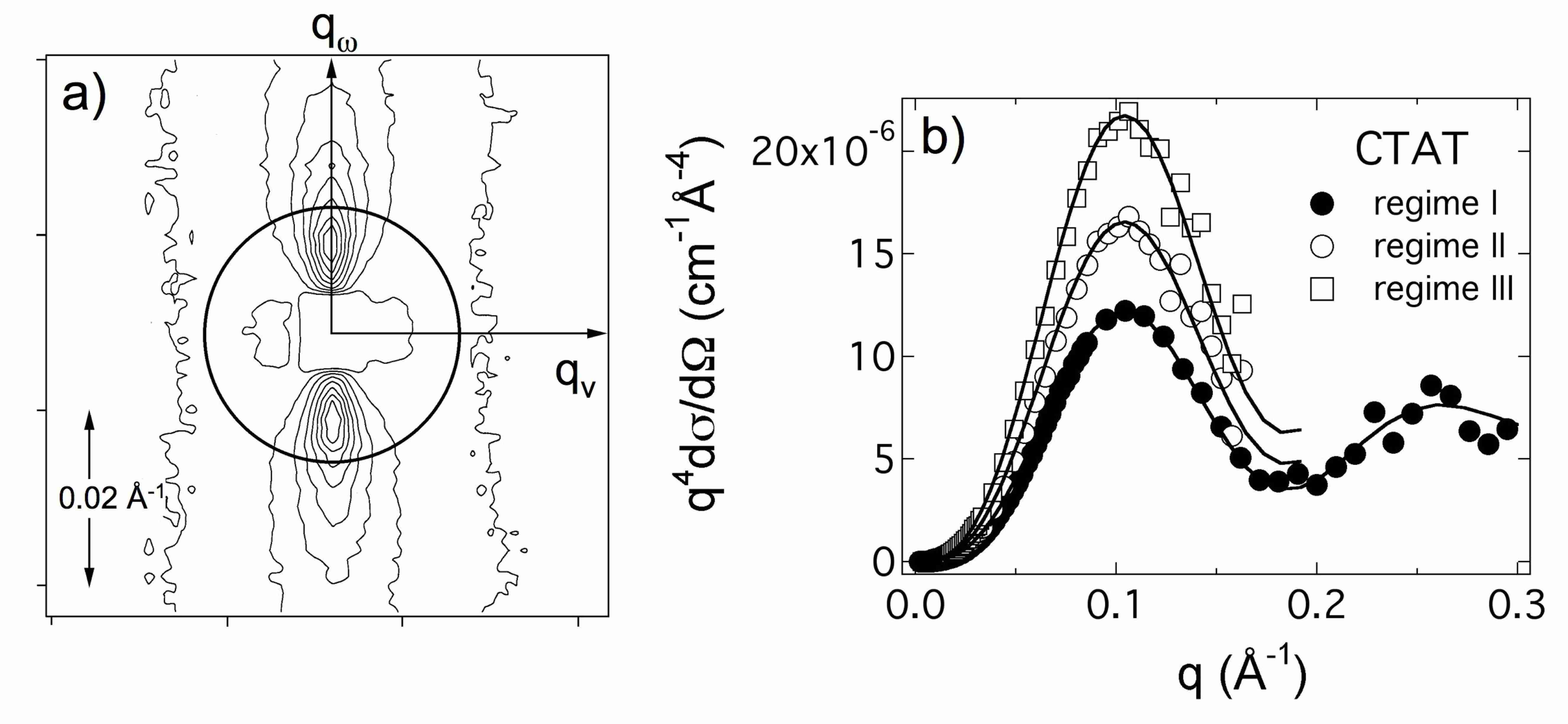}
\caption{(a) Two-dimensional neutron scattering pattern characteristic of the shear-induced phase in hexadecyltrimethylammonium \textit{p}-toluenesulfonate dilute solutions. The solvent here is D$_2$O, the concentration $c=0.26$ wt.~\% and the shear rate  $\dot\gamma=188$ s$^{-1}$ (Regime III in Figs. \ref{figII2}). $\vec q_v$ and $\vec q_w$  are respectively parallel to the velocity and vorticity directions of the flow. The ring shows the wave-vector at which the scattering cross-section is maximum in the quiescent state. (b) Porod representation of the neutron scattering intensities in Regimes I, II and III (intensity in the vorticity direction). The oscillations in the form factors for the three set of data are in agreement with a morphology of cylindrical micelles with radius $R_c=1.95$ nm, with a standard deviation of 0.2 nm (continuous lines). Figures adapted from Ref.~\cite{Berret2001}.}
\label{figII4}
\end{center}
\end{figure*}
During the last two decades, neutron \cite{Bew86,Lin90,Gam99,Tru2002,Berret2001,Hof91,Sch95a,Oda2000,Tru2000,Berret98} and light \cite{Pro97,Web2002,Web2003,Oel2002,Hof81,Oel2002a} scattering under shear were performed repeatedly on dilute thickening systems. As early as in 1986, Bewersdorff and coworkers set up a Couette cell on a neutron spectrometer in order to detect the anisotropy of the scattering induced by the shearing \cite{Bew86}. In a series of runs performed on C$_{14}$TA/Sal, it was shown that under shear, the scattered intensity collected on a two-dimensional detector was highly anisotropic, the scattering being predominantly in the direction perpendicular of the velocity. It was concluded that the shear-induced phase corresponded to a highly aligned state of cylindrical micelles \cite{Bew86}. An illustration of this anisotropy shown in Fig.~\ref{figII4}a for CTAT ($c=0.26$ wt.~\%, $\dot\gamma=188$ s$^{ - 1}$ cor\-res\-pon\-ding to Regime III) is representative for this class of materials. An approach in terms of orientation distribution function was performed on the neutron spectra by analogy with the data analysis of nematic phases \cite{Deu91} (see Section \ref{part3}). The order parameter of the micellar orientations was then derived and found to be 0.8. This value is close to unity, which designates a perfect alignment. 
On the same CTAT specimen, additional information could be gained from the study of the position of the structure factor as functions of surfactant concentration and shear rate. Below c*, the structure factor peak of the SIS was found to shift to lower wave-vectors by 30~\% as compared to its value in the Newtonian regime (Fig.~\ref{figII4}a). It actually moved down to the semidilute $q^{1/2}$-scaling law that was determined from solutions above c*. This shift was interpreted as an indication of a shear-induced growth of the micelles, from short rodlike to wormlike aggregates. Similar shifts of the structure factor were observed for the gemini 12-2-12 surfactants \cite{Sch95a,Oda2000}.

Concerning the shear-thickening transition, the question was raised about a possible transition of morphology, a transition where the original microstructure would be changed into heterogeneous patterns showing stippled or spongelike textures \cite{Kel97}. It was also suggested by others that the micellar threads would eventually undergo a transition towards a bundle state \cite{Bar2001}. Fig.~\ref{figII4}b shows the scattering intensities in the direction perpendicular to the flow velocity in regimes I, II (shear-thickening) and III (shear-thinning), again for the CTAT system \cite{Berret2001}. There, the Porod representation ($q^{4}\times d\sigma (q)/d\Omega $ \textit{versus} $q$) has been used in order to emphasize the local morphology at rest and under shear. As a result, all three data sets exhibit oscillations consistent with a cylindrical micelles with radius $R_{c}=1.95$ nm. Qualitatively, the fact that the position of the first maximum remains unchanged under shear supports the conclusion that the local morphology remains rod-like. Similar results were found in gemini 12-2-12 dilute solutions, although they were not interpreted using the Porod representation \cite{Sch95bis}. 

More recently, on  surfactants, Weber and Schosseler have investigated the light scattering properties under shear, in order to probe the sheared fluid at length scales larger than those accessible by SANS \cite{Web2002}. For a 18.3 mM solution ($c=1.0$ wt.~\%), an intense streak pattern perpendicular to the velocity direction was observed in the shear-thickening regime. The patterns exhibited strong fluctuations in amplitude, as well as a spatial modulation along the vorticity axis. A correlation length of the order of 30 $\mu $m was derived from this modulation. It was finally argued that these 30 $\mu $m were not compatible with the intermicellar distance, estimated in this solution at a few tens of nm. The light scattering data were interpreted in terms of a strongly aligned and heterogeneous gel-like layers in the gap of the Couette cell. 
\subsubsection{\label{stfb}Flow birefringence}
\indent

A remarkable property of the shear-induced phase is its flow birefringence. Flow birefringence experiments on shear-thickening surfactant solutions were introduced by Hoffmann and coworkers more than two decades ago 
\cite{Ohl86,Wun87}. Wunderlich \textit{et al.} have shown for instance that for C$_{14}$TA/Sal \cite{Wun87}, the onset of flow birefringence coincided with the increase of viscosity. Such results were found on all shear-thickening systems  studied by this technique. 
Flow birefringence was measured by transmission using a Couette geometry. With this configuration, the polarized light propagates along the vorticity direction and the transmitted light reads: 
\begin{figure*}[b]
\begin{center}
\includegraphics[scale=0.1]{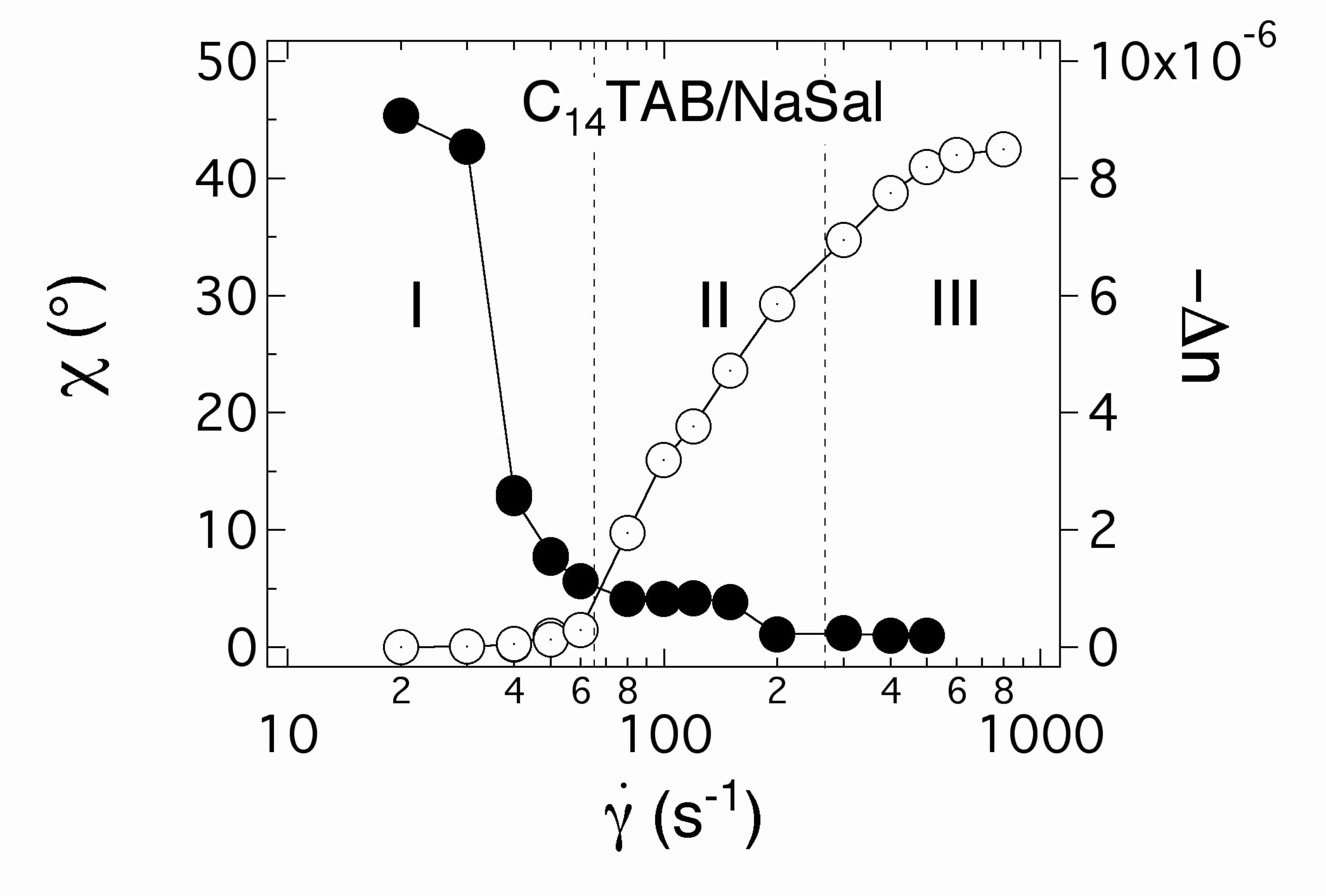}
\caption{Shear rate dependences of the flow birefringence $\Delta n$ (right scale, open symbols) and of the alignment angle $\chi$ (closed symbols, left scale) for 3/3 mM tetradecyltrimethylammonium bromide and sodium salicylate (C$_{14}$TAB/NaSal), corresponding to a total weight concentration $c=0.15$ wt.~\%. Regimes I, II and III were determined from steady shear viscosity. Figure adapted from Ref. \cite{Deh2007}, courtesy J.P. Decruppe.}
\label{figII5}
\end{center}
\end{figure*}
\begin{equation}
I=\frac{I_0}{2}\sin^{2}\frac{\delta}{2}\sin^{2}(2(\chi-\theta))
\label{eqn2}
\end{equation}

\noindent
where $I_{0}$ is the incident light intensity, $\delta=2\pi h\Delta n/\lambda$ the phase angle, $\chi $ the extinction angle and $\theta $ the angle made by the polarization of the incident beam with the flow velocity (in the expression of the phase angle, $h$ is the height of the Couette cell and $\lambda$ the wavelength of light). The values of the birefringence $\Delta n$ were found to be negative, comprised between -10$^{ - 5}$ and -10$^{ - 7}$, depending on the weight concentration. The flow birefringence was essentially measured at steady state as a function of the shear rate and under various conditions of temperature, concentration and ionic 
strength \cite{Wun87,Deh2007,Berret2002,Ber2000a,Hof91,Oda97,Hof81}. 

The two main results of birefringence are illustrated in Fig.~ \ref{figII5} for the 3/3 mM C$_{14}$TAB/NaSal solution ($c=0.15$ wt.~\%)  \cite{Deh2007}. They are : 
\begin{itemize}
\item An increase of $\Delta n$ in regime II, followed by a saturation in regime III.
\item An abrupt decrease of the extinction angle at the onset of thickening toward a value close to 0\r{}, both in regimes II and III. 
\end{itemize}
Fig.~ \ref{figII5} illustrates that as soon as the new phase is induced, it is strongly oriented by the flow. In some reports, $\chi $-data shown as a function of the shear rate have revealed that the extinction angle undergoes a discontinuity from $\chi=45^{\circ}$ to $\chi\sim$ 0$^{\circ}$ at $\dot\gamma_c$ \cite{Berret2002}. Recent measurements have confirmed this feature \cite{Deh2007}. Time- and space-resolved studies of the flow birefringence were also attempted \cite{Berret2002}. In CTAT, these studies have revealed that once the SIS is initiated, it spread over the whole gap of the cell, and no regime of coexisting states (such as birefringent and non birefringent) could be detected.
\subsubsection{\label{stpiv}Particle Image Velocimetry}
\indent
\begin{figure*}[b]
\begin{center}
\includegraphics[scale=0.12]{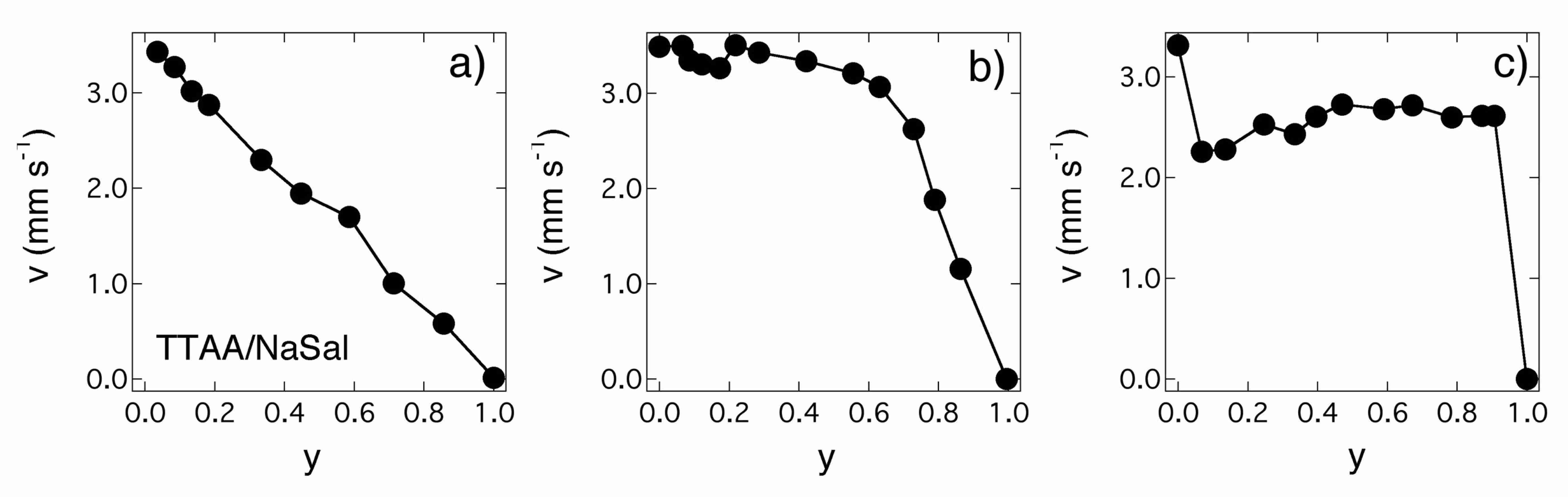}
\caption{Time development of the velocity profile in a 4 mm-gap Couette cell for a 1.7/1.7 mM TTAA/NaSal solution. The time frames a), b) and c) correspond to 6 s, 438 s and 3071 s respectively after the inception of shear ($\dot\gamma=1$ s$^{-1}$). The coordinates are referenced with respect to the inner cylinder (gap position 0) and to the outer cylinder (gap position 1). Figures adapted from Ref. \cite{Hu98a}.}
\label{figII6}
\end{center}
\end{figure*}
Flow velocimetry measurements on shear-thickening include works by Koch and coworkers on C$_{14}$TA/Sal and C$_{16}$TA/Sal \cite{Koc98} and by Hu \textit{et al.} on TTAA/NaSal micelles \cite{Hu98a}, both using particle image velocimetry (PIV). One of the reasons for the few PIV studies lies in the fact that the critical shear rates are high (in general some tens of s$^{ - 1})$ and that in such conditions, measurements of flow velocities using seeding particles remain challenging.
Fig.~ \ref{figII6} shows three cartoons of the velocity profiles determined at different times during a start-up experiment. The system placed under scrutiny was TTAA/NaSal at 1.7 mM and at 1:1 ratio between surfactant and aromatic counterions (weight concentration 0.10 wt.~\%) \cite{Hu98a}. The gap of the Couette cell was 4 mm and the inner cylinder (gap position 0) was moving. After the inception of shear (Fig.~ \ref{figII6}a), the linear velocity profile for a homogeneous shear flow was observed. As the SIS began to grow from the inner wall (Fig.~ \ref{figII6}b), a progressively thicker region of uniform velocity developed, with a steeper velocity gradient near the outer cylinder. At long times (Fig.~ \ref{figII6}c), the velocity field remained uniform over most of the gap, with two thin and fast layers near the walls. Hu and coworkers concluded that at steady state, the SIS fills most of the center of the gap and behaved as a \textquotedblleft solid\textquotedblright body in rotation (plug flow) \cite{Hu98a}. Although less documented than the Hu \textit{et al}.'s paper, the data from Koch \textit{et al.} displayed typically the same effect, namely that the shear-induced structure was associated to a highly non-homogeneous flow, with slippage at the walls \cite{Koc98}. Convincing evidences of wall slip were also reported by Sung \textit{et al.} \cite{Sun2003} in CPCl/NaSal solutions from direct rheological measurements.
\subsection{\label{stconclusion}Conclusion}
\indent

Although the surfactants in Table \ref{tabstk} have not all been investigated with the techniques described in this section, it can be assumed that these systems share the same elementary properties when submitted to shear : above a critical shear rate, a structure that is more viscous than the suspending solvent is induced, yielding an increase in the apparent viscosity of the fluid. In the following, we recapitulate the milestones that are important in the present experimental context, and suggest a minimal scenario for the transition.
\begin{enumerate}
\item \textit{The local micellar structure does not change under shear} 

By restricting ourselves to the dilute case, it can be concluded that the rodlike micelles are unentangled at rest and in the Newtonian domain. There, the viscosity is close to, or a few times that of water. Small-angle neutron scattering shows conclusive evidences that the cylindrical structure of the rods is preserved at all shear rates. The hypothesis suggested in the past  according to which the shear-thickening could originate from a modification of the local structure of the surfactant assemblies can reasonably be ruled out \cite{Kel97,Bar2001}. 

\item \textit{Shear-thickening is associated with micellar growth} 

The structural modifications of the aggregates occur on the contrary at a larger scale, namely at the scale of their length. 
Using light and  neutron scattering, it was demonstrated that the shear-thickening transition is accompanied by a uniaxial growth of the micelles, which hence undergo a transition from rodlike to wormlike aggregates \cite{Pro97,Berret2001,Oda2000}.

\item \textit{The shear-induced structure is viscoelastic} 

The second indication of the wormlike micellar character of the SIS is its viscoelasticity. The viscoelasticity of the shear-induced phase was observed in stop flow experiments in a series of systems \cite{Hof81a,Ohl86,Wun87,Ber2000a,Hof91,Oda97,Berret98}. According to the definition of viscoelasticity \cite{Larson99}, the long time seen in these experiments can be ascribed to the intrinsic relaxation of the shear-induced state. With this in mind, the Weissenberg number (the product of the shear rate and relaxation time) for these micellar fluids can be estimated. In regimes II and III, the Weissenberg numbers reach values comprised between 10 (at $\dot\gamma_c$) and 1000. Depending on the system, this range can go even higher, as in gemini 12-2-12 \cite{Oda97} or CTAT \cite{Berret98}. Hence, once the micelles have grown in size, they are directly brought to a state that is strongly sheared on the time scale of the fluid. At high Weissenberg numbers, the sheared solutions could undergo elastic instabilities, that could then generate more complex flows such as flows along the vorticity direction \cite{Larson92}. 3-dimensional flows associated with the shear-thickening in dilute regime have not been reported so far.
\end{enumerate}

\begin{figure*}[t]
\begin{center}
\includegraphics[scale=0.12]{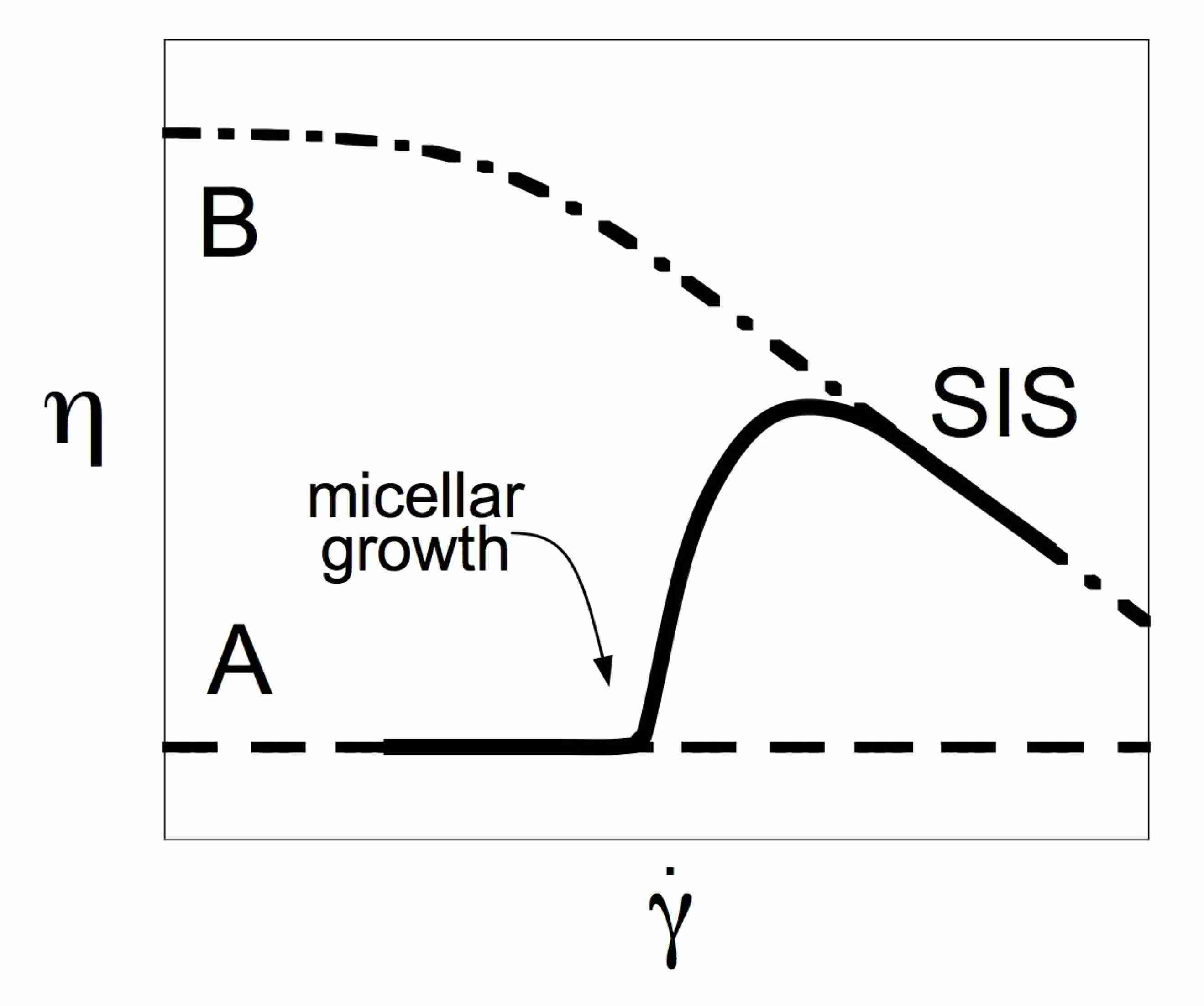}
\caption{Schematic diagram accounting for the shear-thickening transition in dilute surfactant solutions}
\label{figII7}
\end{center}
\end{figure*}

In Fig.~ \ref{figII7}, a schematic diagram accounts for a possible scenario of the shear-thickening transition. There, curve A denotes the apparent viscosity of a dilute surfactant solution containing non-overlapping rod-like micelles (Newtonian), whereas curve B corresponds to the flow curve of the same solution, but for which the micelles are long and entangled (shear-thinning). At the transition rate, the fluid jumps from branch A to branch B. With decreasing shear rate, starting from the induced phase,  the SIS vanishes reversibly as the micelles disassemble into short rod-like aggregates. From this minimal scenario, it can be understood that the micelles are strongly aligned in the flow, or that the flow becomes non homogeneous \cite{Koc98,Hu98a,Hu98} or turbulent in the shear-thinning regime \cite{Hu98a,Bol97}. The hydrodynamic instabilities of dilute wormlike micelles and in turbulent flows remain one of the most promising issues of this field.

Concerning the mechanism of growth induced by shear, many theories and models were developed during these last three decades, and none of them were fully satisfactory. Most models were based on the assumption that the increase of viscosity was related to a shear-induced \textquotedblleft gelation\textquotedblright. Many phenomenological models were constructed assuming a banded state of coexisting gel and fluid phases \cite{Ajdari98,Gov99}. Some microscopic theoretical attempts had anticipated that \textquotedblleft gelation\textquotedblright could be connected to a shear-induced micellar growth \cite{Cates90,Bruinsma92,Tur92}. Concerning these earlier models however, the predicted critical shear rates were too large as compared to the experimental values \cite{Berret98}. It is out of the scope of the present review to survey the theoretical treatments of the shear-thickening transition. We rather refer to recent and exhaustive reviews by Cates and Fielding \cite{Cat2006} and by Olmsted \cite{Olm2008}. 

The structural memory effects found in different systems (such as CTAT, gemini 12-2-12 and in the fluorocarbon surfactant C$_8$F$_{17}$), and discussed in the transient rheology section suggest that the aggregation in the quiescent state and the thickening transition are interrelated. It is certainly not easy to conceive that dilute and very dilute solutions could exhibit exotic behaviors, in particular in reference to the self-assembly mechanism. One possible explanation would be that the surfactant solutions are in a metastable self-assembled state, due for instance to the long range electrostatic interactions. This metastable state could then be described as a coexistence state of short rodlike aggregates and slowly evolving supramolecular structures, such as huge micelles or pieces of entangled network. This additional and unexpected populations of large micelles have been recently observed in two systems, again the gemini 12-2-12 studied by Schosseler and coworkers and in the fluorcarbon surfactant by Oehlschager \textit{et al}. \cite{Oel2002}. Light scattering performed on quiescent solutions have shown the coexistence of short, intermediate and very large micelles, which respective populations varied with the thermal and shear histories. It remains now to demonstrate that these large structures are playing the role of initiators for the shear-thickening transition, as well as to understand the metastability of the different self-assemblies.
\section{\label{part2}Shear banding transition in semi-dilute and concentrated giant micelles}
\subsection{\label{introsb}Introduction}
\indent

This part is devoted to the nonlinear rheology of semi-dilute and concentrated giant micelles systems. In the semi-dilute regime, characterized by concentrations ranging typically from 0.1 wt.~\% to $\simeq$ 10 wt.~\%, wormlike micelles form a viscoelastic network and, are supposed, by analogy with polymers, to follow simple scaling laws \cite{Cat90a,Cat88}. In the concentrated regime, corresponding typically to weight concentrations between $\simeq$ 10 wt.~\% and $c_{\scriptscriptstyle{I-N}}$, the isotropic-to-nematic phase boundary, the mesh size of the  entangled micellar network becomes of the order of or shorter than the persistence length (see Fig.~\ref{figI1}).

When submitted to simple shear flow, giant semi-dilute and concentrated micelles show original nonlinear responses. A number of experimental publications suggest that micellar solutions undergo a shear-banding transition. This transition, due to the coupling between the internal structure of the fluid and the flow is usually associated with a new mesoscopic organization of the system. In turn, the modification of the supramolecular architecture of the fluid affects the flow itself and generates shear localization effects generally characterized by a splitting of the system into two macroscopic layers bearing different shear rates and stacked along the velocity gradient direction. \\
This transition from a homogeneous towards a non homogeneous flow has been reported in complex fluids of various microstructure such as lyotropic micellar and lamellar phases \cite{Berret,Sal2003b,Sal2003c}, triblock copolymers solutions \cite{Ber2001,Man2007}, viral suspensions \cite{Let2004}, thermotropic liquid crystal polymers \cite{Puj2001}, electro-rheological fluids \cite{Vol99}, soft glassy materials \cite{Cou2002}, granular materials \cite{Los2000,Mue2000}  or foams \cite{Deb2001,Lau2004,Gil2006}. 
\begin{figure*}[t]
\begin{center}
\includegraphics[scale=0.13]{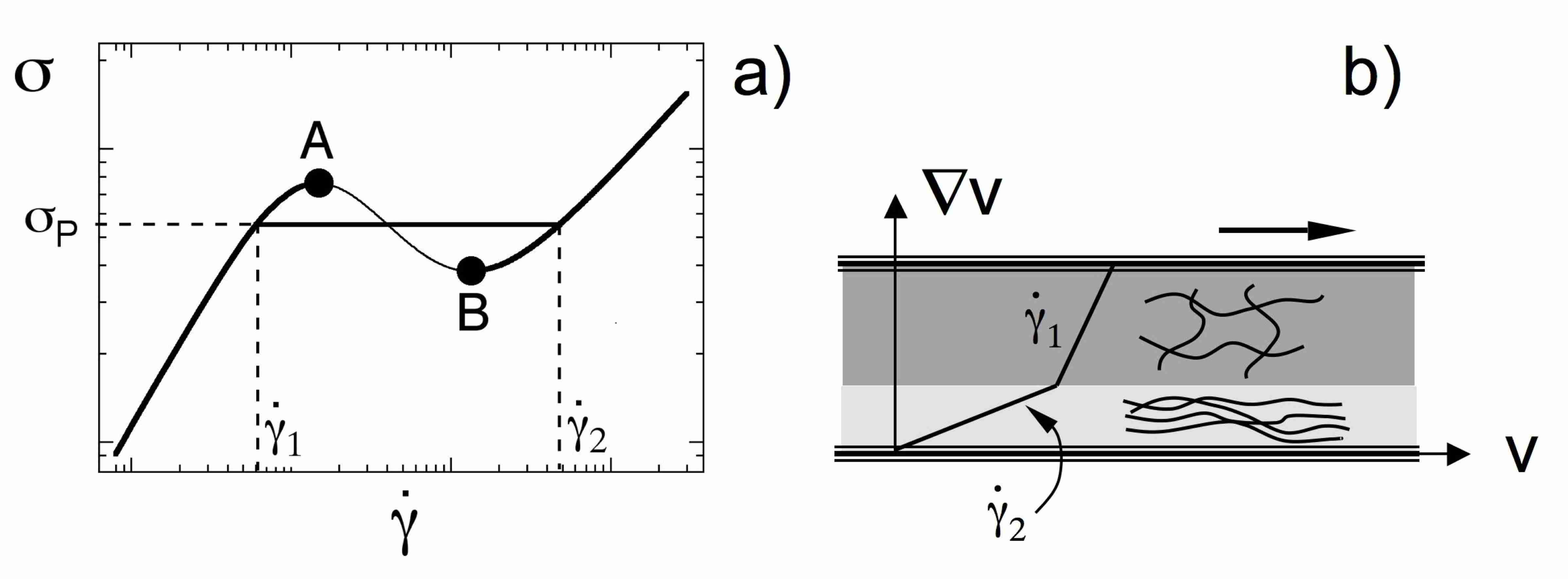}
\caption{(a) Non-monotonic constitutive relation for giant micelles composed of two stable branches separated by an unstable region AB. The corresponding steady-state flow curve presents a stress plateau at $\sigma=\sigma_p$, extending between two critical shear rates $\dot\gamma_1$ and $\dot\gamma_2$, and associated with the shear-banding transition. (b) Scheme of the shear-banding scenario in giant micelles systems.}
\label{figIII1}
\end{center}
\end{figure*}

Among these systems, the shear banding flow of reversible wormlike micelles is particularly well documented \cite{Berret}. The rheological signature of this type of flow has been observed for the first time in the pioneering work of Rehage \textit{et al} \cite{Reh91}: the measured flow curve $\sigma(\dot\gamma)$ is composed of two stable branches respectively of high and low viscosities separated by a stress plateau  at $\sigma=\sigma_{p}$ extending between two critical shear rates $\dot\gamma_{1}$ and $\dot\gamma_{2}$ (see Fig.~\ref{figIII1}.a). When the imposed shear rate $\dot\gamma$ is lower than $\dot\gamma_{1}$, the state of the system is described by the high viscosity branch which is generally shear-thinning~:~the micellar threads are slightly oriented with respect to the flow direction and the flow is homogeneous. For macroscopic shear rates above $\dot\gamma_{1}$, the flow becomes unstable and evolves towards a banded state where the viscous and fluid phases coexist at constant  stress $\sigma_{p}$ (see Fig.~\ref{figIII1}.b). The modification of the control parameter is supposed to only affects the relative proportions $f$ and $1-f$ of each band according to a simple lever rule that results from the continuity of the velocity at the interface:
\begin{equation}
  \dot\gamma=f\dot\gamma_{1}+(1-f)\dot\gamma_{2}
  \label{eqn3}
\end{equation}  
Above $\dot\gamma_{2}$, the system is entirely converted into the fluid phase : the induced structures are strongly aligned along the flow direction and the homogeneity of the flow is recovered. This scenario, due to the existence of a non-monotonic relation between the shear stress and the shear rate as schematized in figure \ref{figIII1}, has been predicted by Cates and coworkers more than fifteen years ago  \cite{Cat93}. Since then, it has been the subject of intense experimental and theoretical studies. From an experimental point of view, shear banding has been identified unambiguously in wormlike micelles using various techniques probing either the local flow field or the structure of the system.

In the following, we review the phenomenology of shear banding flow in semi-dilute and concentrated wormlike micelles. This part is organized as follows. In section \ref{rheo}, we describe the mechanical signature of the shear-banding transition. Section \ref{vel} is devoted to the characterization of the local flow field, while in section \ref{ori}, we focus on the structural properties of the banded state. 
\subsection{\label{rheo}Nonlinear rheology}
\subsubsection{\label{ss}Steady-state rheology}
\paragraph{\label{sb}Standard behavior}
\indent 

In order to illustrate the typical nonlinear mechanical response of wormlike micelles under steady shear flow, we chose to focus on the cetylpyridinium (CPCl)/sodium salicylate (NaSal) system. It is often considered as a model system since it follows the right scaling laws for the concentration dependence of the static viscosity and plateau modulus \cite{Ber93}. Moreover, for concentrations ranging from 1 wt.~\% to 30 wt.~\%, the samples behave, in the linear regime, as almost perfect Maxwellian elements with a single relaxation time $\tau_{\scriptscriptstyle{R}}$ and a plateau modulus $G_{0}$. This system has been extensively studied during the last two decades \cite{Reh91,Ber94,Cal96,Reh88,Ber97a,Ber97,Bri99a,Gra97,Men2003} and the description of its mechanical behavior is certainly one of the most complete.
 
\begin{figure*}[b]
\begin{center}
\includegraphics[scale=0.6]{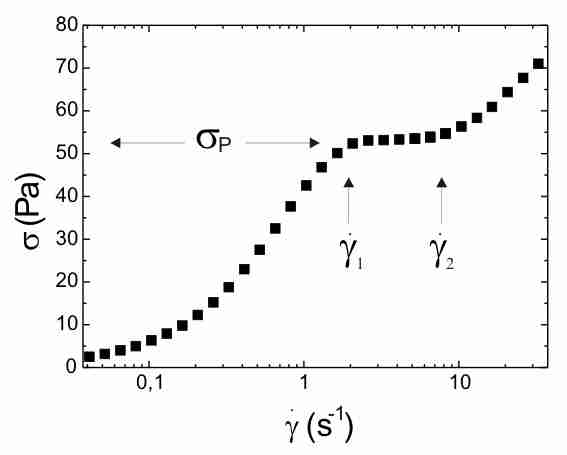}
\caption{ Experimental steady-state flow curve of a semi-dilute binary mixture made of cetylpyridinium chloride/sodium salicylate diluted in 0.5 M NaCl-brine at a temperature of 25\r{}C. The total weight fraction is 6.3\% and the molar ratio $R=$ [Sal]/[CPCl] $=0.5$. The shear stress is measured under strain-controlled conditions in a cone and plate geometry.}
\label{figIII2}
\end{center}
\end{figure*}

Figure \ref{figIII2} displays on a semi-logarithmic plot, the evolution of the shear stress $\sigma$ as a function of the shear rate $\dot\gamma$ for a sample at a total weight fraction $\phi=6.3$ \% obtained under strain-controlled conditions in a cone and plate geometry \cite{Ber94}.
This flow curve is made up of two increasing branches separated by a stress plateau extending between two critical shear rates $\dot\gamma_{1}$ and  $\dot\gamma_{2}$. The high viscosity branch is Newtonian at very low shear rates and becomes shear-thinning when approaching the first threshold $\dot\gamma_{1}$, whereas the low viscosity branch above the second critical shear rate $\dot\gamma_{2}$ is usually purely shear-thinning indicating that the constitutive behavior of the induced structures is non-Newtonian. At the critical shear rate $\dot\gamma_{1}$, the shear stress  reaches a value $\sigma=\sigma_{p}$ and the flow curve exhibits a strong change of slope followed by a stress plateau that can extend over several decades in shear rates, depending on the composition of the sample. In some cases, the stress plateau presents a significant slope and is generally well fitted by a power-law  $\sigma\sim\dot\gamma^{\alpha}$ with exponent $\alpha$ between 0.1 and 0.3. This shear rate dependence is usually explained by the coupling between flow and concentration fluctuations \cite{Sch95,Fie2003a}.

Various shear histories have been applied in order to test the robustness of the stress plateau. The latter has been found to be unique and history independent. This reproducibility is a crucial feature of the nonlinear rheology of wormlike micellar systems \cite{Ber94,Ber97,Gra97,Ler2000}.

The mechanical behavior described above concerns most of semi-dilute wormlike micelles. The situation for concentrated samples is analogous with minor changes~:~the low shear rate branch is purely Newtonian and the transition towards the stress plateau is more abrupt \cite{Ber97a}. 

Hence, the stress plateau in the flow curve $\sigma(\dot\gamma)$ is the central feature of the nonlinear rheology of semi-dilute and concentrated giant micelles systems and appears as the mechanical signature of the shear-banding transition. The first experimental evidence for such a behavior is due to Rehage and Hoffmann \cite{Reh91} on the semi-dilute CPCl (100 mM)/NaSal (60 mM) ($c=4.5$\%) solution. From that time, stress plateau in wormlike micelles has generated an abundant literature. It is now reported, using various  flow geometries such as cylindrical Couette, cone and plate, plate and plate or vane-bob and capillary rheometer, in many other surfactant systems with or without additive and/or salt as illustrated in Table \ref{tabplateau}. 
\begin{table}
\caption{Systems of wormlike micelles known to exhibit a stress plateau in their steady flow curve. Column 5 lists the experimental techniques that were used to study shear-banding. The abbreviations are \textquotedblleft sd\textquotedblright for semi-dilute and \textquotedblleft c\textquotedblright for concentrated. The letters in the last column denote sets of references detailed below. (a) \cite{Reh91,Cal96,Reh88,Gra97,Men2003,Mai96,Mai97,Bri97a,Bri97b,Fis97a,Bri99,Por97,Men2003a,Lee2005,For2005}, (b) \cite{Ber94,Ber97a,Ber97,Bri99a,Bri99,Por97,Man2004a,Sal2003,Hu2005,Mil2007,Hol2003,Lop2004,Lop2006,Bal2007,Bal2005,Bal2007a,Mas2008}, (c) \cite{Cap95,Dec95,Cap97,Fis2000,Fis2001,Bec2004,Bec2007}, (d) \cite{Sch94}, (e) \cite{Shi88,Kha93a,Kim2000,For2005,Shi94,Shi94b,Fis97,Ler98,Kad96,Whe96,Kad97,Kad97a,Kad98,Azz2005,Has2005,Ino2005,Dra2006,Dec2006}, (f) \cite{Kha93a,Dec97,Hum98a,Rad2003,Ler2004}, (g) \cite{Ler2000,Mas2008,Cap97a,Cap98,Dec2001,Ler2006,Ler2008},  (h) \cite{Ber94a,Ber98,Ber99},(i) \cite{Ban2000,Gan2006,Sol95,Sol96,Sol99,Her99,Bau2002}, (j) \cite{Gan2006,Ban2003,Gan2006a}, (k) \cite{Mak95,Ali97,Ali97a,Dec2003}, (l) \cite{Esc2003}, (m) \cite{Ang2003}, (n) \cite{Yes2006}, (o) \cite{Sch2004}, (p) \cite{Sch2004,Rag2001}, (q) \cite{Oda98,Has96}, (r) \cite{Koe2000,Sch2003}, (s) \cite{Mu2002}, (t) \cite{Pim2006}, (u) \cite{Sol2007}.}

\label{tabplateau}       
\begin{tabular}{cccccc}
\hline\noalign{\smallskip}
Surfactant & Additive & Salt & conc. 	& Experiment 							& Refs. \\
 					 & 					& 		 & regime &  						& 			\\
\noalign{\smallskip}\hline\noalign{\smallskip}
CPCl 			 & NaSal 		& 		 & 	sd		&  NMR, FB, PIV, SANS, SALS		&	 (a)			 \\
CPCl 			 & NaSal 		& NaCl &	sd/c  &	NMR, FB, DLS, PTV, PIV, USV, FI	& (b) \\
C$_{16}$TAB 			 &				  &			 &	c	  	& NMR, FB, USV, SANS 		& (c) \\
CPClO$_{3}$ & 				& NaClO$_{3}$ & c &	SANS							& (d) \\
C$_{16}$TAB 			 & NaSal 		& 		 &	sd		& FB, USV, SANS, LSI, SALS					& (e) \\
C$_{16}$TAB 			 & 					&	 KBr &	sd/c	& FB									& (f)\\
C$_{16}$TAB 			 & 					& NaNO$_{3}$ &	sd/c &		FB, LSI, SALS		&  (g) \\
CPCl 			 & Hex 			& NaCl & 	c			&					SANS, FB     & (h) \\
CTAT 			 &					& 		 & 	sd/c	&		...							& (i) \\
CTAT  		 &    			& NaCl &	sd		&		...						  & (j) \\
C$_{16}$TAC 			 & NaSal 		&			 &	sd/c		&	FB									& (k) \\
C$_{12}$TAB 			 & NaSal		&      & 	sd		&					...						& (l) \\
SDS 			 &					& Al(NO$_{3}$)$_3$ &sd	& 			...							& (m) \\
EHAC 			 &					& NH$_{4}$Cl & sd	& 	...    & (n) \\
EHAC 			 &					& NaCl & sd	& 	FB, SANS, SALS   & (o) \\
EHAC 			 &	NaSal		&      & sd	& 	FB, SANS, SALS    & (p) \\
CTAHNC			&         &      & sd &  ... & (q) \\
CTAT				& SDBS		&				&sd & FB, SANS & (r) \\
SDES				&					& AlCl$_3$&sd&...& (s) \\
SDS				& LAPB  & NaCl & sd & ... & (t) \\
CTAVB     &						&			& sd & ... & (u)\\
\noalign{\smallskip}\hline
\end{tabular}
\end{table}

\begin{figure*}[t]
\begin{center}
\includegraphics[scale=0.62]{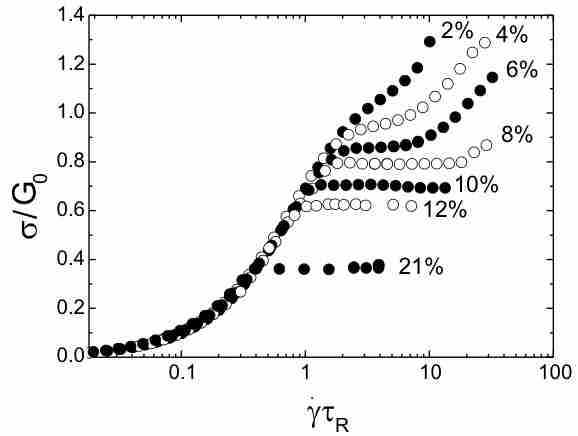}
\caption{ Ge\-ne\-ra\-li\-zed \textquotedblleft flow phase diagram\textquotedblright obtained for CPCl/NaSal system derived from a superimposition between flow curves at different concentrations and temperatures, using normalized coordinates $\sigma/G_{0}$ and $\dot\gamma\tau_{\scriptscriptstyle{R}}$. No stress plateau is observed beyond the critical conditions $\sigma_{p}/G_{0}>0.9$ and $\dot\gamma\tau_{\scriptscriptstyle{R}}\simeq 3\pm 0.5$. From Berret \textit{et al}.~\cite{Ber97a}.}
\label{figIII3}
\end{center}
\end{figure*}

If normalized shear stress $\sigma/G_0$ and shear rate $\dot\gamma\tau_R$ are introduced, it is possible to summarize the overall nonlinear rheological behavior measured at different concentrations and temperatures on a master dynamic phase diagram as shown in Figure \ref{figIII3} \cite{Ber97a}. The flow curve at 21 wt.~\%, a concentration close to the I-N transition, makes the link with the concentrated regime. As concentration decreases, stress plateaus are still observed, but the normalized stress and shear rate at which the discontinuity occurs are shifted to larger values. At 6 wt.~\% and below, the transition between the high viscosity branch and the stress plateau becomes much smoother and the shear stress levels off without discontinuity. Beyond the following critical  conditions $\sigma_{p}/G_{0}>0.9$ and $\dot\gamma\tau_{\scriptscriptstyle{R}}\simeq 3\pm 0.5$, the stress plateau is replaced by an inflexion point. In other words, the above critical conditions suggest that, by choosing the concentration, temperature or salt content adequately, it is possible to find a stress plateau comprised between $\sigma_{p}/G_{0}\simeq 0$ and 0.9, and of onset $\dot\gamma\tau_{\scriptscriptstyle{R}}$ between 0 and 3. A striking point in Figure \ref{figIII3} is that the set of normalized flow curves is strongly reminiscent of the phase diagram of a system undergoing an equilibrium phase transition. 

All the data presented and discussed until now have been gathered with the shear rate as control parameter. However, numerous studies dealing with the effect of an imposed shear stress have been performed both on semi-dilute and concentrated wormlike micellar systems \cite{Kha93a,Gra97,Cap95,Dec97,Cap97a,Mak95,Rag2001,Koe2000,Mu2002,Whe98}. Steady-state flow curves obtained in stress-controlled mode have been found to coincide with flow curves measured under strain-controlled conditions. However, there is a major difference for systems with flat plateaus : it is not possible to reach a stationary coexistence state at imposed stress, since the system directly switches from the low to the high shear rate branch. 

Finally, it is also important to emphasize that the nonlinear rheology of viscoelastic surfactant solutions is characterized  by the existence of normal stresses of non negligible magnitude. In steady-state flow, a non-zero first normal stress difference $N_1$ has been detected once the first stable branch becomes shear-thinning.  $N_1$ was found to increase with $\dot\gamma$ and a slight change of slope was observed at the onset of the banding regime ($\dot\gamma>\dot\gamma_{1}$)  \cite{Reh91,Fis97a,Lop2006,Pim2006}.
Normal stresses in shear-banded flows are much less documented than their shear counterpart. However, they are well-known to drive elastic instabilities for sufficiently high shear rates \cite{Larson92}. Their role is probably essential to explain some fluctuating behaviors observed in shear-banded flows of wormlike micelles \cite{Fie2005,Fie2007,Fie2007a}.

The steady-state mechanical behavior described in this paragraph is representative for entangled wormlike micelles  solutions. In the semi-dilute concentration range however, a few exceptions to this standard behavior have been reported, as briefly discussed below.\\

\paragraph{\label{nsb}Non-standard behaviors}
\indent

In this paragraph, we mention some marginal rheological behaviors encountered in semi-dilute wormlike micelles. This list is  not exhaustive but allows the illustration of the rheological diversity in these systems.

If the stress plateau is the most encountered feature in the rheology of giant micelles, it is also possible to find solutions for which the Newtonian branch is followed by shear-thinning where the flexible chains simply align along the flow direction as in the case of classical polymer solutions. Such a phenomenology has been reported for samples with low or high concentrations of strongly binding counterions \cite{Reh91,Fis97a,Dec2003}.

Another system showing a non-standard behavior is the equimolar solution made of cetylpyridinium chloride and sodium salicylate, the concentration of each component being fixed to 40 mM. This corresponds to a total weight fraction of 2.1 \%. This peculiar system has been extensively studied especially by Fischer's group during the last years \cite{Fis2000,Whe98,Fis2002,Her2005,Her2007a,Her2008} and more recently by Marin-Santibanez \textit{et al}.~\cite{Mar2006}. Its nonlinear rheology has been investigated in various flow geometries~:~at very low shear rates, the solution is Newtonian and then shear-thins, the stress smoothly reaching a pseudo-plateau but without evidence of shear-banding. This regime is followed by a pronounced shear-thickening behavior above a reduced shear rate $\dot\gamma\tau_{\scriptscriptstyle{R}}\simeq 3$ associated with vorticity banding and complex dynamics. We will come back on that point in the section dedicated to time-dependent evolutions. 

At the lowest concentrations in the class of semi-dilute systems, typically ranging from 0.1 to 1 wt.~\%, a simple shear flow can lead to strong thickening above a critical stress \cite{Bol97a,Hu98a,Bol97,Hu98}. This is the case of the TTAA/NaSal solutions, already discussed in the part devoted to the shear-thickening transition (Section \ref{part1}). The overall rheological behavior of such systems, resembles in some respect that of the equimolar CPCl/NaSal solution 40 mM just evoked above. At concentrations around 10 mM, the micellar network is entangled and the static viscosity is larger than that of the solvent by a factor $10-1000$. As the shear rate is increased, there is first shear-thinning and an abrupt shear-thickening at the critical stress. In addition, concomitantly with the shear-thickening transition, shear-induced structures grow from the inner cylinder, as in the shear-banding transition.

Finally, Hoffmann and coworkers \cite{Jin90} have investigated a binary mixture of hexadecyloctyldimethylammonium bromide (C$_{18}$-C$_8$DAB) in water at a concentration of 2.3 wt.~\%. The flow curve of this solution does not present a stress plateau. However, using small angle neutron scattering experiments under simple shear flow, the authors argued that this system undergoes an isotropic-to-hexagonal transition, where cylindrical micelles of different lengths coexist. The short ones contribute to the isotropic phase, while the long ones ensure the long range hexagonal order. This flow-induced transition presents strong similarities with the I/N transition under shear in concentrated wormlike micelles but without the mechanical signature described in the previous section.
\subsubsection{\label{rheotransient}Time-dependent rheology}
\indent

During the past decade, many authors have paid close attention to the evolution of the shear stress as a function of time in systems exhibiting a stress plateau. The aim was to identify the mechanisms responsible for the shear-banding transition. In most cases, shear stress time series in response to steady shear rate consists of a slow transient (compared to the relaxation time of the system) before reaching steady state. Nonetheless, more complex fluctuating behaviors such as erratic oscillations suggestive of chaos or periodic sustained oscillations of large amplitude have been observed in peculiar systems.

\paragraph{\label{sbtrans}Standard transient behavior}
\indent

The time-dependent mechanical response is collected from start-up of flow ex\-pe\-ri\-ments~:~at $t=0$, a steplike shear rate is suddenly applied to the sample at rest and the evolution of the shear stress as a function of time is recorded until steady-state is achieved. For imposed shear rates below $\dot\gamma_1$ and belonging to the Newtonian part of the high viscosity branch, the shear stress follows a monoexponential growth towards steady-state, with a characteristic time corresponding to the Maxwell time of the system \cite{Reh91,Ber97}. When the applied shear rate lies in the shear-thinning region of the high viscosity branch, the stress response shows an overshoot at short time before reaching steady-state, a feature classically observed in concentrated solutions of entangled polymers \cite{Larson88,Menezes82}.
\begin{figure*}[t]
\begin{center}
\includegraphics[scale=0.8]{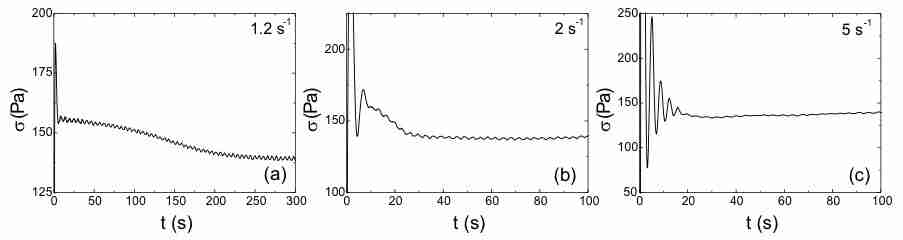}
\caption{Transient shear stress recorded after different step shear rates (a) $\dot\gamma=$ 1.2 s$^{-1}$, (b) 2 s$^{-1}$ and (c) 5 s$^{-1}$ for a semi-dilute sample of CPCl/NaSal (12 wt.~\%) in 0.5 M NaCl-brine at a temperature $T=20.3$ \r{}C. All the applied shear rates belong to the plateau region. Reprinted from Berret \cite{Ber97}.}
\label{figIII4}
\end{center}
\end{figure*}

\begin{figure*}[t]
\begin{center}
\includegraphics[scale=0.6]{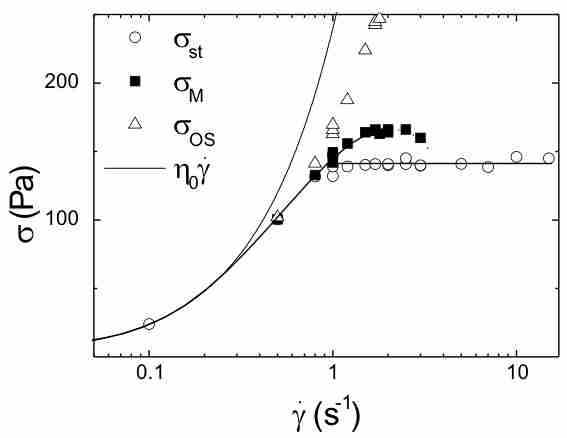}
\caption{Stress overshoot $\sigma_{os}$, initial shear stress  before the onset of the long-time sigmoidal relaxation $\sigma_{\scriptscriptstyle{M}}$ and steady-state shear stress $\sigma_{st}$ gathered from start-up of flow experiments on the semi-dilute sample of CPCl/NaSal (12 wt.\%) in 0.5 M NaCl-brine. The purely Newtonian behavior ($\eta_0\dot\gamma$) has been added for comparison. Reprinted from Berret \cite{Ber97}.}
\label{figIII5}
\end{center}
\end{figure*}

The start-up curves for various imposed shear rates in the plateau region are displayed in Figure \ref{figIII4}. 
For all investigated shear rates, the shear stress exhibits an overshoot at short times, the amplitude ($\sigma_{os}$) of which increases significantly with $\dot\gamma$, followed by a slow relaxation towards  a stationary value $\sigma_{st}$. This relaxation process comprises a latency period during which the stress remains practically constant at a value $\sigma=\sigma_{\scriptscriptstyle{M}}$ and then, a decay of sigmoidal shape whose time scale greatly exceeds the terminal relaxation time of the solution (Fig.~\ref{figIII4}a). The characteristic time  $\tau_{\scriptscriptstyle{N}}$ of this slow relaxation diminishes with $\dot\gamma$ while $\sigma_{\scriptscriptstyle{M}}$ increases (Fig.~\ref{figIII4}b). 
When the mean shear rate is incremented, $\sigma(t)$ shows oscillations at short times that preceed the long sigmoidal decay. The variations of $\sigma_{os}$, $\sigma_{st}$ and $\sigma_{\scriptscriptstyle{M}}$ with the mean shear rate are given in Figure \ref{figIII5}. The $\sigma_{\scriptscriptstyle{M}}(\dot\gamma)$ curve provides evidence for the existence of a metastable branch in which the system is trapped on time scales much longer than the relaxation time $\tau_{\scriptscriptstyle{R}}$. At higher strain rates, the stress response is dominated by damped oscillations (Fig.~\ref{figIII4}c). The  period of the oscillations has been found to decrease with $\dot\gamma$ but, in contrast to nematic wormlike micelles, it does not scale with the inverse shear rate (see Section~\ref{part3}). For concentrated samples, a purely monoexponential decay has been observed \cite{Ber99}. Note that, such transients are often prolonged by a small undershoot before the steady-state is achieved \cite{Ber94,Gra97,Ler2000,Hu2005,Ler2004,Dec2001,Ler2006,Ler2008}.

This type of time-dependent behavior has been observed in various semi-dilute  \cite{Shi88,Ber94,Ber97,Gra97,Ler2000,Bri99,Hu2005,Lop2006,Shi94,Ler98,Kad97,Ler2004,Ler2006,Ler2008,Sol99,Pim2006} and concentrated systems \cite{Ber94,Bec2004,Bec2007,Ber99}. The sigmoidal decay has been modeled by a stretched exponential of the form \cite{Ber94,Ber97,Ber99}~:
\begin{equation}
\sigma(t)=\sigma_{st}+(\sigma_{\scriptscriptstyle{M}}-\sigma_{st})\exp{\left[-\left(\frac{t}{\tau_{\scriptscriptstyle{N}}}\right)^{\alpha}\right]}
\label{eqn4}
\end{equation}
Depending on the system and on the applied shear rate, $\alpha$ has been found to vary between 1 and 4 \cite{Ber94,Ber97,Gra97,Ler2000,Ler2004,Ber99}. 
Such kinetics suggests metastability reminiscent of equilibrium first-order phase transitions and has been originally interpreted by Berret and coworkers \cite{Ber94,Ber97} in terms of nucleation and one-dimensional growth of a fluid phase containing highly ordered entities. Other mechanisms involving the slow drift of a sharp interface to a fixed position in the gap of the cell have also been advanced to explain this slow kinetics \cite{Rad2003,Spe96,Olm97}.

Up to now, we described the time-dependent behavior of the shear stress as a transient towards a steady-state value $\sigma_{st}$. However, the notion of steady-state shear stress has to be made clear. Strictly speaking, at long times, $\sigma(t)$ is not rigorously stationary since it presents fluctuations around an average value defined as $\sigma_{st}$. The relative amplitude of the fluctuations never exceeds 1\% in most of cases. However, some authors have reported  fluctuations of stronger amplitude (typically between 5 and 25\% of the steady-state signal), revealing complex stress dynamics \cite{Ban2000,Hol2003,Fis2000}, that we address in the next paragraphs.
\paragraph{\label{rheochaos}Rheochaos}
\indent

Bandyopadhyay \textit{et al}.~ focused on the time-dependent behavior of semi-dilute solutions of hexadecyltrimethylammonium \textit{p}-toluenesulfonate (CTAT) at weight fractions around 2 wt.~\% in water, with and without addition of sodium chloride (NaCl). This system is well-known to exhibit stress plateau or pseudo-plateau in the flow curve for concentrations ranging between 1.3 and 20 wt.~\% \cite{Ban2000,Sol99,Ban2003,Gan2006a,Gan2008}.

\begin{figure}[t]
\begin{center}
\includegraphics[scale=0.65]{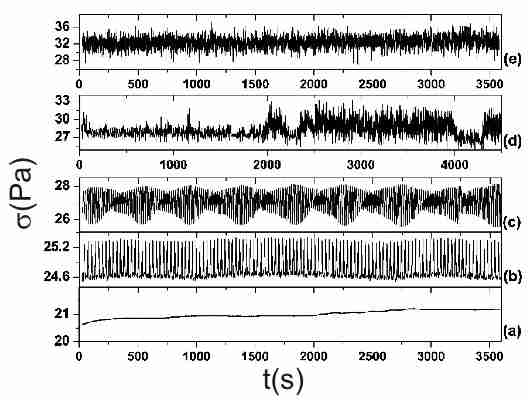}
\caption{Stress time series recorded after start-up of flow experiment at a fixed shear rate $\dot\gamma=25$ s$^{-1}$ for different temperatures (a) $T=31.5$\r{}C, (b) $T=28.8$\r{}C, (c) $T=27.2$\r{}C, (d) $T=26.5$\r{}C, (e) $T=26$\r{}C. The sample under scrutiny is made of hexadecyltrimethylammonium \textit{p}-toluenesulfonate (CTAT) at 2 wt.~\% mixed with 100 mM NaCl in water. The time sequences are found to be (a) time-independent, (b) periodic, (c) quasi-periodic, (d) intermittent and (e) chaotic. Reprinted with permission from R. Ganapathy and A.K. Sood  \cite{Gan2006a}.}
\label{figIII6}
\end{center}
\end{figure}

Typical time sequences observed in this system are illustrated in Fig.~\ref{figIII6}. In this data set, the shear rate is kept fixed and the shear stress is recorded as a function of time with the temperature as the control parameter. From the highest temperatures, the stress temporal patterns appear successively periodic and then quasi-periodic with two dominant frequencies (Fig.~\ref{figIII6}b-c). At lower temperature, the time series still present quasi-periodicity but disrupted by chaotic bursts, typical of intermittency (Fig.~\ref{figIII6}d). If the temperature is further decreased, the signal becomes finally  chaotic (Fig.~\ref{figIII6}e). The existence of deterministic low-dimensional chaos generating erratic fluctuations in the time series is proved by positive Lyapunov exponent \cite{Ott93} and fractal correlation dimension greater than 2. The route to rheochaos is \textit{via} type-II temporal intermittency with a Hopf bifurcation \cite{Gan2006a}. Similar time sequences for the shear rate have been gathered by decreasing the temperature at fixed stress. In that case, the route to rheochaos was found to be of type-III intermittency with period doubling bifurcation \cite{Gan2006a}.\\
The route to rheochaos can thus be tuned by varying the temperature and consequently the mean micellar length at fixed stress or shear rate \cite{Gan2006a}. These results agree with recent theoretical predictions due to Fielding \textit{et al}.~\cite{Fie2004} using the Diffusive Johnson Segalman (DJS) model and adding a coupling of the flow variables with the mean micellar length. Coupling between flow and concentration fluctuations is also supposed to play a major role in the observed complex dynamics \cite{Gan2006,Gan2008}.

The same type of irregular time variations of the shear stress (resp. shear rate) at imposed shear rate (resp. shear stress) have been reported for a given temperature. In both cases, the sequence of figure \ref{figIII6} has been reproduced by increasing the control parameter \cite{Ban2000,Sol99,Ban2003}. 

Note that rheochaos has also been observed in solutions of mixed anionic-zwitterionic surfactants \cite{Pim2006} and in other systems including shear-thickening wormlike micelles \cite{ban2001}, lamellar, onion and sponge surfactant phases \cite{Sal2003,Sal2002,Man2004b,Wun2001,Cou2004} and dense colloidal suspensions \cite{Loo2003}.
\paragraph{\label{vb}Case of the vorticity banding}
\indent

In section \ref{nsb}, we made reference to a complex constitutive behavior for an equimolar solution of CPCl/NaSal  experiencing a shear-thinning to shear-thickening transition. The dynamics of this system has been extensively studied both in strain- and stress-controlled modes. Huge sustained oscillations of the measured quantities (shear and normal stresses or shear rate) as a function of time  have been observed in the shear-thickening regime  \cite{Fis2000,Her2005}. The authors showed that these oscillations are correlated with the existence of a banding pattern organized along the vorticity direction and exhibiting a complex dynamics (see section \ref{orivb}). Note that, unlike the classical shear-thickening transition encountered in dilute surfactant systems and discussed previously, there is no induction time for the shear-induced structures to grow. 

Similar stress dynamics has been observed in a semi-dilute solution of C$_{16}$TAB (50 mM) and NaSal (100 mM) ($c=3.4$\%) that also exhibits apparent shear-thickening \cite{Azz2005}. However, for this system, the phenomenology is different insofar as the shear-thinning region preceding the apparent thickening transition is characterized by a stress plateau and gradient shear-banding. Besides, Decruppe \textit{et al}.~ did not observe vorticity structuring in that case, the flow remaining homogeneous \cite{Dec2006}.
\subsection{\label{vel}Structure of the flow field : velocimetry}
\indent

To elucidate the shear-banding scenario in wormlike micelles, different velocimetry techniques with high spatial resolution, typically between 10 and 50 $\mu$m such as nuclear magnetic resonance (NMR) velocimetry, particle image velocimetry (PIV), particle tracking velocimetry (PTV), photon correlation spectroscopy (DLS) and ultrasonic velocimetry (USV) have been developed. All provide the velocity component along the flow direction taken from a 1D slice across the gap. For details on these techniques, the reader is invited to refer to ref.~\cite{Man2008,Cal2008}. 
\subsubsection{\label{tavp} Long-time response~:~time-averaged velocity profiles}
\paragraph{Semi-dilute systems}
\indent

 The early velocimetry studies of shear-banding flow in wormlike micellar systems have been performed by Callaghan's group using NMR imaging of the semi-dilute CPCl (100 mM)/NaSal (60 mM) ($c = 4.5$ wt. \%) solution in different flow geometries \cite{Cal96,Mai96,Mai97,Bri97a,Bri97b,Bri99}. The method is based on a combination of magnetic field gradient pulses and resonant radio-frequency pulses to encode the NMR signal both from the nuclear spin position as well as translational displacement. The typical spatial resolution is 30 $\mu$m. The acquisition times typically vary from 30 min to 4 hours, the observed banding structures resulting from long time averages. In addition, the use of a specific encoding method allows the determination of pointwise velocity distribution \cite{Cal2008}. 

In millimetric pipe flow, the increase of the flow rate from the Newtonian to the plateau regime was characterized by a transition from a nearly parabolic Poiseuille profile to an almost flat velocity profile with high shear bands near the tube walls, clearly distinguishable from slip \cite{Cal96}. The thickness of this high shear rate band was found to grow with the apparent shear rate \cite{Bri99a}. Broadening of the velocity distribution revealed fluctuations of the flow field on time scales larger than the encoding time (50 ms) and shorter than the total duration of the experiment  \cite{Bri99a,Mai97}. Similar velocity profiles were also obtained from PIV measurements in capillary flow \cite{Men2003a}. From this local description, the authors were able to reconstruct a complete macroscopic flow curve, consistent with the one gathered from bulk rheology. Note that for the PIV technique, a radial laser sheet illuminates a cross section of the sample in a plane ($\vec v,\vec\nabla v$). Images are taken from 90$^{\circ}$ to the laser sheet and velocity profiles are extracted from spatial correlation between pairs of images. 

\begin{figure*}[t]
\begin{center}
\includegraphics[scale=0.5]{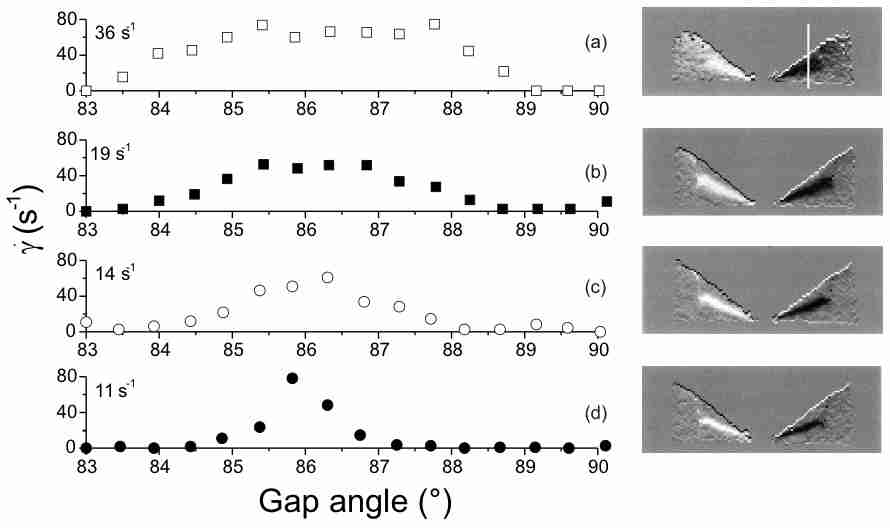}
\caption{Time-averaged shear rate profiles and corresponding grey scale images for the CPCl (100 mM)/NaSal (60 mM) ($c=4.5$\%) solution at $T=25$\r{}C, measured using NMR velocimetry in a 7$^{\circ}$ cone-and-plate at different applied macroscopic shear rates (a) 11 s$^{-1}$, (b) 14 s$^{-1}$, (c) 19 s$^{-1}$ and (d) 36 s$^{-1}$. The profiles are extracted along a line (in white) at fixed radius from the cone axis as illustrated on the images at right. The grey scale indicates the shear rate in arbitrary units. Note the opposite sign shear for the receding and advancing segments of fluid on opposite sides of the gap. The \textquotedblleft free\textquotedblright surface of the fluid is in contact with a containment jacket in Teflon, leading to vanishing of the gradient banding structure beyond a critical radius. Each measurement requires an acquisition time of about 2 hours. Reprinted with permission from Britton \textit{et al}.~ \cite{Bri99}.}
\label{figIII7}
\end{center}
\end{figure*}

In cone-and-plate geometry, where the stress distribution is homogeneous in the small gap approximation, Britton \textit{et al}.~ showed that the flow field along the plateau region was organized in three bands as displayed in figure \ref{figIII7} \cite{Bri99}.
The gray scale images clearly show the presence of a central high shear band flanked by two adjacent low shear regions for all applied shear rates. Shear rate profiles computed along the white line point out the mid-gap position of the high shear band. As the mean applied shear rate is incremented, the high shear band expands in width at a constant maximum shear rate around 60 s$^{-1}$. This value is not consistent with the critical shear rate $\dot\gamma_{2}\simeq$ 100 s$^{-1}$ at the upper limit of the stress plateau. This discrepancy has been ascribed to local fluctuations of the flow field. Velocity fluctuations of different time constants have been observed, depending on sample composition \cite{Bri99a}. 

In Couette geometry, a thin high-shear band ($\simeq$ 30 $\mu$m) near the inner cylinder but in the bulk of the fluid as well as wall slip have been detected \cite{Mai96,Mai97}. A different picture emerged from the NMR-study of the 10 wt.~\% CPCl/NaSal in brine \cite{Hol2003}~:~the shear-banding structure was composed of two macroscopic layers bearing different shear rates, the lower one being compatible with the $\dot\gamma_{1}$ value. The width of the high shear rate band was found to increase significantly with the applied shear rate while wall slip was observed at the moving wall. Enhanced local velocity fluctuations have been reported in the high shear rate region at the vicinity of the inner rotating cylinder.

\begin{figure*}[t]
\begin{center} 
\includegraphics[scale=0.5]{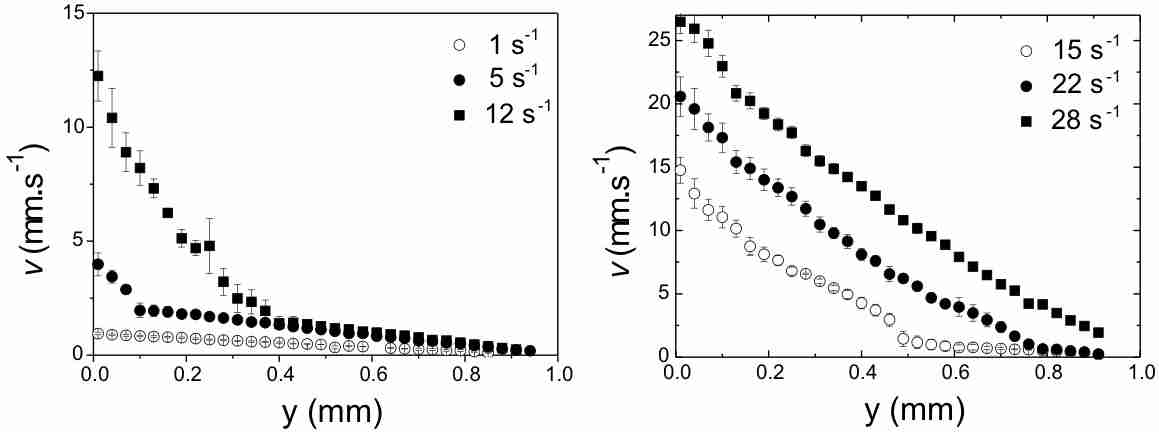}
\caption{Velocity profiles obtained using heterodyne dynamic light scattering for different mean shear rates along the flow curve~:~$\dot\gamma=1$ s$^{-1}$, $\dot\gamma=5$ s$^{-1}$, $\dot\gamma=12$ s$^{-1}$, $\dot\gamma=15$ s$^{-1}$ , $\dot\gamma=22$ s$^{-1}$ and  $\dot\gamma=28$ s$^{-1}$. For reference, the stress plateau extends from $\dot\gamma_{1}\simeq 2.5$ s$^{-1}$ and  $\dot\gamma_{2}\simeq 26$ s$^{-1}$. The sample is 6 wt.~\% CPCl/NaSal in brine and is sheared in a Couette device at a temperature of 21.5\r{}C. The inner rotating and outer cylinders are marked respectively by the positions $y=0$ and $y=1$ mm. The errors bars are representative of temporal fluctuations. Reprinted with permission from Salmon \textit{et al}.~\cite{Sal2003}.}
\label{figIII8}
\end{center}
\end{figure*}

Very similar time-averaged velocity profiles in Couette geometry have also been determined using heterodyne dynamic light scattering (DLS). This technique, developed by Salmon and coworkers, is based on the analysis of the correlation function of the interference signal generated by the mixing of a reference beam and the light scattered by a small volume of sample (typically (50 $\mu$m$)^{3}$) \cite{Sal2003d}. The scattering signal is enhanced by  nanoparticles embedded in the fluid. A few seconds are required to yield the mean velocity for each scanned scattering volume, the total duration for the acquisition of a complete profile reaching a few minutes. This technique has been implemented to examine the precise local structure of the flow in the much-studied 6 wt.~\% CPCl/NaSal in brine \cite{Man2004a,Sal2003}. Figure \ref{figIII8} summarizes the typical velocity profiles gathered for various applied shear rates all along the flow curve, the latter being recorded in a simultaneous way. For $\dot\gamma<\dot\gamma_1$, the velocity profile is continuous and nearly linear, with a slight curvature typical of a weakly shear-thinning fluid, consistent with the evolution of the high-viscosity branch. For $\dot\gamma_1<\dot\gamma<\dot\gamma_2$, the velocity profiles become discontinuous, and are composed of two linear regions of well-distinct slopes. The flow is then non-homogeneous with two coexisting layers supporting very differing local shear rates, the values of which are compatible with $\dot\gamma_{1}$ and $\dot\gamma_{2}$. Note that rapid temporal fluctuations of the flow field in the high shear rate band were reported. The increase of the applied shear rate $\dot\gamma$ along the stress plateau only  affects the relative proportions $f$ and $1-f$ of both bands~:~$1-f$ was found to increase linearly with $\dot\gamma$, hence satisfying the classical lever rule (Eq.~\ref{eqn3}). Finally, for $\dot\gamma>\dot\gamma_{2}$, the flow appears homogeneous again~:~the velocity profiles are characteristic of a strongly shear-thinning fluid, consistent with the evolution of the low viscosity branch. These experiments provided evidence for the classical shear-banding scenario invoked in the introduction (section \ref{introsb}).  \\
This type of velocity profiles, showing the coexistence of two differently sheared regions, the relative proportions of which vary with the applied shear rate, has also been measured using PIV and USV in Couette geometry on CPCl/NaSal in brine \cite{Hu2005,Mil2007} and  C$_{16}$TAB/NaSal \cite{Dec2006}. However, other features have sometimes been reported, including organization of the high shear band into multiple bands, variation of the local shear rate in each band with the control parameter and wall slip at the moving inner wall.\\
\paragraph{\label{cs}Concentrated systems}
\indent

 The structure of the flow field has also been explored in the concentrated regime. Fischer \textit{et al}.~ carried out  NMR velocimetry experiments in Couette geometry on a sample of C$_{16}$TAB/D$_{2}$O at 20 wt.~\%, a concentration just below $c_{\scriptscriptstyle{I-N}}$. For the investigated mean shear rates in the plateau region, the authors identified wall slip at the inner moving wall together with the following banding sequence~:~low-high-low shear rate bands stacked from the inner to the outer cylinder \cite{Fis2000,Fis2001}. The shape of the velocity profile near the moving inner cylinder was interpreted as evidence for near rigid body motion. As we shall see later in the section devoted to the microstructure of the coexisting phases, the comparison of these velocity profiles with ordering profiles simultaneoulsy gathered from NMR spectroscopy experiments led the authors to argue that, surprisingly, the induced nematic state is a state of high viscosity, possibly associated with mesoscale ordering. In addition, local fluctuations of the flow field were suggested from the analysis of the velocity distribution at each pixel across the gap.\\
B\'ecu \textit{et al}.~studied the same system using high-frequency USV. This technique, developed by Manneville and coworkers \cite{Man2004}, is based on cross-correlation of high frequency ultrasonic signals backscattered by the moving fluid. The signals consist of ultrasonic speckles resulting from the interferences of echoes from micrometric scatterers suspended in the fluid. The displacement of the scatterers is estimated from the time shift between two successive echoes. Full velocity profiles are recorded every 0.02 - 2 s, depending on the shear rate, with a spatial resolution of 40 $\mu$m. By averaging velocity profiles over 100 s, the authors observed a similar banding sequence, with an apparently unsheared region at the vicinity of the inner moving cylinder where the fluid velocity is slightly larger than the rotor velocity \cite{Bec2007}. Their interpretation of these peculiar profiles differed somewhat from  the previous one, incriminating a three-dimensional flow instability. We will come back on that point in section \ref{velofluc}.\\

Thereby, all these experiments confirm that the stress plateau in semi-dilute and concentrated micellar solutions is effectively associated with non-ho\-mo\-ge\-neous flow. They also reveal the existence of fluctuations of the flow field and address the question of the role played by wall slip. However, a unified picture of shear-banding has not emerged from the measurement of these time-averaged velocity profiles. A spatio-temporal approach with enhanced  resolution is then required to get a better description of the shear-banding transition. Taking into account this need of high temporal resolution to follow the dynamics of the flow field, some groups have improved their velocimetry technique \cite{Cal2006}, while others have developed new powerful probes \cite{Hu2005,Mil2007,Man2004}.
\subsubsection{\label{trvp} Time-resolved velocity profiles}
\paragraph{\label{velotrans}Transient behavior}
\indent

The early stages of banded-state formation have been studied on different solutions of CPCl/NaSal in brine \cite{Hu2005,Mil2007}.
\begin{figure*}[t]
\begin{center} 
\includegraphics[scale=0.8]{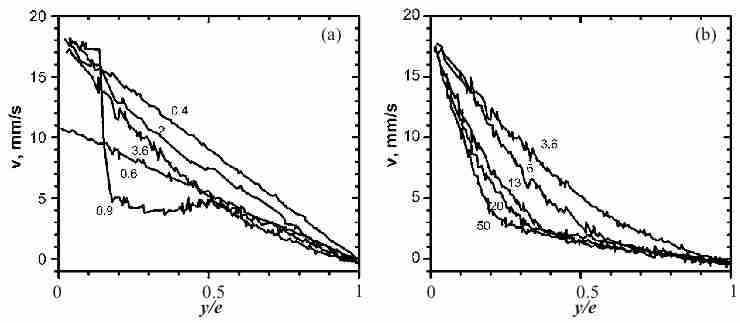}
\caption{Velocity profiles as a funtion of time at $\dot\gamma=10$ s$^{-1}$ obtained in a Couette device of gap $e$ using PTV. The spatial resolution is 10 $\mu$m. The micellar system is the 6.3\% w/v CPCl and NaSal at molar ratio $R$ = [Sal]/[CPCl] = 0.5, in 0.5 M H$_2$O NaCl-brine at $T=23$\r{}C. The kinetics of formation of the banding structure is composed of two main stages (a) A short-time response where the flow stays homogeneous most of the time with increasing  and  decreasing local shear rates respectively at the inner and outer walls. (b) Growth of the low shear rate band from the outer wall. Reprinted with permission from Hu \textit{et al}.~\cite{Hu2005}.}
\label{figIII9}
\end{center}
\end{figure*}
Using a PTV method, Hu \textit{et al}.~ investigated the kinetics of shear-banding on the CPCl/NaSal (6.3 wt. \%) in 0.5 M NaCl-brine \cite{Hu2005}. PTV only differs from PIV by the image processing, which is based on particle tracking rather than spatial correlation. This yields an improved spatial resolution ($\simeq 10$ $\mu$m). Interestingly, the time evolutions of the velocity profiles and the shear stress have been monitored simultaneously, providing information on the correlation between local and global rheology. The transient stress response after start-up of flow of this system has been discussed in section \ref{sbtrans}.\\ Fig.~\ref{figIII9} illustrates the corresponding transient velocity profiles. 
For $t<\tau_{\scriptscriptstyle{R}}$, the velocity profiles are linear across the gap of the Couette device : the flow is homogeneous with no wall slip at the walls (see Fig.~\ref{figIII9}(a)). When the stress overshoot occurs, the local behavior  is found to depend on the applied shear rate. For huge overshoots, significant wall slip is detected at the inner wall, and the velocity profile takes an abnormal shape ($t=0.9$ s). During the short-time relaxation of the stress overshoot, the homogeneous flow is restored. The velocity profile appears then slightly curved ($t=2$ s), the local shear rate increasing towards the inner wall and decreasing towards the outer wall. This process has been ascribed to coupling between stress gradient inherent to the flow geometry and chain disentanglement. When the local shear rate at the outer wall reached the critical value $\dot\gamma_{1}$ ($t\simeq 3.6$ s), the low shear band begins to grow. The expansion of the low shear band towards the inner cylinder together with the increase of the local shear rate in the high shear band are linked to the slow stress relaxation towards steady state including the small undershoot (Fig.~\ref{figIII9}b). The stationary state is characterized by two well-defined shear bands with a relatively broad interface, the position of which fluctuates as a function of time. Contrary to the observations of ref.~\cite{Sal2003}, the local shear rates in each band varies with the imposed shear rate. Under stress-controlled conditions, the development of the banding structure was somewhat different : the kinetics was much slower and the curvature of the initial linear profile was followed by the nucleation and growth of the high shear rate band from the inner rotating cylinder. Note that, using PIV with a high-speed camera, Miller \textit{et al}.~ also identified  multi-stages development of the banding structure \cite{Mil2007}.
\paragraph{\label{velofluc}Fluctuating behaviors}
\indent

\begin{figure*}[t]
\begin{center} 
\includegraphics[scale=1.2]{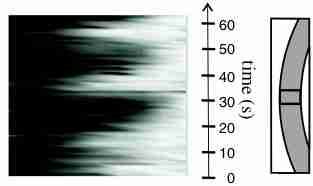}
\caption{Velocity profiles as a funtion of time, recorded at 1 s intervals using rapid NMR imaging in Couette device for an imposed shear rate $\dot\gamma=37$ s$^{-1}$. The left and right sides of the gray scale map correspond respectively to the outer and inner cylinders. The gray levels give the magnitude of the velocity, ranging from 0 to 25 mm.s$^{-1}$. The spatial resolution is 100 $\mu$m. The micellar system placed in the 1 mm gap is made of 10\% w/v CPCl and NaSal at molar ratio $R$ = [Sal]/[CPCl] = 0.5, in 0.5 M H$_2$O/NaCl-brine. The temperature is kept to 25\r{}C. Reprinted with permission from Lopez-Gonzalez \textit{et al}.~\cite{Lop2004}.}
\label{figIII10}
\end{center}
\end{figure*}

Using fast NMR velocimetry imaging, Lopez-Gonzalez \textit{et al}.~ gathered 2D full velocity profiles every second for a semi-dilute sample of  CPCl/NaSal in 0.5 NaCl-brine \cite{Lop2006,Hol2004}. The usable extension of the velocity map along the flow and the gradient directions was 5 mm and the data were averaged over 5 mm along the vorticity direction. 
The long-times flow dynamics was studied for a fixed applied shear rate $\dot\gamma=37$ s$^{-1}$. The corresponding stress time series, recorded independently, exhibited fluctuations by about 5\% around its mean value. Time-resolved velocity profiles across the 1 mm gap extracted from a 1D slice of the velocity map are displayed in figure \ref{figIII10}. In addition to the banding structure described for this system in section \ref{tavp}, the authors demonstrated that the position of the interface between bands strongly fluctuates as a function of time. The high shear rate band (light gray zone in figure \ref{figIII10}) supports a roughly constant shear rate ($\dot\gamma\simeq 70$ s$^{-1}$) and  its width is strongly correlated to the degree of slip at the inner moving wall. The fluctuations are quasi-random or periodic, depending on sample preparation and seem then to be driven by wall slip. A frequency analysis revealed that the correlation time ($\simeq 10$ s) is of the same order of magnitude as the one observed in stress fluctuations. Moreover, along the 5 mm  of the observation window in the velocity direction, the velocity profiles did not change.

\begin{figure*}[b]
\begin{center} 
\includegraphics[scale=1.2]{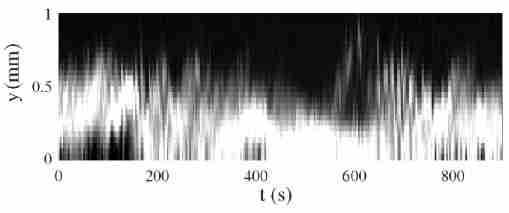}
\caption{Map of the local shear rate $\dot\gamma(y,t)$ determined using USV, for a 20 wt. \% C$_{16}$TAB/D$_{2}$O concentrated micellar system after start-up of flow at a fixed shear rate $\dot\gamma=50$ s$^{-1}$ in the banding regime. The extremum values of the gray scale are associated with $\dot\gamma=0$ s$^{-1}$ (black) and $\dot\gamma>120$ s$^{-1}$ (white). The positions $y=0$ and $y=1$ correspond respectively to the inner moving and outer fixed cylinders. The temperature is kept at 41\r{}C. Reprinted with permission from B\'ecu \textit{et al}.~\cite{Bec2007}.}
\label{figIII11}
\end{center}
\end{figure*}
Similar fluctuations of the banding structure have been observed on the C$_{16}$TAB/ D$_2$O (20 wt.~\%) concentrated system using high frequency USV \cite {Bec2004,Bec2007}. The typical features of the flow dynamics in response to a step shear rate are illustrated in the gray scale map of $\dot\gamma(y,t)$ in figure \ref{figIII11}. First, strong erratic fluctuations of the interface position between the two shear bands from $y \simeq 0.3$ mm to $y \simeq 0.5$ mm occur on time scales ranging from 3 to 200 s. They are correlated to the dynamics of the slip velocities. Second, an unsheared zone appears intermittently in the vicinity of the inner cylinder, with a spatial extension between 200 and 400 $\mu$m. During this process, the time-dependent velocity profiles are highly unstable and composed of three distinct shear bands. The authors emphasized that the local velocity in the unsheared region passes through a maximum that overcomes the rotor velocity, a feature also observable in ref.~\cite{Fis2001}. Taking into account their experimental configuration, they interpreted this particular shape of velocity profiles as the signature of three-dimensional flow. This assumption was reinforced by the organization of the ultrasonic tracers into patterns along the vorticity direction. \\

\paragraph{\label{velovb}Vorticity banding}
\indent

Herle \textit{et al}.~ \cite{Her2008} investigated the equimolar solution CPCl (40 mM) and NaSal (40 mM) ($c = 2.1$ wt. \%) that exhibits a complex flow curve with successively Newtonian, shear-thinning and shear-thickening regimes (see sections \ref{nsb} and \ref{vb}).  They performed pointwise USV velocimetry measurements in Couette geometry with the stress as the control parameter. In the shear-thinning region, the flow was homogeneous. At the onset of the shear-thickening regime, the authors observed the presence of two radial bands supporting different shear rates. At higher imposed shear stresses, coexistence of both radial and vorticity bands has been identified. Large temporal oscillations of the velocity profiles have been reported, related to periodic appearance and disappearance of the vorticity bands.

Note that the same system has been explored using PIV in capillary flow \cite{Mar2006}. The flow curve built from the local velocity profiles was consistent with the one obtained  from the bulk rheology. Above the critical wall stress, shear-thickening zones characterized by convexity in the velocity profiles were observed, extending roughly on half of the capillary diameter. Spatio-temporal oscillations of these profiles together with stick-slip were mentioned, leading to continuous creation and breakage of the induced structures. Similar coupling between changes in the microstructure and stick-slip has been also suggested to explain the \textquotedblleft apparent\textquotedblright shear-thickening branch of the C$_{16}$TAB (50 mM)/NaSal (100 mM) ($c=3.4$\%) solution \cite{Dec2006}.
\subsection{\label{ori}Structural characterization of the banded state: rheo-optics, scattering and spectroscopy}
\indent

In the previous section, the existence of shear bands at constant stress has been demonstrated. The shear bands have differing viscosities and consequently different internal structures. A way  to collect additional information is then to probe the microstructural organization of the system. Different tools have been used to study the shear-banding transition in wormlike micelles, including flow birefringence (FB), small angle neutron and light scattering (SANS and SALS) and NMR spectroscopy. As in the case of the velocimetry, most of these ex\-pe\-ri\-men\-tal techniques have been improved to give a space- and time-resolved description of the transition. In the following, we focus on the local structure of the banded state and on the potential connections with the bulk rheology and the local flow field.
\subsubsection{\label{cbs}Characterization of the shear-bands under steady flow}
\paragraph{\label{vbs} Direct observation using flow birefringence}
\indent

\begin{figure*}[b]
\begin{center} 
\includegraphics[scale=0.9]{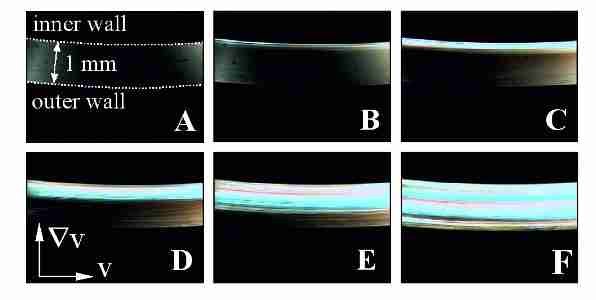}
\caption{ Snapshots of the 1 mm gap of a Couette cell containing a concentrated solution of CPCl-Hex at respective concentrations $c_{\textrm{CPCl}}=28$ wt.~\% and $c_{\textrm{Hex}}=3.9$ wt.~\% for different applied shear rates along the flow curve. The sample is illuminated with white light and observed between crossed polarizer and analyser. The orientation of the polarizer is chosen to match the mean orientation of the micelles in the band adjacent to the fixed cylinder. This configuration makes the latter dark and improves the contrast between bands. Exposure times are typically of the order of milliseconds. With increasing shear rates (from (A) to (F)), the bright induced band fills up the gap progressively. Reprinted from Berret ~\cite{Berret}.}
\label{figIII12}
\end{center}
\end{figure*}

The first evidence for band separation in wormlike micellar solutions came from flow birefringence experiments performed on a concentrated solution by Decruppe and coworkers \cite{Dec95}. The experimental configuration was as follows~:~the sample, placed in the gap of a Couette geometry, was illuminated with a white light source and visualised between crossed polarizers. The transmitted intensity was recorded on a digital camera, the spatial resolution reaching 15 $\mu$m. Figure \ref{figIII12} shows six photographs of a 1 mm-gap filled with a system CPCl/Hex in 0.2 M brine, illustrating the typical scenario observed in concentrated giant micelles \cite{Berret}. Figure \ref{figIII12}A is representative of the optical behavior along the high viscosity branch~:~the gap of the Couette cell appears dark and the birefringence intensity and the average orientation of the micellar medium are homogeneous. Once the first critical shear rate $\dot\gamma_{1}$ is reached, a highly birefringent bright band nucleates against the moving inner cylinder and coexists with a dark band of differing optical properties. The location of the induced band at the inner cylinder is explained by the stress gradient inherent to the Couette geometry \cite{Gre97,Olm2000}. The stress varies as $1/r^{2}$ in the gap, implying that the state of higher stress is reached at the inner wall. When the applied shear rate is further increased along the stress plateau, the bright band broadens (see Fig.~ \ref{figIII12}B-E), the orientation state of the system remaining unchanged in each band as demonstrated by pointwise FB \cite{Lee2005,Mil2007,Ler2004}. Finally above $\dot\gamma_{2}$, the gap is entirely filled by the highly birefringent and oriented shear-induced structures (Fig.~\ref{figIII12}F). 

This scenario was originally interpreted as evidence of the shear-banding transition, the states of low and high shear rates being deduced from the molecular alignment in each band. From a general point of view, it has been observed in concentrated solutions such as C$_{16}$TAB in water and C$_{16}$TAB/D$_2$O \cite{Dec95,Cap97} and in several semi-dilute samples, e.g. in C$_{16}$TAC/NaSal \cite{Mak95,Dec2003}, C$_{16}$TAB/KBr/H$_2$O and -D$_2$O \cite{Dec97,Rad2003,Ler2004}, CPCl/NaSal \cite{Ber97a,Hu2005}, C$_{16}$TAB/NaSal \cite{Ler98,Dra2006} and C$_{16}$TAB/NaNO$_3$ \cite{Ler2000,Dec2001}. Notwithstanding, careful examination of the \textquotedblleft  steady-state\textquotedblright banding structure revealed important differences in the spatial organization from a sample to another with for example, heterogeneities of the induced bright band or multiple bands \cite{Ber97a,Ler2000,Lee2005,Hu2005,Mil2007,Ler98,Dec97,Ler2004}.

In this context, the question of the correspondence between shear and birefringence bands has been raised by some authors  \cite{Fis2000,Fis2001}. In recent studies of semi-dilute CPCl/NaSal solutions, simultaneous measurements of velocity profiles and optical visualisations were performed, showing a good agreement between shear and birefringence bands \cite{Hu2005,Rau2008}. However, this correlation did not seem so obvious for other systems \cite{Fis2000,Fis2001}. We will see later (section \ref{oritdb}) using time-resolved birefringence measurements that all these features could certainly be explained by the existence of fluctuations already revealed by velocimetry experiments.

Quantitative measurements of the birefringence intensity $\Delta n$ and extinction angle $\chi$ (Eq.~\ref{eqn2}) carried out with convenient arrangements of optical components \cite{fuller} have shown a general steady evolution compatible with the one of the shear stress~:~the absolute value of $\Delta n$ first increases linearly with $\dot\gamma$ while, in the same time, $\chi$ decreases smoothly from 45$^{\circ}$, indicating a  gradual alignment of the micelles with respect to flow direction. Above $\dot\gamma_1$, both quantities exhibit a discontinuity of slope followed by a plateau, characteristic of the coexistence of the dark and bright bands \cite{Ler2000,Lee2005,Mil2007,Dec95,Shi94,Whe96,Ler2004,Mak95}. 
Typically, dark and bright bands have birefringence intensity of the order of -10$^{-5}$ and -10$^{-3}$ (the negative sign of $\Delta n$ is due to the anisotropy of the polarisability tensor associated with the monomeric surfactant chain \cite{Shi94,Hum98}), and extinction angle ranging from 20$^{\circ}$ to 40$^{\circ}$ for the dark band and of the order of a few degrees for the bright band. This indicates that the induced structures are strongly aligned with respect to the flow direction. Note that, the stress-optical law, which establishes a linear relationship between stress and refractive index tensors, does not hold in the shear-banding regime \cite{Shi94,Hum98a,Dec2003,Hum98}.
\paragraph{\label{nip}Nature of the induced phases}
\indent

\paragraph{i) Concentrated systems}
\indent

Various micellar solutions, at a concentration close to the isotropic-to-nematic transition at rest, have been studied using SANS \cite{Cap97,Sch94,Ber94a,Ber98} and NMR spectroscopy \cite{Fis2000,Fis2001}.\\
SANS gives information on the orientational degrees of freedom of a fluid subjected to a flow and has been used to probe  the structure of these systems for shear rates all along the flow curve. The sample was placed in a Couette device  and two-dimensional scattering patterns were collected in radial configuration, the incident neutron beam passing through the cell along the velocity-gradient direction. For shear rates along the high viscosity branch, the scattering consists of an isotropic ring and exhibits a broad maximum resulting from strong translational correlations between the micellar threads. The order of magnitude of the distance between micelles, estimated from the position of this maximum, is typically 6 - 9 nm. In the plateau regime, the scattering function becomes anisotropic, with crescent-like peaks in the vorticity direction. Finally, along the low viscosity branch, at high shear rates, the ring-like structure vanishes and the scattering is dominated by anisotropic pattern, qualitatively analogous to the one obtained from a micellar solution which is nematic at rest and subjected to a moderate shearing. These results, showing that the macroscopic phase separation observed in flow birefringence corresponds to shear-induced isotropic-to-nematic transition, have been reported in several systems such as CPClO$_3$/NaClO$_3$ \cite{Sch94}, CPCl/Hex \cite{Ber94a,Ber98} and C$_{16}$TAB/D$_2$O \cite{Cap97}. In addition, the nematic order parameter and the proportions and concentrations of each phase as a function of shear rate in the plateau regime could be derived from the SANS spectra. Breakdown of the simple lever rule has been observed \cite{Cap97,Ber98} and for the CPCl-Hex system, the I-N transition was supposed to result from  flow-concentration coupling \cite{Ber98}.\\

\begin{figure*}[t]
\begin{center} 
\includegraphics[scale=0.75]{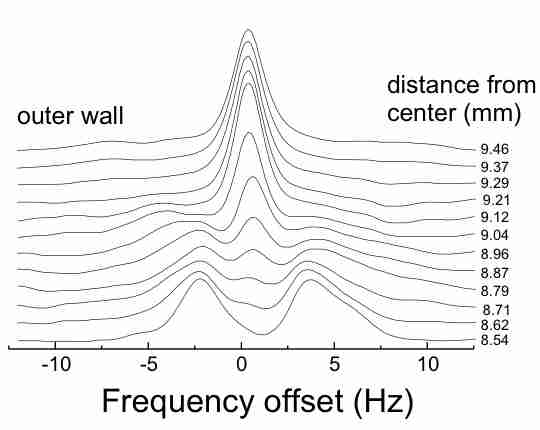}
\caption{$^{2}$H NMR spectra obtained from 20 wt. \% C$_{16}$TAB/D$_2$O (41\r{}C) at different positions across the gap of a cylindrical Couette cell. The applied shear rate is 20 s$^{-1}$. Near the inner wall, a quadrupole splitting is observed, consistent with a nematic phase, while near the outer wall the single peak of an isotropic phase is seen. Both regions are separated by a mixed phase. Reprinted with permission from Fischer \textit{et al}.~\cite{Fis2000}.}
\label{figIII13}
\end{center}
\end{figure*}

NMR spectroscopy is also useful to probe the local microstructure. It allows to resolve spatially the spectral splitting  associated with  the quadrupolar interactions of the deuteron nuclei with the local electric-field gradient. This splitting is actually proportional to the order parameter of the phase that is initiated. Should this splitting be zero, the phase is disordered, should it be non zero, the phase is nematic. The splitting is actually due to the fact that, in an oriented nematic phase of micelles, the D$_2$O molecules of the solvent inherit of the alignment of the cylindrical structures. For a detailed description of this technique, the reader can refer to ref.~\cite{Cal2006,Cal99}. NMR spectroscopy brought confirmation about the nematic order of the induced phase in the C$_{16}$TAB/D$_2$O concentrated sample (20 wt.~\%) as illustrated in figure \ref{figIII13} \cite{Fis2000,Fis2001}.
At a given shear rate, the splitting characteristic of an ordered phase is clearly visible near the inner moving cylinder, while the spectra at the fixed wall are composed of a single peak, indicative of isotropic phase. The completely ordered and isotropic phases are separated by a mixed region. The existence of broad isotropic and nematic bands is compatible with birefringence observations on the same system \cite{Cap97}. Note that, in contrast, the corresponding NMR velocity profiles do not seem to be correlated in an obvious way with this simple picture (see section \ref{tavp}). Notwithstanding, velocity profiles measured using USV seem in reasonable agreement with ordering profiles \cite{Bec2007}. \\
Relative volume fractions of each phase could be computed from the NMR spectra and were found in remarkable agreement with the SANS data. 

\paragraph{ii) Semi-dilute systems}
\indent

For semi-dilute micellar solutions at concentrations far from the isotropic-to-nematic boundary at rest, the situation is less clear, insofar as the set of structural data on different standard systems available in the litterature is reduced, making a definite conclusion about the nature of the induced structures difficult. 

NMR spectroscopy has been used to investigate the local structure of the 10 wt. \% CPCl/NaSal (molar ratio 2:1) in 0.5 M NaCl-brine \cite{Lop2006,Hol2004}. The shear-induced alignment of the micelles was measured by introducing a deuterated dodecane probe molecule into the micellar core. For applied shear rates belonging to the plateau region, splitting of the NMR spectra were observed, indicating a non-zero order parameter, strongly suggestive of nematic order. The volume fraction of the shear-induced nematic phase was found to increase linearly with the mean shear rate and seemed then to follow a simple lever rule. Moreover, proton NMR spectroscopy revealed a strong correlation between molecular orientational dynamics and shear stress.

Shear-induced isotropic to nematic phase transition has also been reported in semi-dilute wormlike micelles solutions   with excess of salt or strongly binding counterions, forming multiconnected networks \cite{Kad98,Sch2004}. Such a string-like phase has been observed using polarized SALS under shear and manifests itself by anisotropic SALS patterns characterized by butterfly or tulip-like shapes with enhanced scattering in the flow direction superimposed to bright streak perpendicular to the flow direction \cite{Kad96,Whe96,Kad97,Kad97a}. These features of the scattering patterns indicate shear-enhanced concentration fluctuations \cite{Hel89} at different length scales and are usually accompagnied by turbidity and flow dichroism \cite{Sch2004}. Recent pointwise SANS experiments on these systems suggested that the high shear rate band, which was turbid and appears strongly striated, was composed of a highly branched concentrated micellar solution coexisting with a nearly isotropic, brine phase \cite{Lib2006}.

Such SALS patterns and shear-induced turbidity have been highlighted in the equimolar CPCl/NaSal system exhibiting vorticity banding (see section \ref{vb}) \cite{Whe98,Fis2002,Her2005}, but also in more classical semi-dilute systems  \cite{Ler2000,Hu2005,Ler98,Dec2001,Ler2006,Ler2008,Gan2006,Gan2008,Rau2008}. In those cases, these phenomena are strongly time- and space-dependent and will be addressed in the following section.
\subsubsection{\label{oritdb}Time-dependent behaviors}
\paragraph{\label{oritran} Transient behaviors}
\indent

\begin{figure*}[b]
\begin{center} 
\includegraphics[scale=0.68]{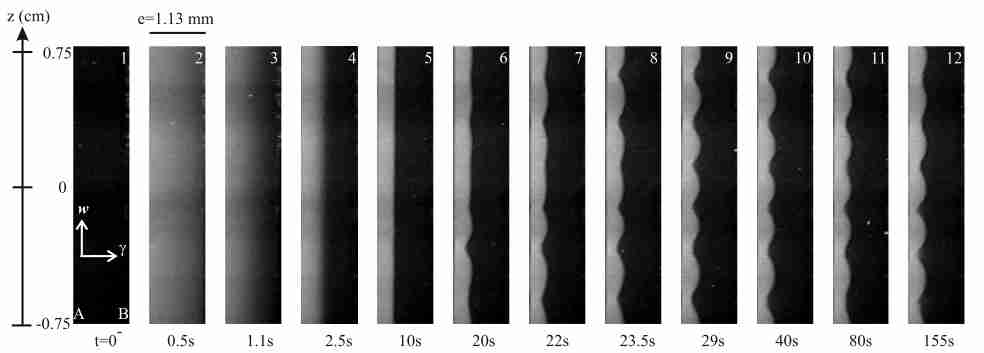}
\caption{Views of the 1.13 mm gap in the plane velocity gradient/vorticity taken out from the recording of the scattered intensity at different times during the formation of the induced band at a shear rate of 30 s$^{-1}$. The sample made of C$_{16}$TAB (0.3 M) and  NaNO$_3$ (0.405 M) ($c=11$\%) at $T=30$\r{}C, is initially at rest and does not scatter the laser light (photo 1). The experimental configuration is as follows~:~a transparent Couette cell is illuminated with a thin laser sheet propagating along the velocity gradient direction and the intensity scattered at 90$^{\circ}$ of the incident beam is recorded on a digital camera. The scattering signal is gathered simultaneously with the temporal stress evolution allowing precise correlation of the structural and mechanical  responses. The left and right sides of each picture correspond respectively to the inner and the outer cylinders. Due to the compromise between the spatial resolution and the size of the field of observation, this latter is limited to 1.5 cm in height (the total height of the inner cylinder is 4 cm) and centered at halfway of the  cell. Reprinted from Lerouge \textit{et al}.~\cite{Ler2006}.}
\label{figIII14}
\end{center}
\end{figure*}

The formation of the banding structure has been explored  using FB, direct visualisations, SALS and turbidity measurements in various semi-dilute wormlike micelles systems including C$_{16}$TAB/NaSal, C$_{16}$TAB/KBr, C$_{16}$TAB/NaNO$_3$, CPCl/NaSal in brine \cite{Ler2000,Lee2005,Hu2005,Mil2007,Ler98,Ler2004,Dec2001,Ler2006,Ler2008}. Note that, in each case, the transient rheology after a sudden step shear rate in the coexistence zone, followed the \textquotedblleft standard behavior\textquotedblright evoked in section \ref{sbtrans}. Interestingly, the temporal evolution of the birefringence intensity and extinction angle, averaged over the gap thickness, showed strong quantitative analogies with that of the shear stress. 

The short-time response was dominated by an overshoot and followed by a stretched exponential or damped oscillations depending on the applied shear rate. The subsequent behaviour was a slow variation towards steady state that could be related to the small undershoot observed in $\sigma(t)$ \cite{Ler2000,Ler2004}.

Figure \ref{figIII14} illustrates the main stages of formation of the banding state for a semi-dilute mixture of C$_{16}$TAB (0.3 M) with NaNO$_3$ (0.405 M) ($c=11$\%) under controlled shear rate. 
\begin{itemize}
\item Step 1~:~At the onset of the simple shear flow, the entire gap becomes turbid (photo 2). The maximum of scattered intensity is reached when the overshoot in $\sigma$, $\Delta n$ and $\chi$ occurs.  The observed turbidity is then supposed to result from the orientation and the stretching of the micellar network \cite{Ler2000,Ler98,Ler2004,Dec2001}, generating concentration fluctuations along the flow direction \cite{Hel89} as suggested by butterfly patterns observed using 2D SALS experiments under shear \cite{Hu2005,Dec2001}. At this time, all the new phase is nucleated but not arranged into a macroscopic band. 
\item Step 2~:~The building of the banding structure starts with the relaxation of the stress overshoot. One can observe the formation of a diffuse interface that migrates from the fixed wall towards its stationary position in the gap and progressively sharpens (see photos 3 to 5). The corresponding  behaviour in the shear stress response is the sigmoidal relaxation or the damped oscillations depending on the magnitude of the averaged shear rate.
\item Step 3~:~When the front is sharp and has reached its equilibrium position, first signs of interface destabilization along the vorticity direction are observed (photo 6). The instability grows with time and finally saturates (photos 7-9). The final state corresponds to the coexistence of a turbid band with a non-turbid band, separated by an undulating interface  with well-defined wavelength and finite amplitude (photo 12). Note that such different scattering properties in each band have also been  reported in ref.~\cite{Lee2005,Hu2005} using polarized SALS. The part of the stress dynamics corresponding to the appearance and the development of the interface instability is the small undershoot that preceeds steady-state \cite{Ler2006,Ler2008}. 
\end{itemize}
A crucial point in this time sequence is that the small undershoot in the $\sigma(t)$ curve appears as the mechanical signature of the interface instability. Since the undershoot has been detected on other semi-dilute systems (see section \ref{sbtrans}), this suggests that the interfacial instability is presumably not inherent to this particular solution.

\paragraph{\label{oridbs} Dynamics of the banding structure}
\indent

If the early stages of formation of the banding structure seem to be common to various semi-dilute solutions, the space and long-time responses strongly differ from a sample to another.
\begin{figure*}[b]
\begin{center} 
\includegraphics[scale=0.5]{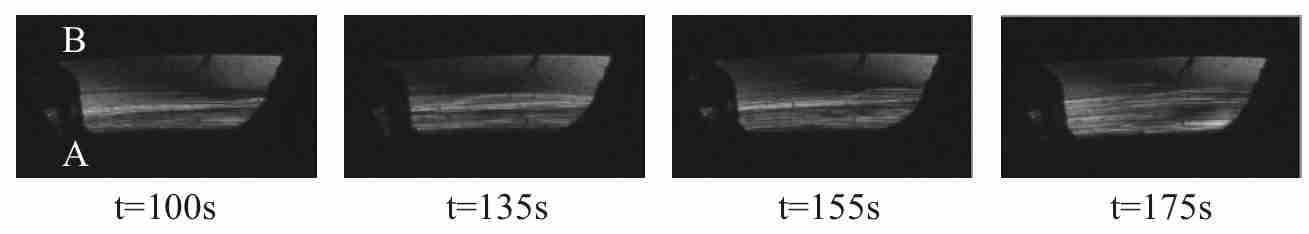}
\caption{Illustration of fluctuating behavior of the banding structure on long time scale. The sample, made of C$_{16}$TAB (0.3 M) and NaNO$_3$ (1.79 M) ($c=11$\%) is sheared at $\dot\gamma=8$ s$^{-1}$ in the annular gap of a Couette cell placed 
between crossed polarizers and illuminated by a source of white light. The temperature is fixed at 30\r{}C. The observation is realized in the ($\vec v, \vec\nabla v$) plane. The inner rotating cylinder and the outer fixed wall of the Couette cell are indicated by the letters A and B respectively. Reprinted from Lerouge \textit{et al}.~\cite{Ler2000}.}
\label{figIII15}
\end{center}
\end{figure*}

Figure \ref{figIII15} displays, for example, the long-time dynamics of a C$_{16}$TAB/NaNO$_3$ solution ($c=11$\%) sheared at $\dot\gamma=8$ s$^{-1}$ in a 1.5 mmm gap of a Couette cell and observed between crossed polarizers. For reference, the critical shear rates for this sample are $\dot\gamma_1=5$ s$^{-1}$ and $\dot\gamma_2\simeq 110$ s$^{-1}$. As expected along the stress plateau, the fluid is splitted into two bands of strongly differing optical properties. The band located against the fixed cylinder is homogeneous. However, a careful examination of the induced band reveals additional features~:~the latter appears striated and is composed of fine sub-bands, the typical thickness of  which is estimated to be 100-150 $\mu$m. These sub-bands are characterized by differing refraction index. They continously nucleate from the inner rotating cylinder and migrate towards the outer fixed wall. The interface between the two macroscopic bands seems unstable and shows fluctuations of position, reminiscent of the pictures observed in velocimetry (see section \ref{velofluc}). Note that the striations extend over approximately half of the gap width. Taking into account the boundaries of the stress plateau, the birefringence bands observed here do not follow a simple lever rule.

This type of dynamical behavior has also been reported in various CPCl/NaSal solutions using FB and SALS \cite{Ber97,Lee2005,Hu2005}. Ref.~\cite{Hu2005} also brought new insights about the fine structuring of the induced band. The simultaneous recording of the velocity profiles showed that the sub-bands support the same local shear rate. Besides, cessation of flow experiments suggested that the sub-bands relaxation time was larger than the macroscopic shear bands lifetime. 
\begin{figure*}[t]
\begin{center} 
\includegraphics[scale=0.6]{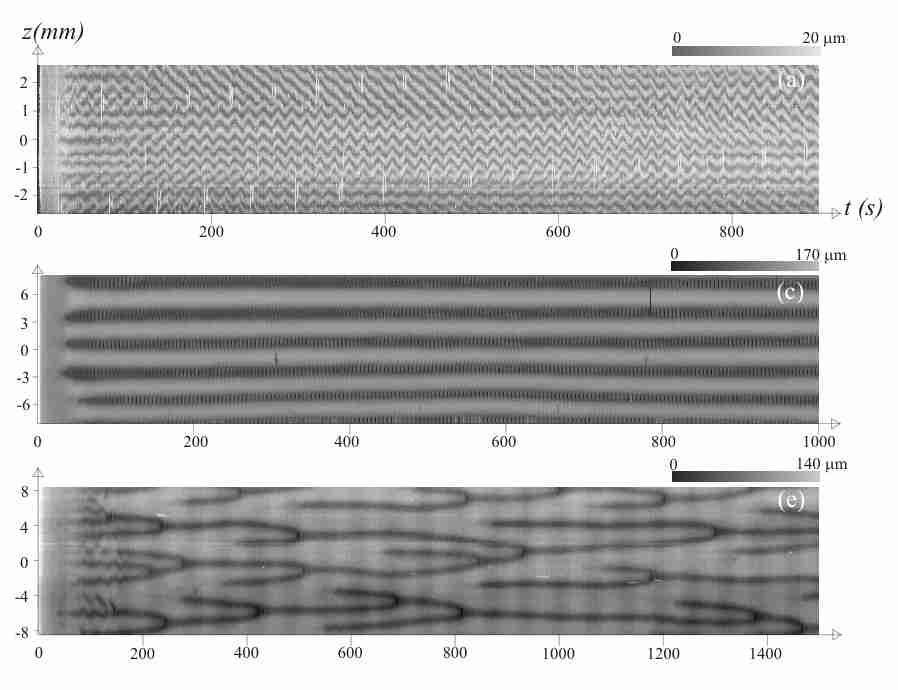}
\caption{Spatiotemporal evolution of the interface position in the gap of the Couette cell during a step shear rate from rest to (a) $\dot\gamma=6.5s^{-1}$, (b) $\dot\gamma=30s^{-1}$, (c) $\dot\gamma=70s^{-1}$. The position of the interface in the gap is given in grey scale, the origin being taken at the inner moving wall so that dark gray regions correspond to positions of the interface nearer to the inner cylinder than light gray regions. The $z$ axis represents the spatial coordinate along the cylinder axis. The sample is made of C$_{16}$TAB (0.3 M) and  NaNO$_3$ (0.405 M) ($c=11$\%) at $T=28$\r{}C. Reprinted from Lerouge \textit{et al}.~\cite{Ler2008}.}
\label{figIII16}
\end{center}
\end{figure*}

The fluctuating character of interface between bands, that comes out from  space- and time-resolved velocimetry and optical experiments, appears then as a robust feature of the shear-banding flow in giant micelles systems. With regard to 1D velocimetry measurements, the experimental configuration of figure \ref{figIII14} enables to follow the spatio-temporal dynamics of the interface using 2D scattering in the velocity gradient-vorticity plane. \\
Three main regimes of dynamics have been highlighted along the stress plateau as illustrated in figure \ref{figIII16} where the grey levels materialize the position of the interface in the gap. Figure \ref{figIII16}a displays a typical spatiotemporal sequence at low shear rates. After a transient including  construction, sharpening and migration of the interface, the pattern exhibits, on large scale, a well-defined wavelength of approximately half of the gap width. At smaller scale the dynamics is more complex~:~the pattern oscillates along the vertical direction of the cell and waves propagate towards the bottom and the top of the cell. At intermediate imposed shear rates in the coexistence zone, the interface keeps a spatially stable profile over very long times. The wavelength of the interface undulation is about three times the gap width (Fig.~\ref{figIII16}b). The amplitude of the interface profile is modulated in course of time.\\
For the highest shear rates in the plateau region, emergence of complex dynamics of the interface is observed (cf. Fig.~\ref{figIII16}c). After a transient slightly under 200 s, the amplitude of the interface profile saturates while the wavelength continuously evolves over time~:~two neighbouring minima have a tendency to merge and when the distance between a pair of minima increases, several other minima close to this pair nucleate and finally merge again with a  minimum of longer lifetime. The system does not seem to tend toward a stationary situation and the spatio temporal diagram strongly suggests chaotic dynamics.
\paragraph{\label{orivb}Case of vorticity banding}
\indent

In section \ref{vb}, we mentioned the particular time-dependent rheological behavior of an equimolar solution of CPCl/NaSal ($c=2.1$ \%). The shear-thickening transition was characterized by strong oscillations as a function of time in the shear and normal stresses at fixed shear rate (or \textit{vice versa}). Using direct visualisations and SALS experiments, the  authors showed that these oscillations were correlated with the existence of a pattern made of clear and turbid rings stacked along the vorticity direction, the position and intensity of which alternate in course of time \cite{Fis2000,Whe98,Fis2002,Her2005}. This complex dynamics, where formation and destruction of shear-induced structures couple with flow instabilities, has been observed in parallel plate, cone and plate and Taylor-Couette cells. Although the development of these  spatio temporal patterns was robust in reference to the flow geometry, the frequency and the amplitude of the oscillations in the bulk rheology were found to depend on the gap width \cite{Her2005}, suggesting that the shear induced structures need some space to fully develop. Recent pointwise SANS and video imaging experiments revealed that both clear and turbid bands contains strongly aligned structures \cite{Her2007a}. Note that vorticity structuring has also been observed in shear-thickening polymer solutions \cite{Hil2002} and in shear-thinning viral suspensions \cite{Let2004,Dho2003,Kan2006}.
\subsection{\label{conc}Conclusion}
\indent

In this part, we described the nonlinear behavior of semi-dilute and concentrated wormlike micelles under simple shear flow. Taking into account the list established in table \ref{tabplateau}, we can argue that most of surfactant wormlike micelles present a \textquotedblleft standard rheological behavior\textquotedblright at steady state, characterized by the existence of a stress plateau in the flow curve. The results accumulated during the last two decades using different velocimetry and optical techniques show unambiguously that the stress plateau is associated with a shear-banding transition. Probing the flow field and the local structure in the plane ($\vec v,\vec\nabla v$) leads to a minimal one-dimensional scenario for the base flow at least when time-averaged measurements are considered~:~all the velocimetry techniques (NMR imaging, DLS, PIV, PTV, and USV) show that, once the stress plateau is reached, the sample splits into two macroscopic layers bearing different shear rates and stacked along the velocity gradient direction. When the control parameter is increased, the high shear rate band progressively invades the gap of the shearing cell. Birefringence experiments also supports this basic picture. The degree of orientation in each layer is very different and the induced structures in the high shear rate band are strongly aligned along the flow direction. For concentrated samples, SANS and NMR spectroscopy under shear indicate that the new phase exhibits long range orientational order of nematic type. 

In addition to this simple picture, strongly fluctuating behaviors have been reported in many semi-dilute and concentrated surfactant systems, including irregular time variations of the bulk rheological signals indicative of chaotic dynamics and fluctuations of the flow field, mainly localized in the high shear rate band. New insights recently emerged from the development of rheo-optical and velocimetry tools with improved spatial and temporal resolution. 1D NMR and USV velocimetry measurements in the ($\vec v,\vec\nabla v$) plane have evidenced, on some well-known systems, that the position of the interface between bands exhibits periodic or erratic fluctuations as a function of time, correlated with the slip at the moving wall. 2D optical visualizations in the ($\vec v,\vec\omega$) plane of one particular semi-dilute sample revealed that the interface between bands is unstable with respect to wave-vector in the vorticity direction. Interestingly, this behavior is associated with a time-dependent mechanical signature shared by a great number of semi-dilute and concentrated systems, suggesting that such interfacial instability is potentially common to different wormlike micellar systems. 

Note that, in a more marginal way, vorticity structuring has also been observed in semi-dilute giant micelles showing a shear-thinning to shear-thickening transition. This particular spatial organization also presents a complex dynamics.

From a theoretical point of view, the shear-banded base flow is a consequence of a non-monotonic constitutive relation between the shear stress and the shear rate. Such a relation has been formulated by Cates using a microscopic approach mote than 15 years ago \cite{Cat90}. Substantial advances in the understanding of the shear-banding transition have been realized using phenomenological models. Such models derived with inclusion of non-local (or spatial gradient) terms in the equation of motion of the viscoelastic stress lead to a unique stress plateau, independent of flow history. They also allow a more realistic description of the interface between bands, taking into account its finite width \cite{Spe96,Olm97,Olm2000,Olm90,Dho99,Olm99,Rad99,Yua99,Lu2000,Rad2000}. 
In the framework of such models, other issues have been addressed such as, the impact of the flow geometry \cite{Olm2000,Rad2000} and its interplay with the boundary conditions \cite{Ada2008} on the banding structure, the role of flow-concentration coupling \cite{Sch95,Fie2003a,Olm99a}, and the effect of the control parameter \cite{Fie2003a,Dho99,Olm99a}.

Recently, a strong effort has also been made to rationalize the complex time-dependent behaviors observed experimentally. Using the diffusive Johnson-Segalman (DJS) model, Fielding \textit{et al}.~ demonstrated that a coupling between mechanical and structural variables such as the concentration or the micellar length can qualitatively reproduce the irregular time-variations of the stress (or shear rate) leading to rheochaos \cite{Fie2007a,Fie2004}. Note that complex dynamics and rheochaos have also been predicted in phenomenological models of vorticity banding \cite{Ara2005,Ara2006}.

The stability of the 1D planar shear-banding flow has also been examined within the DJS model. The interface between bands is found to be unstable with respect to small perturbations with wave-vector in the plane made by the flow and the vorticity directions \cite{Fie2005,Fie2007,Wil2006}. In the asymptotic state, the interface presents undulations along the velocity and vorticity directions. Jumps in normal stresses and shear rate across the interface  are supposed to be the parameters driving the instability.  The nonlinear analysis reveals a complex spatio temporal dynamics of the interface with a transition from travelling to rippling waves depending on the ratio between the thickness of the interface and  the length of the cell \cite{Fie2006}. 
For details on the theoretical state of the art and exhaustive bibliography, the reader is invited to refer to recent reviews on shear-banding \cite{Cat2006,Olm2008,Fie2007a}. 

To conclude, the shear-banding transition in wormlike micelles exhibits complex features beyond the basic one-dimensional scenario. The recent results brought a large number of new information and also open promising perspectives that will require strong experimental and theoretical efforts to fully elucidate the underlying mechanisms. The open issues concern, among others, the organization of the induced structures in the semi-dilute case, the origin of the turbidity and consequently of microscopic length scales in these systems, the role of wall slip in the complex dynamics, the effect of the cell geometry and boundary conditions, the microscopic origin of non local terms, the mechanisms driving the interfacial instability. 
Determination of the complete 3D velocity profiles is also a very challenging but exciting task for the near future. 
\section{\label{part3}Nematic phases of wormlike micelles}
\subsection{\label{introlq}Introduction}
\indent

At surfactant concentrations above 20 - 30 wt. \%, wormlike micellar solutions undergo equilibrium transitions from an isotropic state to nematic and hexagonal liquid crystalline states. Nematic phases exhibit long-range orientational order with no positional order, whereas hexagonal phases show both orientational and translational long range orders of the centers of mass of the micelles. From their orientation and texture properties, these lyotropic phases bear strong similarities with thermotropic liquid crystals \cite{Osw2000}. However with lyotropics, the concentration remains the control parameter for the transitions between the orientationally disordered and ordered states. The nematic phase is characterized by a non-zero order parameter noted $S$, which describes the orientation state of the director \cite{Osw2000,Ons49,Lea79}~: 
\begin{equation}
S=\left\langle\frac{1}{2}\left(3\cos^{2}\psi-1\right)\right\rangle
\label{eqn5}
\end{equation}
In Eq.~\ref{eqn5}, the brackets indicate the averaging over the orientational distribution function, and $\psi$ is the angle between the orientation of a micelle and that of the director. The order parameter is zero in the isotropic phase and equals unity in the fully aligned state. In contrast to the isotropic state, nematic and hexagonal phases are birefringent. Observed by optical microscopy between crossed polarizers, solutions exhibit a strong static birefringence associated with Schlieren (nematic) and fan-like (hexagonal) textures. 

Since the work by Lawson and Flaut on sodium decylsulfate/decanol/water (SdS/ Dec) \cite{Law67}, 
numerous surfactant solutions were found to display long range orientational order at high concentrations \cite{Por86,Gom87,Ama88,Her88,Cap95,Bec2007,Sch94,Ber94a,Itr90,Qui92,Itr93,Ber95,Thi2001,Cla2006,De2007}. 
From magnetic susceptibility measurements \cite{Che77}, two different nematic phases were evidenced. One phase, noted N$_{\scriptscriptstyle\textrm{C}}$ for \textit{nematic calamitic}, was found to be made of rod-like aggregates, whereas the second phase, noted N$_{\scriptscriptstyle\textrm{D}}$ for \textit{discostic nematic} displayed disk-like aggregates. As shown by structural studies on SdS/Dec, the anisotropy ratio in the N$_{\scriptscriptstyle\textrm{C}}$ phase was estimated around 3, yielding a micellar length around 10 nm \cite{Hen81,Hen83}. Interestingly, Porte and coworkers mentioned that the nematic phases could also be obtained with long and flexible aggregates, such as in the system cetylpyridinium chloride/hexanol/brine 0.2 M NaCl (CPCl/Hex) \cite{Por86,Gom87}. There, for a fixed cosurfactant/surfactant ratio, the nematic phase formed a small island located between large isotropic and hexagonal areas. Hexadecyltrimethylammonium bromide (C$_{16}$TAB) and sodium dodecylsulfate/decanol (SDS/Dec) were also shown to display nematic wormlike micelles. These phases were later investigated by rheology \cite{Cap95,Cap97,Bec2007,Thi2001,Cap2002}. 

For semiflexible chains characterized by contour length much larger than persistence length, Semenov and Kokhlov demonstrated that the values of the phase boundaries between the isotropic and nematic states, as well as the order parameter of the nematic phase could be predicted \cite{Sem88}. Following Onsager \cite{Ons49}, it was shown that the boundary $c_{\scriptscriptstyle{I-N}}$ depends only on the ratio between the radius and the persistence length \cite{Ons49,Sem88,Kho81}. For nematic wormlike micelles, this limit was found to vary between 20 wt.~\% and 45 wt.~\%, in fair agreement with the theoretical values \cite{Berret}.\\ 

At still higher concentration, around 40 wt.~\%, hexagonal phases occur. In hexagonal phases, the surfactant aggregates are assumed to be very long and  arranged according to a 6-fold symmetry structure. The translational order confers to these phases a property that is not present in isotropic and nematic phases. Hexagonal phases are strong elastic gels, and as such possess a yield stress \cite{Larson99}. The rheology of hexagonal phases will not be presented in the present review. One reason is the relative low number of papers dealing with shear-induced instabilities and transitions in hexagonal phases. The second reason concerns the nature of the surfactant systems that were examined. These were on one hand nonionic surfactants such as penta(ethylene glycol) monododecyl ether (C$_{12}$E$_5$) \cite{Ric94,Luk95,Mul97,Sch98,Sch2008}, and on the other microemulsions in which the cylinders are swollen with an apolar solvent \cite{Ram2000,Ram2001,Ram2004}. In terms of size, charge and microstructure, these systems are different from those discussed so far. Moreover, the rheology of their dilute or semi-dilute phases was not extensively addressed. We refer to the above citations for details. 
\subsection{\label{rheolq}Rheology}
\subsubsection{\label{rheosslq}Steady-state}
\indent

\begin{figure*}[t]
\begin{center} 
\includegraphics[scale=0.11]{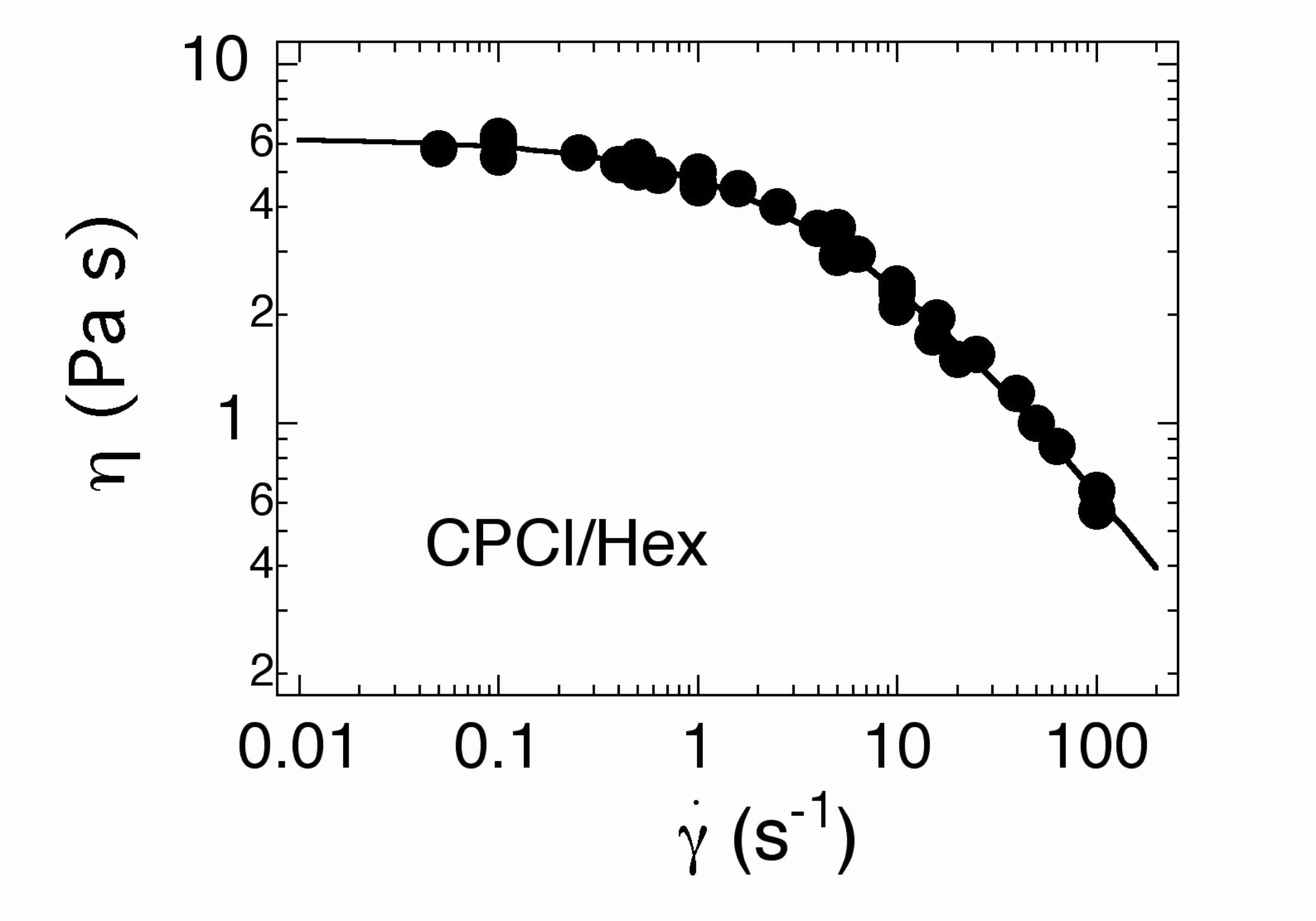}
\caption{Variation of the apparent shear viscosity $\eta=\sigma_{st}(\dot\gamma)/\dot\gamma$ as a function of the shear rate for a solution of CPCl/Hex wormlike micelles in 0.2 M NaCl-brine. The total concentration is $c=36$ \% and the molar ratio [Hex]/[CPCl] = 0.49. The continuous line between the data points is a guide for the eyes. At high shear rates, the viscosity decreases according to a power law with exponent -0.73.}
\label{fig4_1}
\end{center}
\end{figure*}

Fig.~\ref{fig4_1} displays the shear rate dependence of the viscosity at the stationary state for a CPCl/Hex nematic sample at concentration $c=36$ wt.~\% and at molar ratio [Hex]/[CPCl] = 0.49 \cite{Ber95,Rou95}. Up to a shear rate of 1 s$^{-1}$ in Fig.~\ref{fig4_1},  $\eta(\dot\gamma)$ exhibits a \textquotedblleft Newtonian\textquotedblright plateau around 6 Pa s, followed by a shear-thinning regime. This shear-thinning behavior is representative for this class of materials \cite{Cap95,Thi2001,Cla2006,Cap2002}. 
From the early surveys, it was noted that the viscosity of the nematic phase was lower than that of the isotropic solutions located below the transition concentration, a result which is in agreement with theory \cite{Doi86}. In Fig.~\ref{fig4_1}, the shear-thinning behavior was found to be weaker than that of the isotropic solutions. In its asymptotic high shear rate range, the continuous line in Fig.~\ref{fig4_1} corresponds to a shear-thinning behavior of the form $\eta\sim\dot\gamma^{-0.73}$ \cite{Rou95}. 
\subsubsection{\label{rheotlq}Transient, flow reversals and scaling laws}
\indent

Shearing nematic surfactant solutions has revealed complex transient responses \cite{Bec2007,Ber95,Thi2001,Rou95,Ber95a}. When submitted to a step shear rate, nematic samples usually exhibit a transient regime characterized by oscillations of the stress, this regime being followed by a stationary state. The viscosity data in Fig.~\ref{fig4_1} were obtained from these steady state stress values. The main results were obtained on CPCl/Hex \cite{Ber95,Rou95,Ber95a}, C$_{16}$TAB \cite{Bec2007} and on SDS/Dec \cite{Thi2001,Cap2002,Cap99} nematics. They can be summarized as follows~:
\begin{enumerate} 
\item The transient regime depends on the history experienced by the solution prior to the actual measurement. The stationary stress on the contrary is history independent. A convenient way to control the sample history consists in applying a shearing to the sample for a time that is long enough. History controlled procedures made use of preshearing rate $\dot\gamma_{Presh}$. 
\item The time needed to reach the stationary state, $t_{st}$, varies inversely with the shear rate, indicating that whatever $\dot\gamma$, steady state is reached after a constant deformation. For both CPCl/Hex and SDS/Dec, this deformation was of the order of 300 strain units \cite{Thi2001,Rou95}. Nematic surfactant solutions of micelles can be considered to have forgotten their shear history after having been sheared for more than a few hundreds of strain units.
\end{enumerate}
\begin{figure*}[b]
\begin{center} 
\includegraphics[scale=0.1]{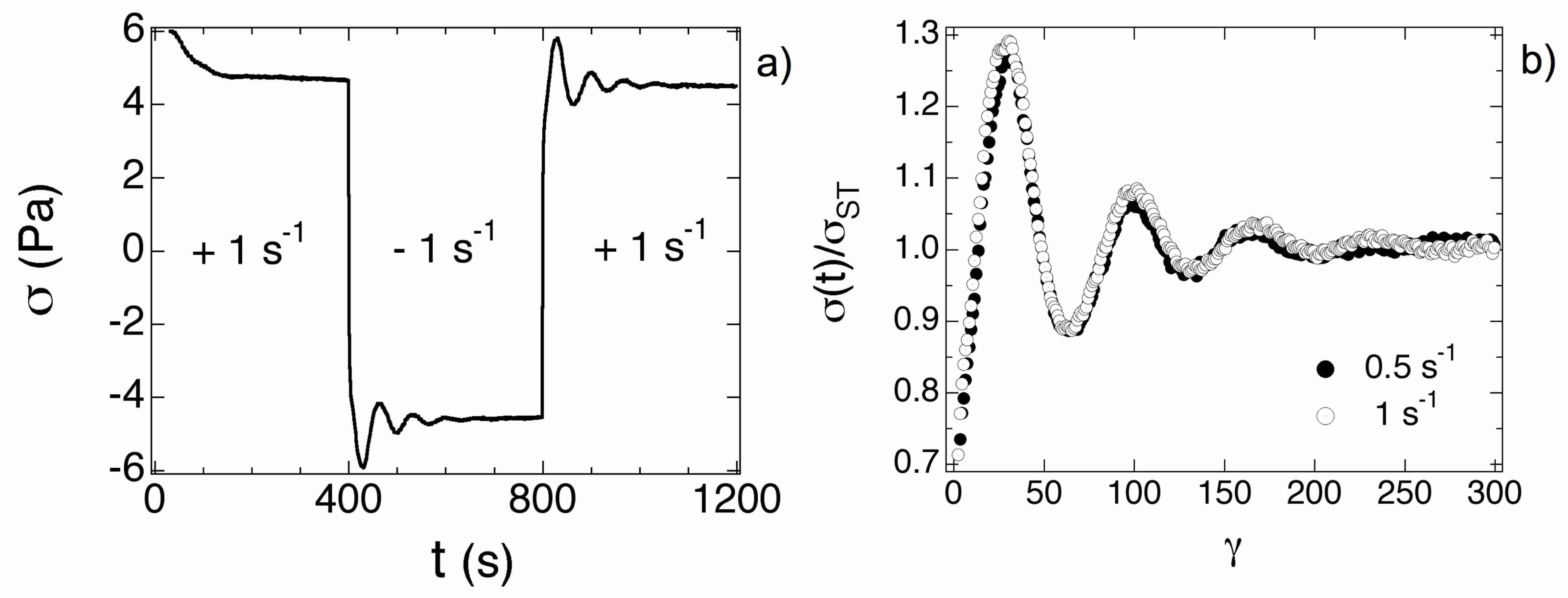}
\caption{(a) Shear stress response obtained from a flow reversal experiment in CPCl/Hex nematic wormlike micelles at total concentration $c=37$ \% and molar ratio [Hex]/[CPCl] = 0.49 \cite{Ber95a}. In the flow reversal, the shear rate is changed stepwise from $+1$ s$^{-1}$ (after sample loading) to $-1$ s$^{-1}$ and then to $+1$ s$^{-1}$. (b) Variation of the normalized shear stress $\sigma(t)/\sigma_{st}$ as a function of the total applied deformation for the same solution. Flow reversal experiments were carried out at  $\dot\gamma= 0.5$ s$^{-1}$ (close circles) and 1 s$^{-1}$ (open circles) and were found to be superimposed.}
\label{fig4_2}
\end{center}
\end{figure*}
The transient responses of nematic wormlike micelles were examined by Berret and coworkers as functions of $\dot\gamma$ and $\dot\gamma_{Presh}$ in the \textquotedblleft Newtonian\textquotedblright and shear-thinning regimes. In this context, flow reversal \cite{Ber95a} and step shear rate \cite{Ber95} experiments were evaluated. In a flow reversal, the sample is first presheared at  fixed rate $\dot\gamma_{Presh}=\dot\gamma_0$ until the steady state condition are reached. The direction of the shear flow is then switched abruptly from $+\dot\gamma_0$ to $-\dot\gamma_0$. In step shear rate experiments, only the amplitude of the shearing is changed, not the sense of rotation. Raw data of two consecutive flow reversals obtained on CPCl/Hex nematic solutions ($c=37$ wt.~\% and [Hex]/[CPCl] = 0.49) are displayed as a function of time in Fig.~\ref{fig4_2}a ( $\dot\gamma_0=1$ s$^{-1}$). Upon flow inception, for a freshly loaded sample, the shear stress exhibited a large overshoot, decreasing then monotonously toward its stationary value. After reversing the flow direction, damped oscillations were observed in the stress responses, either in the negative or in the positive torque ranges. Similar oscillations were found in step shear rates \cite{Ber95}.\\

The approach consisting in preshearing the solutions was actually inspired from the research on liquid crystalline polymers (LCP) \cite{Bur91,Mol91}. LCPs have attracted much attention during the past decades because of their remarkable flow and mechanical properties. In LCPs, the competition between the local molecular relaxation of the polymers and the structure and orientations predicted by classical nematodynamics results in a very complex rheology, for which only mesoscopic theories exist \cite{Lar91}. It is out the scope of the present review to enter into the details of the mesoscopic approaches developed for LCPs, and the reader should refer to the seminal papers by Larson and Doi \cite{Lar91} and Srinivasarao \cite{Sri92}. The first experimental evidence of damped oscillations in the transient shear stress was due to Moldenaers \textit{et al}. on poly(benzyl-L-glutamate solutions) (PBLG) of high molecular weights \cite{Mol91}. PBLG is a lyotropic liquid crystal, where nematogens arise from the hydrogen bonding helix conformation of the backbone. The stress oscillations were related by Larson and Doi to the tumbling instability of the nematic director in the flow \cite{Mol91}. A major property of the constitutive equations derived by Larson and Doi was their scaling behavior. The model predicted that the time evolution of the shear stress is a function of only two parameters, the strain $\dot\gamma t$ applied to the sample  and the ratio  $\dot\gamma/\dot\gamma_{Presh}$ \cite{Lar91,Tak94}, \textit{i.e.}~: 
\begin{equation}
\frac{\sigma(t,\dot\gamma,\dot\gamma_{Presh})}{\sigma_{st}}=F\left(\dot\gamma t,\frac{\dot\gamma}{\dot\gamma_{Presh}} \right)
\label{eqn6}
\end{equation}
where $\dot\gamma/\dot\gamma_{Presh}=-1$ for flow reversals. \\
Fig.~\ref{fig4_2}b displays the results of flow reversals obtained at two different shear rates ($\dot\gamma=0.5$ and $\dot\gamma=1$ s$^{-1}$), where, to obey Eq.~\ref{eqn6}, the stress has been divided by the stationary value, and the time replaced by the deformation. The data at 0.5 s$^{-1}$ and 1 s$^{-1}$ are found to be superimposed, demonstrating the scaling of the stress responses. Scaling laws were also obtained with nematic calamitic and discotic phases of SDS/Dec, however with more complex patterns \cite{Thi2001,Cap2002,Thi98}. It is important to realize that the mechanical responses in Figs.~\ref{fig4_2} are very different from those of the isotropic phases, for which the scaling with deformation or the dependence on the shear history has not been observed \cite{Ber97}.\\
As for CPCl/Hex, the conclusions are twofold~:
\begin{enumerate} 
\item Although polymers and wormlike micelles are very different in nature, the flow properties of their nematic phases are similar. This property was ascribed to the existence of textures at a mesoscopic scale, and to the fact that the dynamics of the textures dominate the mechanical responses of these fluids \cite{Doi91}.
\item One of these similarities concerns the possibility for the nematic director to tumble in the flow.  This tumbling is associated to the periodic oscillations of the stress, as seen in Figs.~\ref{fig4_2}. 
\end{enumerate}

For tumbling nematics, the director is assumed to find no preferred orientation and hence to rotate indefinitely in the flow. Tumbling conditions are met for Ericksen numbers larger than a critical value of about 10 (the Ericksen number is defined as the ratio between the viscous and elastic torques) \cite{Lar93}. 

A second requirement for this instability to occur is that the two Leslie viscosity coefficients $\alpha_2$ and $\alpha_3$ are of opposite signs \cite{Osw2000,Mar85}. Would this ratio between the two viscosities be positive, the director exhibits a different dynamics : it aligns with respect to the velocity at an angle $\theta_0$ such that $\tan^{2}(\theta_0)=\alpha_2/\alpha_3$. Note finally that despite a complex microstructure, the classification in terms of flow-aligning and tumbling nematics, as it is defined for low molecular weight liquid-crystals still applied for lyotropic systems. 

\subsection{\label{miclq}Textures and microscopy}
\indent

A great amount of experimental data using optical microscopy has shown the existence of textures in thermotropic, as well as in lyotropic liquid crystals \cite{Bur91,Lar92,Vermant94}. These textures are in the micrometer range and are interpreted in terms of spatial variations of the nematic director through the sample. At rest, the textures translate into a polydomain structure where the distribution of nematic directors remains constant within each domain. Few optical data have been gathered on surfactant nematics however. For the calamitic phase at rest, Schlieren textures with characteristic line and point defects were observed in various systems \cite{Por86,Ama87,Gom87,Bec2007,Thi2001,Hen81,Rou95}. As samples were allowed to relax for some hours, the Schlieren textures coarsened and an alignment resulted due to the interactions with the glass surfaces of the cell. \\

Under shear, texture refinement was predicted, and observed for LCPs as resulting from the competition between the viscoelastic and the elastic Frank stresses. Here, the Frank stresses arise from the orientational heterogeneities of the director occurring at the transitions or walls from domain-to-domain. The balance between the elastic and viscous energy density allowed Marrucci to derive that the texture lengths should decrease with the shear rate as $\dot\gamma^{1/2}$. Roux \textit{et al}. have performed rheo-optical measurements on CPCl/Hex nematics in order to assess these predictions \cite{Rou95}. At low shear rates, striped textures parallel to the flow could be identified. The texture lengths were estimated of the order of 5 - 10 $\mu$m, depending on the applied rate. A refinement of the stripes was also observed with increasing shear rate, but not estimated quantitatively. Stop-flow experiments on the same system were also carried out and allowed to confirm the predictions of the mesoscopic domain theory \cite{Lar91}. At the cessation of the flow, dark and bright stripes perpendicular to the flow velocity were found, with a band spacing  growing as $\sim t^{1/2}$. Similar results were observed in LCPs \cite{Vermant94}. In conclusion of this section, the rheo-optical data confirm a polydomain structure, the coupling of the textures to the flow and the close analogy with liquid crystalline polymers. 
\subsection{\label{dirlq}Director orientations under shear~:~scattering and NMR}
\indent

Small-angle neutron and X-ray scattering were undertaken in order to retrieve the local orientation of the micellar threads at rest and under shear. As shown on thermotropic \cite{Lea79} and lyotropic \cite{Por86,Gom87,Hen81,Hen83}  liquid crystals at rest, these two techniques provide unambiguous signatures of nematic long-range orientational order through the generation of a diffuse scattering patterns. For experiments at rest, the alignment of the nematic director was achieved using an externally applied magnetic field.

\begin{figure*}[t]
\begin{center} 
\includegraphics[scale=0.13]{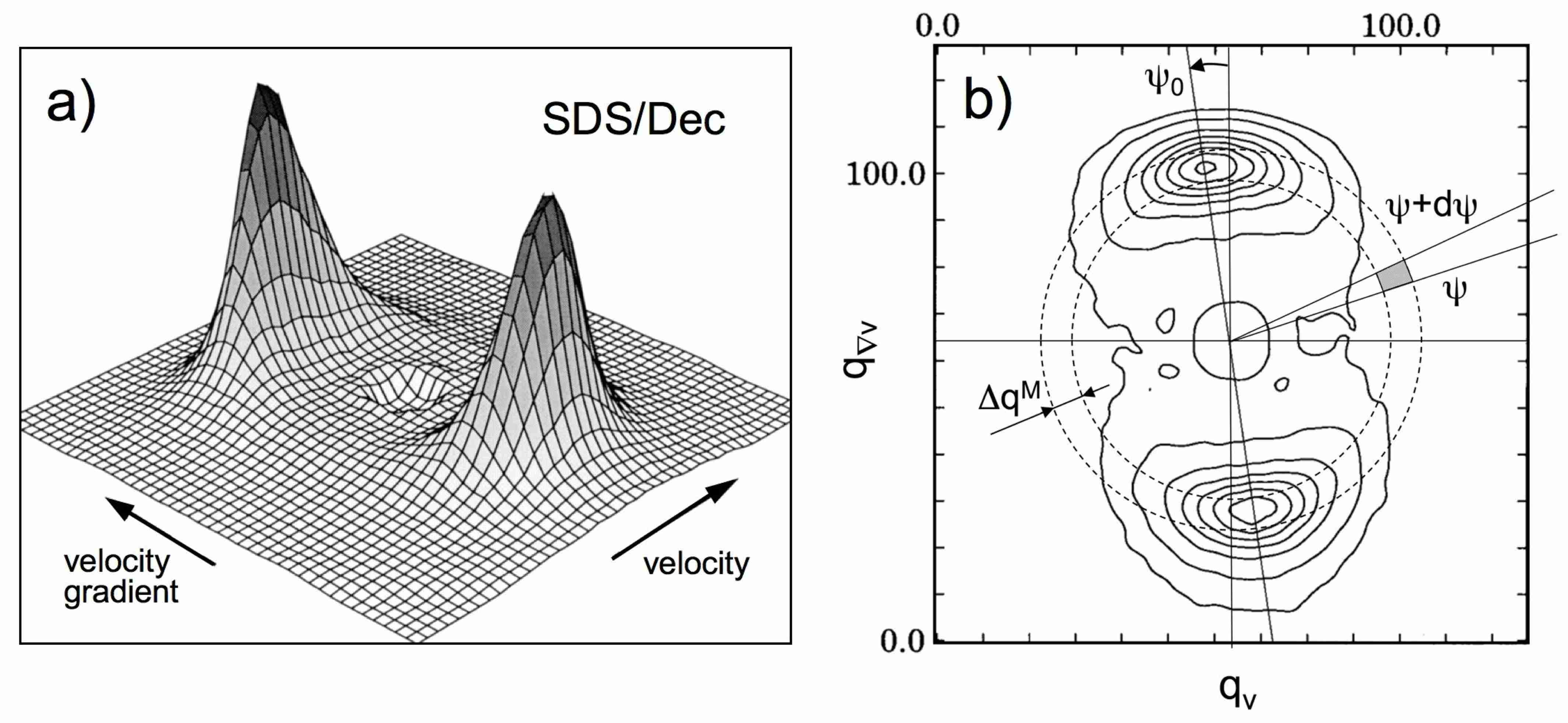}
\caption{(a) 3D-plot of the neutron scattering cross-section measured in the $(\vec q_v,\vec q_{\nabla v})$-vorticity plane for the SDS-Dec nematic calamitic phase. The SDS/Dec solution was prepared in D$_2$O for contrast reasons at total concentration $c=29.5$ wt.~\% and molar ratio [Dec]/[SDS] = 0.33 \cite{Berret2000b}. The shear rate of the experiment was $\dot\gamma=50$ s$^{-1}$. (b) Iso-intensity contours obtained for the same conditions as in a). Along the $\vec q_v$- and $\vec q_{\nabla v}$-axis, the wave-vector ranges from $-0.16$ \AA $^{-1}$ to $+0.16$ \AA $^{-1}$. The dashed circles set the limits of the domain of integration for the data treatment (see text).}
\label{fig4_3}
\end{center}
\end{figure*}
Under shear, Couette or cone-and-plate cells specifically designed for small-angle scattering \cite{Lin90,Noi97} were used at large scale facilities by many authors. With the help of two-dimensional detectors, the anisotropy in the forward scattering was targeted. This combined configuration of shearing and detection allowed to observe the scattering in the three planes of the reciprocal space,
 ($\vec q_v$,$\vec q_{\nabla v}$), ($\vec q_v$,$\vec q_{\omega}$) and ($\vec q_{\nabla v}$,$\vec q_{\omega}$), where we recall that $\vec q_v$, $\vec q_{\nabla v}$  and $\vec q_{\omega}$ are the wavevectors parallel to the velocity, velocity gradient and vorticity. Most of the data on nematic wormlike micelles were collected in the ($\vec q_v$,$\vec q_{\omega}$) and ($\vec q_v$,$\vec q_{\nabla v}$)-planes \cite{Ber94,Cap97,Bec2007,Sch94,Rou95,Ber95a,Cap99,Zip99}. 
Fig.~\ref{fig4_3}a displays a typical 3-dimensional plot of the neutron intensity scattered by a nematic lyotropic solution in the ($\vec q_v$,$\vec q_{\nabla v}$)-plane. The data were obtained on the SDS/Dec calamitic phase at 50 s$^{-1}$  (concentration $c=29.5$ wt.~\% and [Dec]/[SDS] = 0.33). As shown in the iso-intensity contour plot (Fig.~\ref{fig4_3}b), the patterns are characterized by two crescent-like peaks aside from the velocity axis. The maximum scattering corresponds to the first order of the structure factor, from which the distance between the center-of-mass of the micelles can be estimated (here 6 nm for a radius of $\sim$ 2 nm). The modulation of the azimuthal intensity is also of interest since it reflects the distribution of micellar orientations. The spectra were analyzed in terms of angular distribution of the scattered intensity. The scattering was integrated over an elementary surface  $dq_vdq_{\nabla v}= q^{M}\Delta q^{M}\Delta\psi$, where $\Delta q^{M}$ corresponds typically to the half width at half maximum of the peak. 
Plotted as a function of the azimuthal angle $\psi$, two parameters could be retrieved~: 
\begin{figure*}[t]
\begin{center} 
\includegraphics[scale=0.1]{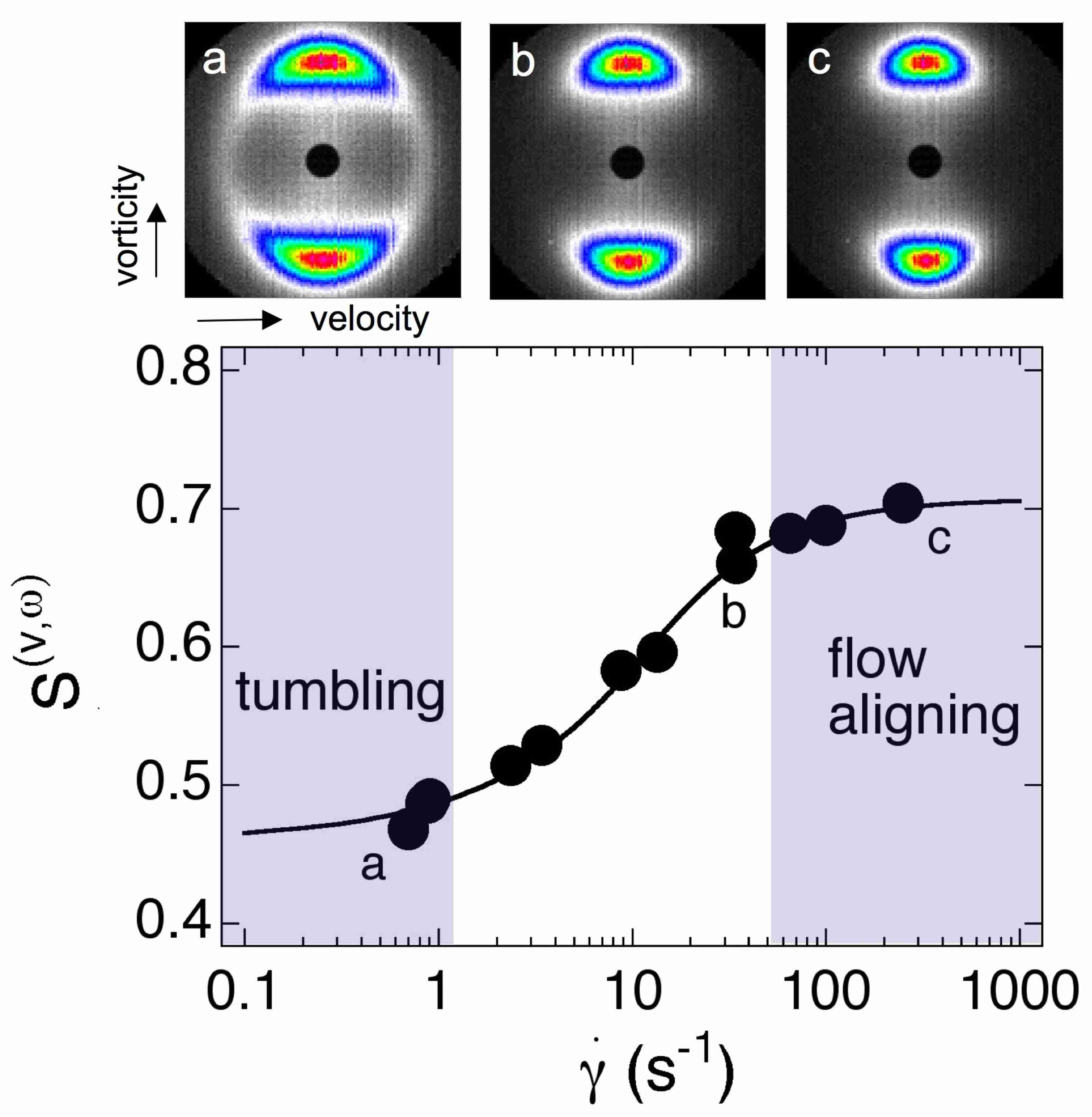}
\caption{Upper panels~:~Neutron scattering intensities obtained from a CPCl/Hex nematic micellar solution in the $(\vec q_v,\vec q_{\nabla v})$-plane at different shear rates~:~a) $\dot\gamma=0.94$ s$^{-1}$; b) $\dot\gamma=34.5$ s$^{-1}$ and c)  $\dot\gamma=250$ s$^{-1}$. The nematic phase was made in deuterated water at concentration $c=35.2$ wt.~\% and molar ratio [Hex]/[CPCl] = 0.49 \cite{Rou95}. Lower panel~:~Shear rate dependence of the orientational order parameter $S^{v,\omega}$ (Eq.~\ref{eqn5}) obtained from the SANS cross-sections shown in the upper panel. The continuous line is a guide for the eyes. The increase of the order parameter was interpreted in terms of a transition between the tumbling and the flow-alignment regimes \cite{Cap2002,Cap99}.}
\label{fig4_4}
\end{center}
\end{figure*}

\begin{itemize}
\item the order parameter $S$ of the nematic phase under shear (Eq.~\ref{eqn5}). The analytical technique to transform the azimuthal intensity into an orientational distribution has been quoted in several papers \cite{Deu91,Osw2000,Lea79}.
\item the tilt angle $\psi_0$ of the scattering pattern with respect to the vertical axis. It was shown that the tilt angle angle $\psi_0$ was actually the orientation angle between the nematic director and the flow. By symmetry, it is zero in the ($\vec q_v$,$\vec q_{\omega}$)-plane, non-zero in ($\vec q_v$,$\vec q_{\nabla v}$)-plane \cite{Cap2002,Ekw69} and changes sign by flow reversal \cite{Berret2000b}.\\
\end{itemize}
In Fig.~\ref{fig4_4}, the order parameter $S^{v,\omega}$ is shown as a function of the shear rate, together with characteristic SANS spectra obtained on CPCl/Hex nematic solution. It illustrates that shear orients the nematic phase steadily and that this process is concomitant to the shear-thinning behavior (Fig.~\ref{fig4_1}). As recognized by Burghardt and coworkers \cite{Cap2002,Cap99}, bulk measurements of $S$ reflect the distribution of micellar orientations around the local director as well as the distribution of director orientations in the polydomain sample. The progressive increase from $S^{v,\omega}=0.45$ to $S^{v,\omega}=0.70$ in Fig.~\ref{fig4_4} was ascribed to the transition between the polydomain tumbling regime toward the flow-alignment monodomain regime \cite{Cap2002,Rou95,Cap99}. 
Using synchroton X-ray radiation Caputo \textit{et al}. reported the time dependence of the order parameter $S^{v,\nabla v}$ in the ($\vec q_v$,$\vec q_{\nabla v}$)-plane \cite{Cap2002,Cap99}. SAXS spectra were collected every second during step shear rate and flow reversal testing. It was shown by these authors that the orientational response of CPCl/Hex nematics was consistent with the Doi-Larson model developed for the LCPs, \textit{i.e.} that the order parameter decreased after reversing the flow, a result that was not seen on classical LCPs. \\

The second surfactant nematics that was studied thoroughly, SDS/Dec, shows a different behavior. 
Although it is a textured material with a bulk shear-thinning rheology similar to CPCl/Hex or C$_{16}$TAB, this system exhibits what appears to be common characteristics of aligning nematics~:~\textit{i)} a constant orientation state and order parameter in steady shear, \textit{ii)} a single undershoot of long duration in average orientation upon flow reversal and \textit{iii)} no significant orientation change upon step increase or decrease in shear rate or upon flow cessation \cite{Thi2001,Cap2002,Thi98,Berret2000b}.  Caputo \textit{et al}. concluded that this systems followed the polydomain model predictions of transient orientation for aligning nematics \cite{Cap2002}.\\
\begin{figure*}[t]
\begin{center} 
\includegraphics[scale=0.11]{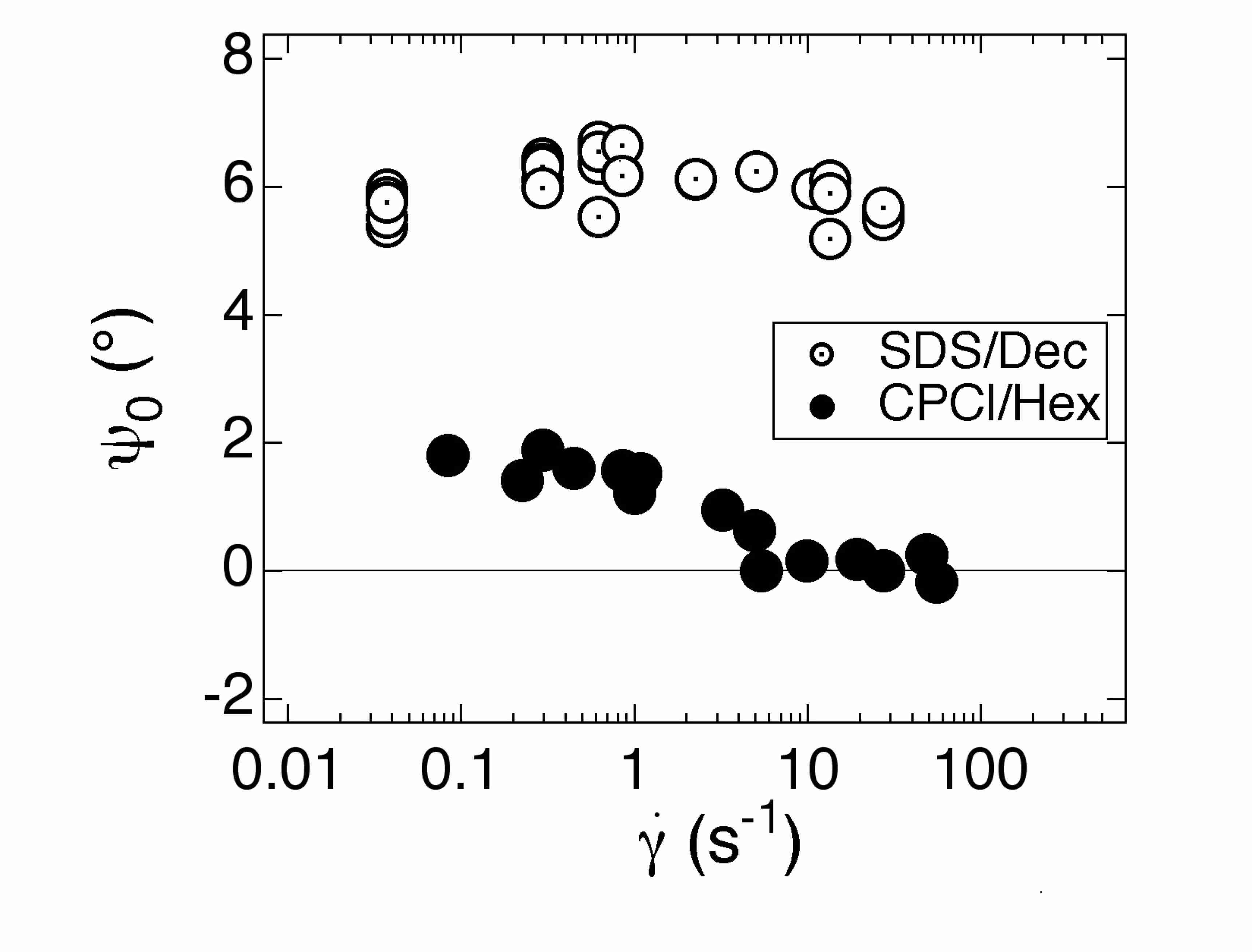}
\caption{Evolution of the orientation angle $\psi_0$ as a function of the shear rate for the SDS/Dec and CPCl/Hex nematics. Samples characteristics are those of Figs.~\ref{fig4_3} and \ref{fig4_3} \cite{Berret2000b}.}
\label{fig4_5}
\end{center}
\end{figure*}
As an illustration, Fig.~\ref{fig4_5} displays the evolution of the orientation angle $\psi_0$ as a function of the shear rate for both SDS/Dec and for CPCl/Hex \cite{Berret2000b}. For SDS/Dec, the orientation of the director with respect to the velocity remains unchanged at $\psi_0=6^{\circ}$ over three decades in shear rates. For the tumbling nematic micelles, the tilt angle has a non zero value at low shear rates, $\psi_0=1.8^{\circ}$, but decreases to zero with increasing shear rates. 
The shear rates at which $\psi_0$ deviates from its low shear rate value corresponds to the upper limit of the tumbling regime. As the system enters into the flow alignment regime, the orientation angle $\psi_0$ decreases to zero, yielding a fully symmetric scattering pattern. 
These results are in qualitative agreement with those obtained on tumbling LCPs by flow birefringence \cite{Hon94}.\\
Using $^{2}$H NMR-spectroscopy under shear, the flow-aligning properties of SDS/Dec calamitic phase prepared in D$_2$O could be confirmed. In this experiment, the anisotropic motions of the D$_2$O molecules in contact with the micellar surfaces result in a small residual quadrupole coupling, which is related to the angle  between the nematic director and the magnetic field \cite{Gra94}. The line splitting $\Delta\nu$ corresponding to this coupling is given by $\Delta\nu$=3/4 $\delta(3\cos^{2}\theta-1)$ where $\delta$ is the quadrupole coupling constant. In the configuration adopted, the magnetic field was parallel to the velocity gradient, \textit{i.e.} $\theta=\pi/2-\psi$, where $\psi$ is the tilt angle defined in Fig.~\ref{fig4_3}b. 
Fig.~\ref{fig4_6}a illustrates a series of spectra obtained for the nematic calamitic sample at $\dot\gamma= 0.32$ s$^{-1}$ and at different deformations $\gamma$ after inception of shear. With increasing deformation, a decrease of the splitting is first observed. At $\gamma\sim$ 1.4, the two resonance lines overlap at the magic angle ($\theta=54.7^{\circ}$). When the deformation is further increased, the splitting stabilizes at the steady state value, which is slightly lower than half of the initial splitting. Beyond $\gamma=15$, no change is observed, neither in the splitting nor in the line shape. The strain evolution in Fig.~\ref{fig4_6}b shows an increase and a saturation at  $\theta_0$ = 78 $\pm$ 2$^{\circ}$. This result is again a strong indication that the SDS/Dec nematic is flow-aligning. The flow-alignment behavior can be checked by comparing the director orientation as a function of strain, $\theta(\dot\gamma)$ as predicted by the Leslie-Ericksen continuum theory \cite{Larson99,Osw2000}. Neglecting the effect of the magnetic field \cite{Sie97} the evolution of the director angle reads~:
\begin{equation}
\theta(\gamma)=\arctan\left[\sqrt\frac{\alpha_2}{\alpha_3}\tanh\left(\frac{\sqrt{\alpha_2\alpha_3}}{\alpha_3-\alpha_2}\gamma\right)\right]with \;\alpha_2,\alpha_3<0	
\label{eqn7}
\end{equation} 
where $\alpha_2$ and $\alpha_3$ are Leslie coefficients. The continuous line in Fig.~\ref{fig4_6}b has been calculated using Eq.~\ref{eqn7} with $\alpha_2/\alpha_3 =\tan^{2}(\theta_0)=25$ as unique fitting parameter. The good agreement between the experimental and the calculated $\theta(\gamma)$ is a further indication of the flow-alignment character of the N$_{\scriptscriptstyle\textrm{C}}$ phase. For a tumbling system a more complex transient evolution of the director orientation is expected, and was indeed observed for the tumbling PBLG \cite{Sch2008}. 
\begin{figure*}[t]
\begin{center} 
\includegraphics[scale=0.13]{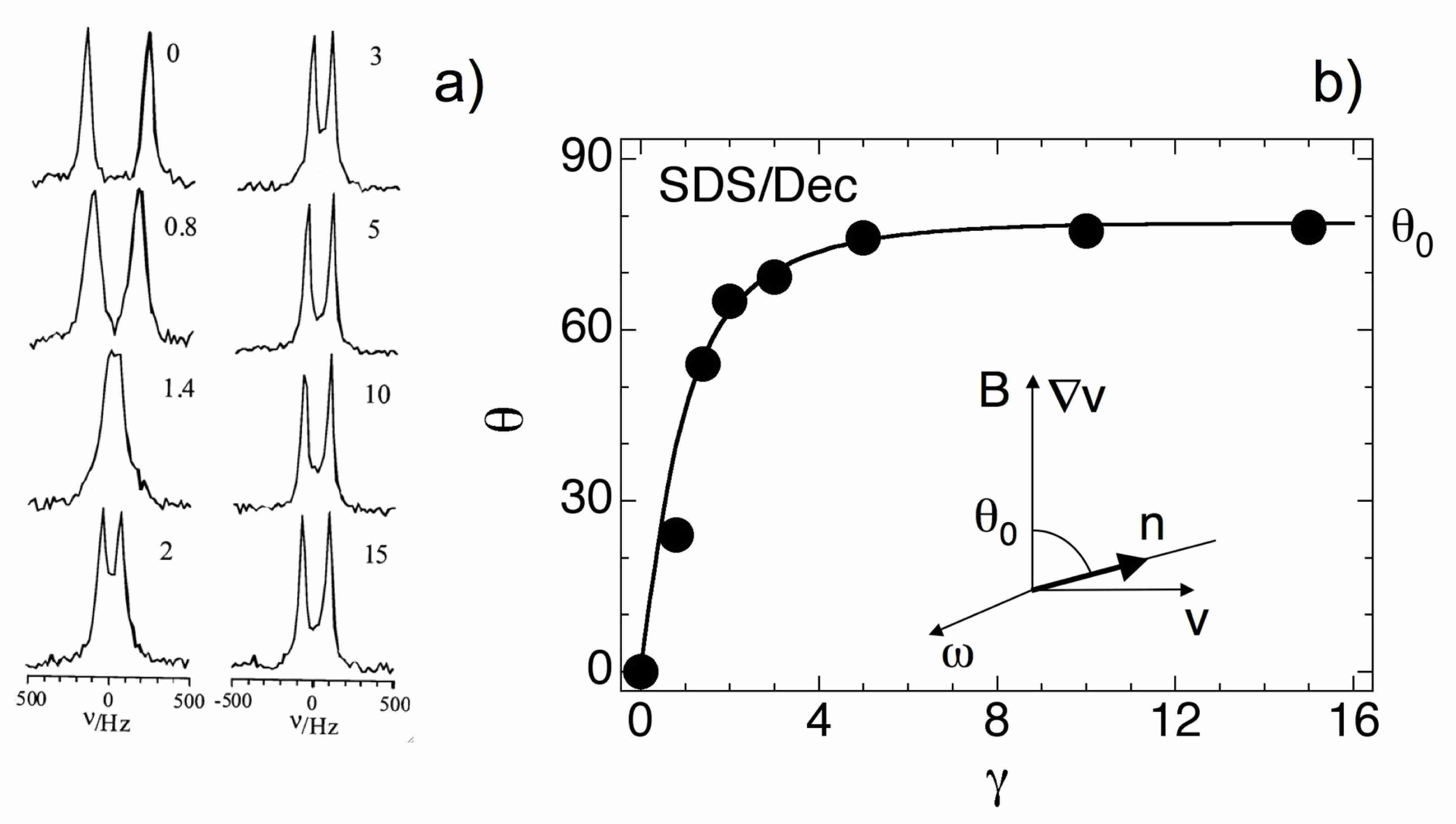}
\caption{(a) Transient behavior of the NMR splitting for a SDS/Dec nematic calamitic solution at concentration $c=29.5$ wt.~\% and molar ratio [Dec]/[SDS] = 0.33 and sheared at  $\dot\gamma= 0.32$ s$^{-1}$ \cite{Thi2001}. (b) Evolution of the orientation angle of the nematic director with respect to the magnetic field as function of the strain. The symbols correspond to the spectra in a), and the continuous line is determined from Eq.~\ref{eqn7}, yielding an asymptotic angle of  $\theta_0=78$ $\pm$ 2 $^{\circ}$. Inset~:~Orientation of the nematic director in presence of the flow and of the magnetic field.}
\label{fig4_6}
\end{center}
\end{figure*}
\subsection{\label{conlq}Conclusion}
\indent

Using scattering and spectroscopy experiments, it has been shown that the physical quantities characterizing the wormlike micellar nematics such as the order parameter, Leslie viscosities ratio or alignment angles can be determined. The main result of this section is the analogy between the wormlike micelles and the liquid-crystalline polymers, as far as their nematic states are concerned. Because these fluids are textured, their rheology is unique. This rheology is determined by the coupling between the spatial heterogeneities in the nematic director and the flow. Another interesting result concerns the evidence of the director tumbling instability found for CPCl/Hex or C$_{16}$TAB nematics. For a better description of these systems under shear, additional experiments should be performed, such as those on the texture dynamics, and on the measurements of the Franck (elastic) and Leslie (viscosity) coefficients.
\section{\label{part4}Summary}
\indent

In this review, we provided an overview of the nonlinear rheology of surfactant wormlike micelles, from dilute to liquid crystal states. The equilibrium phase diagram of these systems is extremely rich. We showed that this leads, under steady shear flow, to different rheological signatures such as shear-thickening, shear-thinning or a combination of the two. These nonlinear behaviors are associated with strong modifications both of the flow field and of the internal structure of the fluid. The dilute phases of wormlike micelles thicken under shear due to the growth of a viscous shear-induced structure. The semi-dilute and concentrated phases of giant micelles undergo a shear-banding transition,  where phases of different fluidities, spatially organized, coexist. Finally, nematic phases of micelles present a tumbling instability of the director. All these transitions and instabilities share a common feature~:~they are characterized by extremely long transients compared to the intrinsic relaxation time of the system. These specific behaviors are related to rearrangements of the internal structure at the mesoscopic scale. Another crucial point that  emerged from recent experiments is that these transitions result in a complex spatio temporal dynamics of the flow, involving either bulk or interfacial instabilities, the driving parameters of which remain to determine.

Besides their practical interest in today's life, giant micelles are attractive, notably because of the accurate knowledge of their phase behavior and dynamical properties, the simplicity of their linear response and their analogy with conventional polymers. The major interest comes from the large diversity of their flow behaviors that continue to fascinate lots of experimentalists and theoreticians, especially because they are representative of many phenomena encountered in other complex fluids. 

\begin{acknowledgements}
The present review would not have been possible without the extended network of colleagues and friends being, as we are, fascinated by this subject. It is a pleasure to acknowledge the collaborations and the fruitful discussions we had over the years with Jacqueline Appell, Wesley Burghardt, Olivier Cardoso, Jean-Louis Counord, Jean-Paul Decruppe, Marc-Antoine Fardin, Olivier Greffier, Guillaume Gr\'egoire,  Heinz Hoffmann, S\'ebastien Manneville, Fran\c cois Molino, Julian Oberdisse, Peter Olmsted, Gr\'egoire Porte, Ovidiu Radulescu, Jean-Baptiste Salmon, Claudia Schmidt, Jean-Fran\c cois Tassin and Lynn Walker. The Laboratoire L\'eon Brillouin (CEA, Saclay, France), the Institute Laue-Langevin and the European Synchrotron Radiation Facilities (Grenoble, France) are also acknowledged for their technical and financial supports. We have also benefited from research organizations and fundings, such as the GDR 1081 \textquotedblleft Rh\'eophysique des Colloides et Suspensions\textquotedblright, European TMR-Network \textquotedblleft Rheology of Liquid Crystals\textquotedblright  contract number FMRX-CT96-0003 (DG 12 - ORGS), Agence Nationale pour la Recherche (ANR JCJC-0020). We are finally very grateful to S\'ebastien Manneville for his comments on the first version of the manuscript. 
\end{acknowledgements}

\bibliography{rheology,shearbanding2,shearthickening,part4,intro}   
\bibliographystyle{unsrt}

\end{document}